\documentclass[preprint,amsmath,amssymb,superscriptaddress,aps,final,dvipsnames,Table,final]{revtex4-1}
\usepackage{rotating}
\usepackage[dvipsnames]{xcolor}
 \usepackage[title]{appendix}
 \usepackage{ragged2e}
 \usepackage{float}
 \usepackage[labelfont=bf,labelsep= newline,justification=raggedright]{caption}
\usepackage{subcaption}
\usepackage{tabularx}
    \newcolumntype{L}{>{\raggedright\arraybackslash}X}
    \newcolumntype{C}{>{\centering\arraybackslash}X}
    \newcolumntype{R}{>{\raggedleft\arraybackslash}X}
    \usepackage{siunitx}
\usepackage{dcolumn}
\usepackage{bm}
\usepackage[]{color}
\graphicspath{ {.//} }
\definecolor{ultramarine}{RGB}{0,32,96}

\begin{document}


\title{Electrolytes for Li-ion all-solid-state batteries:
 a first-principles comparative study of Li$_{10}$GeP$_2$O$_{12}$ and Li$_{10}$GeP$_2$S$_{12}$ in the LISICON and LGPS phases
}

\author{Giuliana Materzanini}
\affiliation{
Theory and Simulations of Materials (THEOS) and National Centre for Computational Design and Discovery of Novel Materials (MARVEL), École Polytechnique Fédérale de Lausanne, CH-1015 Lausanne, Switzerland
}%
\author{Leonid Kahle}
\affiliation{
Theory and Simulations of Materials (THEOS) and National Centre for Computational Design and Discovery of Novel Materials (MARVEL), École Polytechnique Fédérale de Lausanne, CH-1015 Lausanne, Switzerland }

%
\author{Aris Marcolongo}
\affiliation{IBM RSM Zurich Research Laboratory, Zurich, Switzerland}

\author{Nicola Marzari}
\affiliation{
Theory and Simulations of Materials (THEOS) and National Centre for Computational Design and Discovery of Novel Materials (MARVEL), École Polytechnique Fédérale de Lausanne, CH-1015 Lausanne, Switzerland } 
%



%

\date{\today}
\vspace{10cm}
\clearpage
\newpage
\begin{abstract}
In this work we address Li-ion diffusion in thio-LISICON materials and in their oxide counterparts, exploring both the orthorhombic and tetragonal phases of Li$_{10}$GeP$_2$S$_{12}$ (LGPS) and Li$_{10}$GeP$_2$O$_{12}$ (LGPO) through extended 
Car-Parrinello molecular dynamics in the canonical and isobaric-isothermal ensemble.
The (quasi-)orthorhombic and tetragonal phases are studied both for the oxide and for the sulfide, with the aim of comparing their conductivity with the same approach; out of these four case studies, tetragonal LGPO has not been reported and, while dynamically stable, it sits (0.04 Ha/formula unit) above orthorhombic LGPO.
We calculate 
activation energies for diffusion of 0.18 eV and 0.23 eV for tetragonal and orthorhombic LGPS, and of 0.22 eV and 0.34 eV  for tetragonal and orthorhombic LGPO.
In line with experiments, we find orthorhombic LGPO orders of magnitude less conductive,  at room temperature, than the two sulfide systems. However, this is not the case for tetragonal LGPO, which, although less stable than its orthorhombic allotrope, shows at room temperature a conductivity comparable to orthorhombic and tetragonal LGPS, and, if 
synthesized, could make a very attractive 
Li-ion conductor.
\end{abstract}

\pacs{Valid PACS appear here}
\maketitle


\section{\label{Intro}Introduction}
More than two centuries after the beginning of the industrial revolution, a major societal challenge is to limit its reliance on fossil fuels by switching to renewable and green
energy sources
\cite{arunachalam2008global}, 
a goal that cannot be achieved unless
an efficient way to store electrical energy is provided \cite{kyriakopoulos2016electrical}.  Electrochemical storage is particularly suited to meet the needs of electrical grids powered by sun, wind or tides
\cite{yang2010enabling,yang2011electrochemical} and can also enable the electrification of transportation, providing the energy to automotive vehicles that are less dependent on 
fossil fuels, i.e. hybrid electrical, hybrid plug-in electrical and pure electrical vehicles \cite{tie2013review}, as well as the electrification of residential services, as for example in the popular case of the Tesla Home Battery Powerwall \cite{tesla}. 
\\
Among the existing electrochemical technologies, about 30 years after their first introduction in the market by Sony in 1991 \cite{yoshino1991secondary,nagaura1990lithium,blomgren2017development},  Li-ion batteries offer some of the best performance in terms of energy density, memory effects, near-reversibility and longer life cycles \cite{poizot2011clean,goodenough2013li} and 
have enabled the rise of portable electronics and furnished the power supply for safety systems, personal transportation, and digital technologies \cite{winter2004batteries}. Smart-grid integration and mass-market full electrification are the next frontiers, and for that the complex trade-off between energy density (and so range), power density, cost and safety of batteries needs to be addressed \cite{larcher2015towards,kwade2018current}, in addition to the availability of the raw materials \cite{olivetti2018toward}.
\\
A Li-ion battery cell exploits the redox activities of the anode and cathode to generate electricity outside the cell while reversibly intercalating Li ions between the two electrodes through an ionic conducting electrolyte \cite{goodenough2013li}. 
Due to dendrite formation and explosive failure with Li metal anodes \cite{whittingham1976electrical,goodenough2009challenges, blomgren2017development}, the anode is also
an insertion material as the cathode
 \cite{goriparti2014review,rozier2015li}, and   
 non-aqueous electrolytes are usually exploited, due to their wider electrochemical window compared to water-based electrolytes 
 ($\sim$ 4.2 V against $\sim$ 1.2 V \cite{fong1990studies,xu2004nonaqueous}).
 \\
  Energy density has been considered so far the preminent figure-of-merit in Li-ion batteries, and huge efforts have been put on the design of cathodes with high voltages and capacities \cite{melot2013design,whittingham2004lithium}, focusing on three main classes of cathodes, based respectively on the LiCoO$_2$ intercalation structure \cite{mizushima1980lixcoo2}, the LiMn$_2$O$_4$ spinel structure \cite{liu2014spinel} and the polyanion olivine structure \cite{padhi1997phospho}. For example, by partially substituting the Co cations with Ni, Mn or Al in LiCoO$_2$ and successively increasing the Li concentration in the material, the capacity of LiCoO$_2$ 
 was almost doubled \cite{thackeray2007li,wang2007reactivity,rozier2015li}, and partially replacing the Mn$^{3+}$/Mn$^{4+}$ couple with 
 Ni$^{2+}$/Ni$^{4+}$, Cr$^{3+}$/Cr$^{4+}$, Fe$^{3+}$/Fe$^{4+}$ and  Co$^{3+}$/Co$^{4+}$ in LiMn$_2$O$_4$ has given rise to  new generations of high-voltage (5 V) cathodes \cite{liu2014spinel,shigemura2002structural}.
 \\
 However, the effort to design high-energy cathodes cannot come without a parallel effort to address the safety and efficiency challenges that arise when the interaction between the current electrolytes and the electrodes is taken into account. 
 Examples are: the combustion reactions of high-capacity cathodes (Li$_{0.5}$CoO$_2$ and its derivatives) in the presence of ethylene or propylene carbonate electrolytes at temperatures above 180$^o$C \cite{macneil2002reactions, wang2007reactivity}, 
 the Mn dissolution of LiMn$_2$O$_4$ in the electrolyte with subsequent oxidation and degradation of the electrolyte and capacity loss of the cell \cite{amatucci1997materials},
 and in general the oxidations/reductions of the electrolyte if the  
 electrochemical potentials of the cathode/anode are outside the electrolyte stability window \cite{goodenough2013li}. Although the latter issue is spontaneously fixed by the formation of the solid electrolyte interface (SEI) that widens the electrochemical window of the electrolyte \cite{xu2004nonaqueous}, this interface is usually also responsible for the low ionic transport from the electrolyte to the electrode and vice-versa. Usually, electrolyte decomposition occurring at more than 4.2 V {\it{vs.}} Li$^+$/Li is considered the main reason for the capacity fading upon cycling  \cite{dedryvere2010electrode}.
 Finally, when the passivating interface is destroyed at high temperatures (even 80$^o$C), liquid organic electrolytes in contact with the positive material can form highly toxic fluoro-organic compounds  \cite{hammami2003lithium}. 
 All these issues make the use of state-of-the-art Li-ion batteries 
 (where the considerable size and weight of the batteries make these much more exposed to a fire risk during short circuits) 
 a technology still in need of safety and performance improvements, to the point that their original scope, i.e. addressing some core environmental problems,
 is still a matter of debate 
 \cite{larcher2015towards}.
 \\ Some safety challenges can in principle be addressed through an all-solid-state battery strategy, by substituting liquid organic with solid-state (crystalline, glassy or amorphous) inorganic electrolytes \cite{knauth2002solid,knauth2009inorganic,quartarone2011electrolytes,kim2015review,li2015solid,hu2016batteries,manthiram2017lithium,braga2017alternative,jiang2017recent,gao2018promises}. In addition to solving several safety issues and the problem of leakage presented by the currently used organic electrolytes, these materials have a lithium transport number very close to one \cite{quartarone2011electrolytes}, better thermal stability and wider electrochemical windows, and can enable the use of Li metal anodes \cite{bates1993fabrication,braga2017alternative}. Historically, the drawbacks have been a generally lower ionic conductivity with respect to organic electrolytes and a complex engineering of the electrolyte/electrode interface \cite{goodenough2009challenges}. 
 \\
 In this respect, sulfides proved to be an interesting choice, presenting a superior ionic conductivity with respect to many other materials. Moreover, their softness reduces grain boundary resistance and facilitates the manifacture of electrode/electrolyte interface \cite{hayashi2016development,sakuda2013sulfide}.
 \\
 The LISICON family of superionics, thoroughly investigated in the '60s, '70s and '80s \cite{shannon1977new,hong1978crystal,bruce1980phase,bruce1982ionic,rodger1985li+,abrahams1988structure,abrahams1989re,bruce1990defect}, was originally 
 composed of oxide materials, solid solutions between hcp-based $\gamma$-Li$_3$PO$_4$-structure 
 crystals (wurtzite-based edge-sharing tetrahedra with cations at the center and anions at the corners \cite{zemann1960kristallstruktur,west1975crystal,west1982tetragonal,o1978non}) and tetragonal close-packed Li$_4$YO$_4$ crystals (with Y = Ge, Si, Ti) \cite{vollenkle1968kristallstruktur,dittmar1976kristallstruktur,tranqui1979crystal,baur1981three}. Through X-ray and neutron diffraction techniques the structures were determined to be monoclinic \cite{tranqui1979crystal} or orthorhombic, both in the solid solutions and in the parent phases \cite{west1975crystal,o1978non,abrahams1991defect}.
Forming solid solutions such as Li$_{(3+x)}$Y$_x$P$_{(1-x)}$O$_4$ or Li$_{(2+2x)}$Zn$_{(1-x)}$GeO$_4$, i.e. tuning stoichiometries as Li$_3$PO$_4$ \cite{west1975crystal} or Li$_2$ZnGeO$_4$ (where the number of cations and anions is the same) \cite{bruce1982ionic} toward stoichiometries as Li$_4$YO$_4$ \cite{west1973ionic}, introduces Li interstitials in the pristine wurtzite structure   \cite{west1975crystal,bruce1982ionic} and favours ionic conductivity \cite{hong1978crystal}, that usually reaches a maximum at intermediate values of $x$ in the solid solutions (see, e.g., \cite{rodger1985li+}). 
 However, it became clear that replacing oxygen with sulphur, with a larger ionic radius and more polarizable character \cite{murayama2002synthesis}, could improve the conductivity. The thio-LISICON Li$_{(3+x)}$P$_{(1-x)}$Ge(Si)$_x$S$_4$ family, a solid solution between the parent phases Li$_3$PS$_4$ \cite{mercier1982structure,tachez1984ionic,homma2011crystal} and Li$_4$GeS$_4$ \cite{kanno2000synthesis}, was studied for glass, glass-ceramic  \cite{mizuno2006high,minami2006recent,ohtomo2013glass,ohtomo2013all,tatsumisago2013recent} and crystalline systems \cite{kanno2001lithium,murayama2002structure,murayama2004material}, showing superior conductivity with respect to the oxides, with the best conductivity at room temperature ($2\times 10^{-3}$ Scm$^{-1}$) reached for $x=0.75$ \cite{kanno2001lithium}.
 \\
 In an additional effort to improve conductivity, a new phase (tetragonal, space group $P4_2/nmc$, $\#$137)
 of thio-LISICONs at the composition $ x=0.67$ (named thereafter LGPS) was discovered in 2011 \cite{kamaya2011lithium}, with a record conductivity of $12\times 10^{-3}$ Scm$^{-1}$. 
 This inspired a new wave of efforts, both experimental \cite{kuhn2013tetragonal,kuhn2013single,bron2013li10snp2s12,kuhn2014new,kato2014synthesis,seino2014sulphide,whiteley2014empowering,hori2015phase,hori2015structure,hori2015synthesis,kwon2015synthesis,yang2015dual,kato2016high,weber2016structural} and theoretical \cite{mo2011first,adams2012structural,xu2012one,ong2013phase,lepley2013structures,zhu2015origin,wang2015design,zhu2016first,han2016electrochemical}, aiming to understand conduction mechanisms and to push the ionic conductivity even further. By tuning lithium, germanium and phosphorus compositions, a room-temperature conductivity of $14.2\times10^{-3}$ Scm$^{-1}$
 was reached \cite{kwon2015synthesis}, and by substituting germanium with silicon and simultaneously partially replacing sulphur with chlorine a value of
 $25\times 10^{-3}$ Scm$^{-1}$ was obtained \cite{kato2016high}. For an extensive up-to-date overview of LGPS and LGPS-like materials we refer the reader to the recent review by Kanno and coworkers \cite{katoli10gep2s12}. 
 \\
 However, apart from the high ionic conductivity that places these sulfide conductors at the same level of liquid ionic electrolytes, there are important drawbacks that cannot be disregarded when trying to deploy them as electrolytes for all-solid-state batteries. 
 First of all, LGPS and LGPS-like sulfides have very narrow thermodynamical electrochemical stability windows. Results from simulations \cite{zhu2015origin,mo2011first} and impedance spectroscopy \cite{sakuma2016reactions,wenzel2016direct} show that LGPS is chemically unstable below 1.71V vs Li$^+$/Li as a consequence of Ge and P cations reduction at Li metal \cite{wenzel2016direct} or Li alloy \cite{sakuma2016reactions} anodes, giving lithiation products Li$_{15}$Ge$_4$, Li$_3$P and Li$_2$S; it is also unstable above 2.14V due to the sulphur oxidation at the cathode, with delithiation products P$_2$S$_5$, GeS$_2$ and S \cite{han2015battery}. Both calculations and experiments agreed on the electrochemical stability of LGPS and pointed out its claimed wide (5V) electrochemical window \cite{kamaya2011lithium} to be essentially kinetic \cite{wenzel2016direct,zhu2015origin}. Although this interfacial instability 
 can in principle be dealt with coating oxide layers \cite{zhu2016first},
 the solid electrolyte interface, mainly formed due to the chemical instability at the anode of Ge in the LGPS structure \cite{sakuma2016reactions}, shows high resistance (4.6 $\Omega$cm$^2$ after one year \cite{wenzel2016direct}) and the requirement for high Li-ion conductivity across the interface remains a key challenge to be addressed \cite{lotsch2017relevance}. In principle neither a Li metal anode nor high-voltage cathodes (both targeting improved energy density) can be used with LGPS unless coatings are exploited, as is the case for the liquid electrolytes.
Moreover, sulfides show a well-known hygroscopicity as well as instability in air in the potential range
of normal Li battery operation \cite{tatsumisago2013recent,jung2015issues,hayashi2016development,ma2018recent}, they can provoke corrosion of the vacuum chamber \cite{kim2009li} and 
a careful suppression of hydrolysis is mandatory in order to avoid that these materials decompose and generate harmful (and lethal) H$_2$S.
\\
All these issues would suggest the need to turn to a safer electrolyte that does not require the engineering of a coating film protecting the electrodes (which is also the case for liquid electrolytes), while evaluating if this would hamper ionic conductivity. Safety, non-toxicity and no hazard-risks are also the main reasons justifying the interest in solid-state electrolytes, and when seeking for improved materials' performance one shouldn't forget the original goal of large-scale electrochemical storage, i.e. establishing a nontoxic, sustainable-energy economy. 
For the above reasons, oxygen-substituted Li$_3$PS$_4$ \cite{suzuki2016synthesis,wang2016oxygen,takada2005lithium,neveu2020exploration,minami2008electrical} and LGPS-like conductors \cite{sun2016oxygen,hori2016lithium,kim2019structures}, and in general 
oxysulfides \cite{wang2017oxysulfide}, as well as non sulphur-substituted LISICON materials \cite{song2018lithium,gilardi2020li4}, have been recently considered in the literature. Oxides are expected to be more electrochemical stable on the cathode side (for example, the cathode coating for LGPS should be in principle an oxide \cite{zhu2016first,lotsch2017relevance}), and the chemical stability of LGPS at the anode side is expected to improve by substituting sulphur with oxygen, which creates stronger covalent Ge-O bonds \cite{sun2016oxygen}. Indeed, chlorine-substituted LGPS-like oxides show large electrochemical windows up to 9 V \cite{song2015facile}, and oxysulfides show a wide electrochemical window of 5 V \cite{suzuki2016synthesis},
as predicted by  theory \cite{ong2013phase}. In addition, partially-oxygen-substituted LGPS shows room-temperature conductivity only slightly inferior to LGPS, and almost the same activation energy \cite{sun2016oxygen}. The latter result is at variance with calculations, that for the oxide analogue of LGPS,  Li$_{10}$GeP$_2$O$_{12}$ (LGPO), predict almost twice the activation barrier as for LGPS \cite{ong2013phase}. So far there is no experimental evidence for this phase of LGPO, but only for a LISICON material obtained as solid solution of $\gamma$-Li$_3$PO$_4$ and Li$_4$GeO$_4$ \cite{rodger1985li+} and having the same composition as LGPO, i.e. Li$_{4-x}$Ge$_{1-x}$P$_x$O$_{4}$ with $x$ = 0.67, whose structure, reported in Ref. \cite{rabadanov2003atomic}, belongs to the same space group as $\gamma$-Li$_3$PO$_4$ $Pnma$, $\#$62) and for which the moderate ionic conductivity of $1.8\times 10^{-6}$ Scm$^{-1}$ at 40$^o$C is reported in Ref. \cite{ivanov2003growth} (see also Refs. \cite{song2018lithium,gilardi2020li4}).
Thus, an in-depth investigation on structure and conductivity of the oxygen-substituted LGPS (tetragonal) and of its experimentally reported orthorhombic phase becomes important to investigate the feasibility of oxygen substitution to solve the above safety and instability problems of LGPS. In addition, whereas many simulations exist for the tetragonal (LGPS) phase \cite{marcolongo2017ionic,ong2013phase,mo2011first,adams2012structural}, so far there have been no theoretical studies on the less conductive, monoclinic (quasi orthorhombic) thio-LISICON phase \cite{kanno2001lithium}. Comparing oxides and sulfides in different phases would be also important in order to relate structure and anionic substitution to conductivity in these superionics.
\\
In this paper we address Li-ion diffusion in Li$_{10}$GeP$_2$O$_{12}$ and Li$_{10}$GeP$_2$S$_{12}$, both in the LGPS-like structure (tetragonal, space group $P4_2/nmc$, $\#$137)  and in the thio-LISICON structure (orthorhombic, space group $Pnma$, $\#$62) by means of first-principles molecular dynamics simulations. Diffusion coefficients for the four systems are extracted from long ($\sim$200 ps) trajectories, generated by CPMD within the NVE, NVT and NPT ensembles. 
In the remainder of this paper we will refer to Li$_{10}$GeP$_2$S/O$_{12}$ in the tetragonal phase (reported in \cite{kamaya2011lithium,kuhn2013single} for the sulfide material) as t-LGPS and t-LGPO respectively, and to Li$_{10}$GeP$_2$S/O$_{12}$ in the orthorhombic phase (reported in \cite{kanno2001lithium} for the sulfide and in \cite{rabadanov2003atomic} for the oxide material) as o-LGPS and o-LGPO respectively.
\\
The paper is organised as follows:
in Section \ref{Method} we present the details of the computational  methods 
together with the results for the equilibrium geometries and structure relaxation, and details of the electronic structure for the different systems. In Section \ref{sec:FC-MD} we give results for the diffusion from molecular dynamics simulations at fixed cell and constant temperature (NVT).
In Section \ref{sec:VC-MD} results from 
the variable-cell simulations (NPT) at the chosen temperature of 600 K are reported. A general discussion and the conclusions follow in Sec.~\ref{sec:DISC} and Sec.~\ref{sec:CONCL}, respectively.
\\
\section{\label{Method}Methods and static calculations}
\subsection{Methods}
\label{subsec:method}
We use Car-Parrinello (CP) molecular dynamics \cite{car1985unified}, based on Kohn-Sham density-functional theory (DFT) \cite{hohenberg1964inhomogeneous,kohn1965self} in the plane-wave pseudopotential formalism \cite{payne1992iterative,galli1993first}, as implemented in the cp code of the \textsc{Quantum ESPRESSO}  
distribution \cite{giannozzi2009quantum}. 
\\
Instead of solving self-consistently the Kohn-Sham equations at each MD step and following a trajectory on the Kohn-Sham Born-Oppenheimer (BO) energy surface $E^{KS}[\mathbf{R}]$ \cite{kohn1965self}, in CP one follows a trajectory on the fictitious energy surface of the coupled electron-ion Lagrangian,
that is a functional of both ionic degrees of freedom and electronic wavefunctions \cite{car1985unified}: 
\begin{equation}
  \mathcal{L}^{CP} = T_{ions} + T^{fict}_{el} 
  - E^{CP}[\{\psi_j\}, {\mathbf{R}}] 
   - \sum_{jk}\Lambda_{jk}\big(\int d\mathbf{r}{\psi}_j(\mathbf{r}){\psi_k}(\mathbf{r})-\delta_{jk}\big).
  \label{eq_lagrangian}
  \end{equation}
  Compared to the Lagrangian of the physical system, the Car-Parrinello Lagrangian 
  (reported in
  Eq.~(\ref{eq_lagrangian}) for the NVE case) 
   contains  $E^{CP}[\{\psi_j\}, {\mathbf{R}}]$ in place of $E^{KS}[\mathbf{R}]$
   (i.e. the ``instantaneous" Kohn-Sham energy for wavefunctions $\psi_j$ not necessarily on the Born-Oppenheimer surface),
  the Lagrange multipliers $\{\Lambda_{jk}\}$ to ensure orthonormality of the electronic wavefunctions, and the fictitious kinetic energy $T^{fict}_{el}$ of the electronic wavefunctions 
\begin{equation}
   T^{fict}_{el}
   = \mu\sum_{i}\int{\dot{\bigl|{\psi}_{i}\bigr|}}^2d\mathbf{r},
    \label{eq:Telectrons}
\end{equation}
that has no relation to the physical quantum kinetic energy of the electrons \cite{remler1990molecular}, but allows for a dynamical evolution of the Kohn-Sham states following the ionic motion.
The parameter $\mu$, whose dimensionality is [$m \times l^2$], or [$E \times t^{2}$], needs to be chosen sufficiently small in order to ensure that $T^{fict}_{el}$ is small compared to the kinetic energy of the ions $T_{ions} = \displaystyle\sum_{i}^{N_{ions}}\frac{1}{2}m_{i}\mathbf{v}_{i}^{2}$,
so as to avoid an irreversible transfer of energy from the ``hot" ions to the fictitious degrees of freedom,
that would gain
kinetic energy and move away from the Kohn-Sham surface during the dynamics \cite{galli1993first,pastore1991theory}.
However, a smaller $\mu$ implies a smaller time step in the integration of the equations of motion, to ensure accuracy and to keep numerically correct the constant of motion during the dynamics.
In the Supplemental Material we report these quantities for the NVE simulations of t-LGPO, and we show that the choice of $\mu = 500$ a.u. and time step $dt = 4$ a.u. satisfies the criteria discussed above. A similar analysis for t-LGPS is reported in the Supplemental Material of Ref. \cite{marcolongo2017ionic}, giving the same values for $\mu$ and $dt$. 
In analogy with Ref. \cite{marcolongo2017ionic} for t-LGPS, we employ for t-LGPO norm-conserving pseudopotentials \cite{hamann2013optimized,schlipf2015optimization} with a plane-wave cutoff E$_{cut}=80$ Ry. Thanks to the release of version 1.1 of the SSSP library \cite{prandini2018precision,lejaeghere2016reproducibility}, we could choose a lower cutoff E$_{cut}=50$ Ry and ultrasoft pseudopotentials (with 400 Ry cutoff for the electron density) \cite{garrity2014pseudopotentials} using the Standard Solid State Pseudopotential (SSSP)
Efficiency library 1.0 \cite{prandini2018precision} (GBRV \cite{garrity2014pseudopotentials} for Li, O and S,
PSLib \cite{dal2014pseudopotentials} for Ge and P) 
for o-LGPS and o-LGPO.
Brillouin zone integrations are performed using the Gamma point, and the exchange-correlation functional is PBE \cite{perdew1996generalized}. 
\\
In CP \cite{car1985unified,galli1993first} the electronic ground state at a given ionic geometry is first reached by performing 
damped dynamics  \cite{verlet1967computer,frenkel2001understanding} for the electronic wavefunctions  up to a certain kinetic energy threshold, 
that we choose as 1$\times$10$^{-11}$a.u.. 
Successively, damped ionic \cite{verlet1967computer,frenkel2001understanding} and cell  \cite{parrinello1981polymorphic} dynamics  follows to identify the relaxed minimum for the ionic positions and cell parameters. 

\subsection{\label{sec:supercells}Supercells and relaxed structures}
\begin{figure}
\vspace{1cm}
\centering
\begin{subfigure}[t]{0.6\textwidth}
  \includegraphics[width=1.\linewidth]{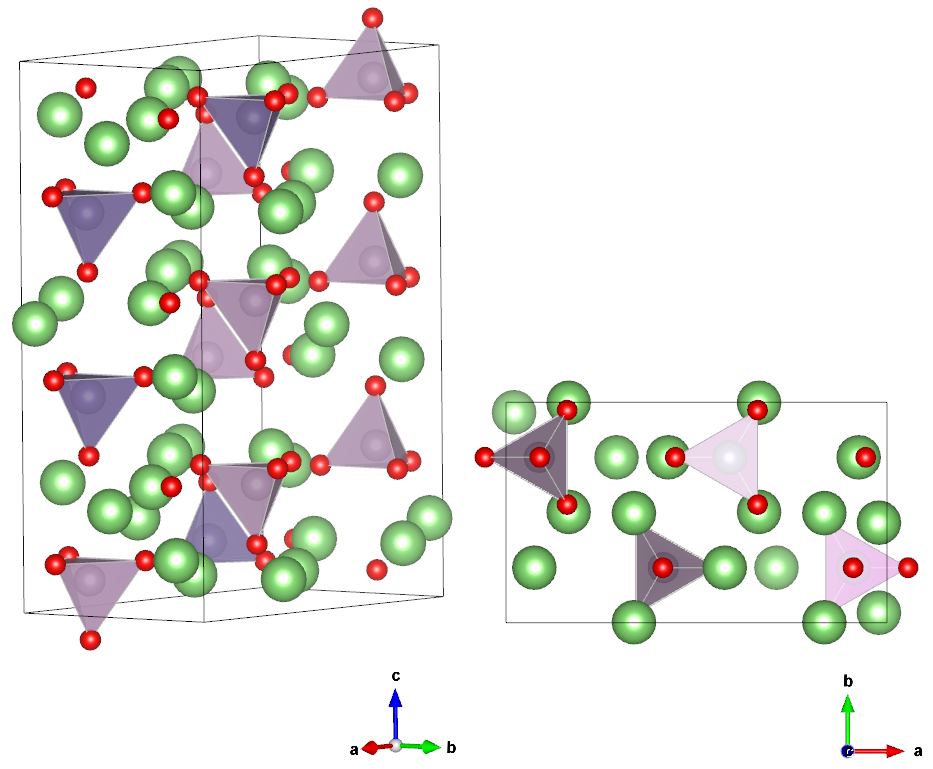}
        \caption{}
         \label{subfig:ortho}%
\end{subfigure} \hfill
       
   %
\begin{subfigure}[t]{0.6\textwidth}
\includegraphics[width=1.\linewidth]
{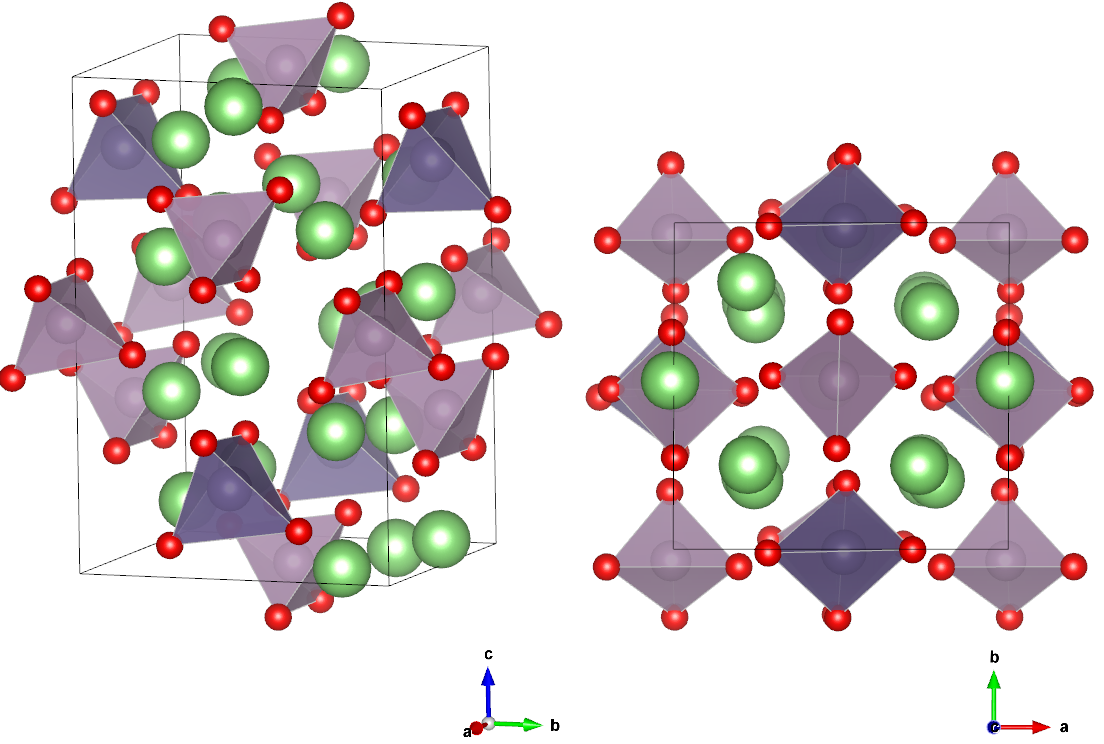}
    \caption{\small }
        \label{subfig:tetra}%
\end{subfigure}
        \vspace{1cm}
\caption{The 100-atom and 50-atom supercells (side and top view) used in the simulations for (a) o-LGPO and (b) t-LGPO from Refs. \cite{rabadanov2003atomic} and \cite{kamaya2011lithium}, respectively (see text). Li atoms are displayed
in green, O atoms are in red, and Ge and P atoms are at the center of the dark and 
light purple tetrahedra, respectively. The analogous LGPS supercells have
sulphur atoms replacing oxygen atoms.}
\label{fig:cells}
\end{figure}

For this work we select two supercells, one for the orthorhombic and one for the tetragonal case, and adapt each of these to the oxide and to the sulfide materials. 
For the orthorhombic structures we build a 100-atom  ($1\times1\times3$) supercell from the unitary cell (orthorhombic {\it{Pnma}}) reported in Ref. \cite{rabadanov2003atomic} for o-LGPO (Fig.~\ref{fig:cells}a), whereas for the tetragonal structures we use the 50-atom supercell already used in previous studies \cite{marcolongo2017ionic,ong2013phase}, starting from the crystallographic positions for the tetragonal cell ($P4_2/nmc$) of t-LGPS \cite{kuhn2013single,kamaya2011lithium} (Fig.~\ref{fig:cells}b). 
We couldn't find in the literature
(see also  \cite{adams2012structural}) space group and atomic positions of the monoclinic thio-LISICON structure reported in Fig. 2 and Table I of Ref. \cite{kanno2001lithium} for the solid solution Li$_{4-x}$Ge$_{1-x}$P$_{x}$S$_4$, but we note that this structure is very similar 
to the
orthorhombic {\it{Pnma}} structure reported for o-LGPO in Ref.~\cite{rabadanov2003atomic}  (Fig.~\ref{fig:cells}a). We thus
adapt the o-LGPO structure \cite{rabadanov2003atomic} to the thio-LISICON volume at $x$ = 0.67 \cite{kanno2001lithium}, to build the o-LGPS supercell.
Furthermore, we use the volume ratio between the two orthorhombic cells \cite{rabadanov2003atomic,kanno2001lithium}
to determine the t-LGPO volume from the experimental t-LGPS volume \cite{kuhn2013single,kamaya2011lithium}.
\\
In Table \ref{tab:lattice_parameters} we report the calculated lattice parameters and angles for LGPO and LGPS in the tetragonal and orthorhombic supercells considered in this work, obtained through CP damped-dynamics relaxations (cfr. Sec. \ref{subsec:method}). 
The convergence criteria used here are an energy difference between two consecutive steps below $5\times 10^{-7}$ a.u. ($\simeq 1.4 \times 10^{-5}$ eV) and forces on the ions below $3\times 10^{-5}$ a.u. ($\simeq 1.5 \times 10^{-3}$ eV/\AA).
Since the CP method is restricted to $\Gamma$-only 
sampling, we also provide the optimized geometry obtained through variable-cell relaxation with the Broyden–Fletcher–Goldfarb–Shanno algorithm (pw code  \cite{giannozzi2009quantum}) with converged  grids of $\mathbf{k}$ points (444 or 343) and the same parameters (pseudopotentials and cutoff energy) used in the CP calculations.
The available experimental results \cite{kamaya2011lithium,kanno2001lithium,rabadanov2003atomic} are also listed, for comparison. 
The agreement with the experimental results is in general very good for the the $\mathbf{k}$-point-converged relaxed cell geometries, but is also  satisfactory 
for the $\Gamma$-sampled  CP-relaxed cell geometries. 
For o-LGPS, we point out that the experimental parameters of Ref. \cite{kanno2001lithium} refer to a (1$\times$3$\times$3) monoclinic cell 
with $\theta = 0.65$ (here $\theta = 0.67$), which might be slightly different from the (1$\times$1$\times$3) orthorhombic supercell of o-LGPO from Ref.~\cite{rabadanov2003atomic}, that we have adapted here to o-LGPS. Finally, in our calculations no symmetry restrictions are imposed to the cell during the optimization, leading to {\it quasi}-orthorhombic and {\it quasi}-tetragonal cells, with angles very close, but not necessarily equal to $\ang{90}$.

\subsection{Density of states and band gaps}
\begin{figure}[tb]
\vspace{1cm}
\centering
\begin{subfigure}[t]{0.7\textwidth}
  \includegraphics[width=1.\linewidth]{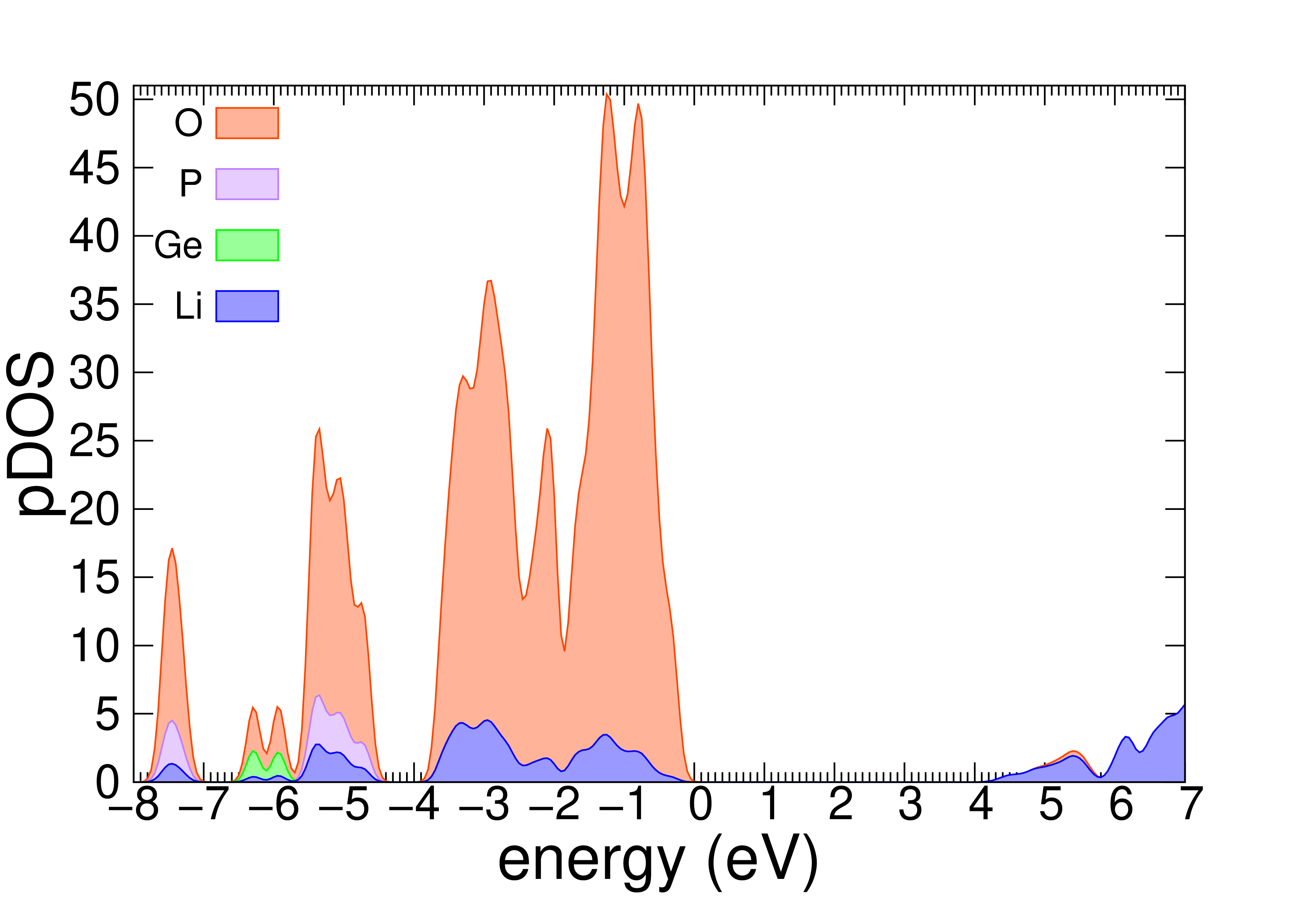}
        \caption{}
         \label{subfig:DOS_LGPO_tetra}%
\end{subfigure} \hfill
\begin{subfigure}[t]{0.7\textwidth}
  \includegraphics[width=1.\linewidth]{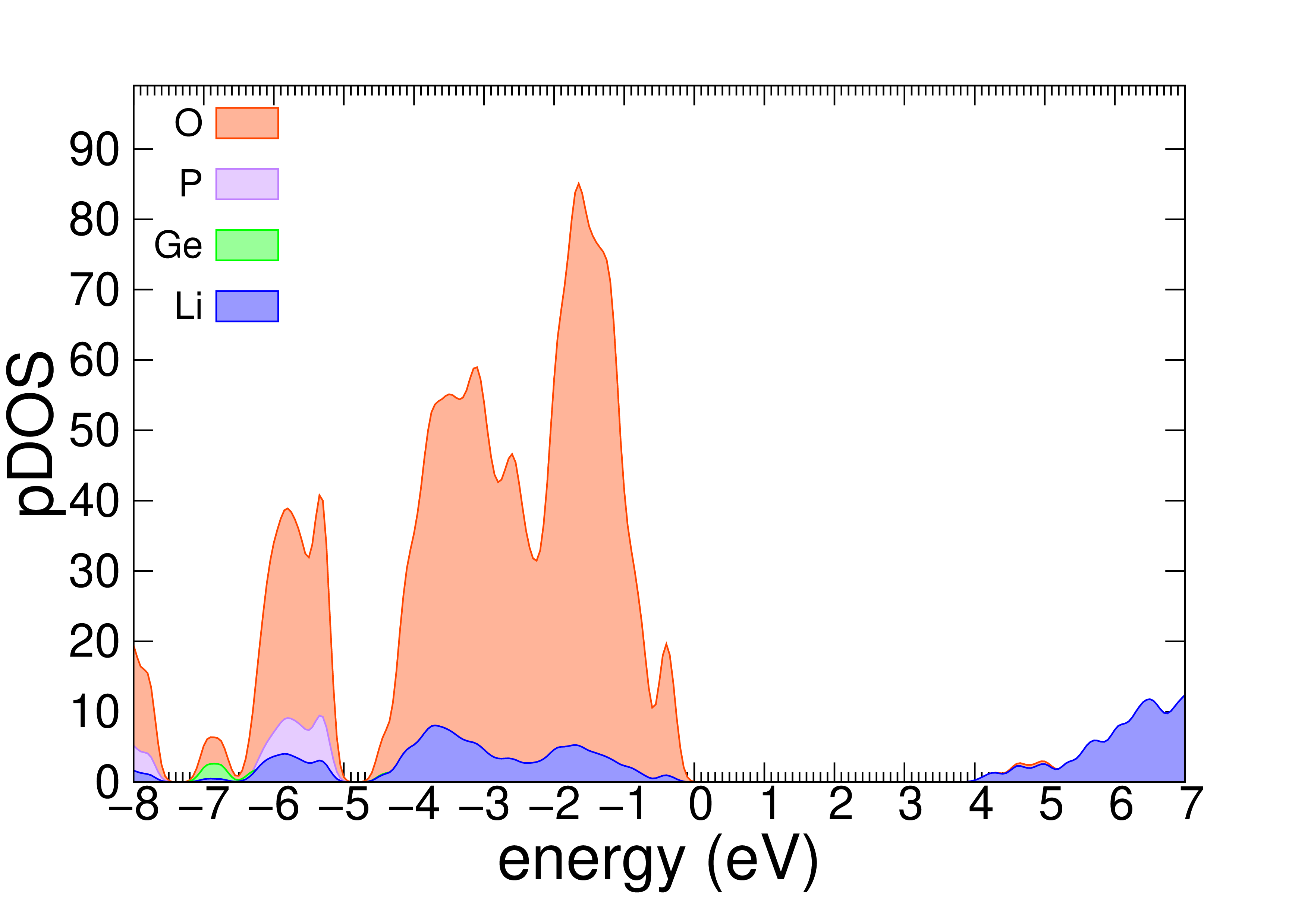}
        \caption{}
         \label{subfig:DOS_LGPO_ortho}%
\end{subfigure}
\vspace{1cm}
 \caption{Projected DOS for a) t-LGPO and b) o-LGPO at the PBE level. Energies are reported with respect to the highest occupied level. A  Kohn-Sham band gap larger than 4 eV is present. }
 \label{DOS_t-LGPO}
 \end{figure}

\begin{figure}[tb]
\vspace{1cm}
\centering
\begin{subfigure}[t]{0.7\textwidth}
  \includegraphics[width=1.\linewidth]{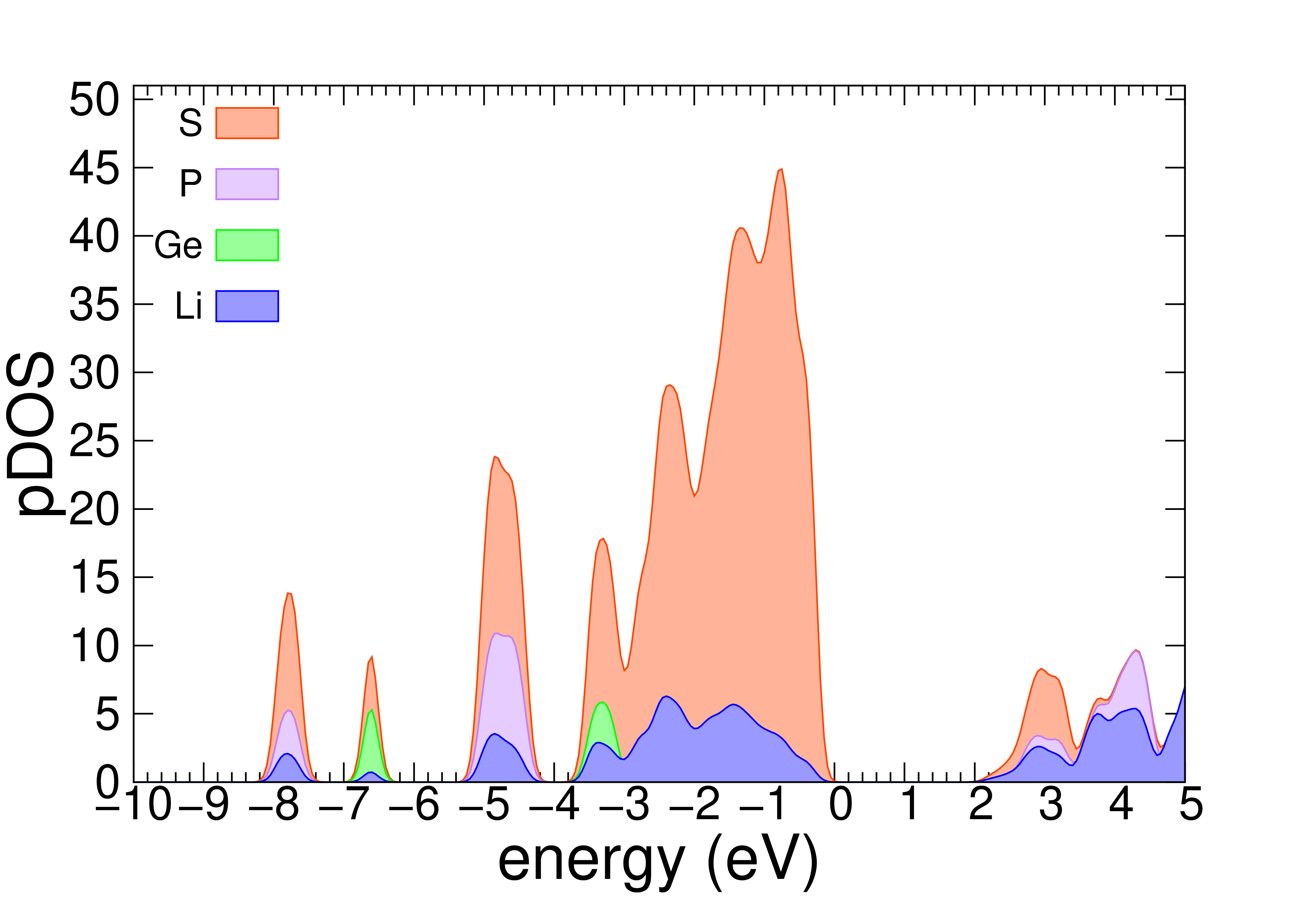}
        \caption{}
         \label{subfig:DOS_LGPS_tetra}%
\end{subfigure} \hfill
\begin{subfigure}[t]{0.7\textwidth}
  \includegraphics[width=1.\linewidth]{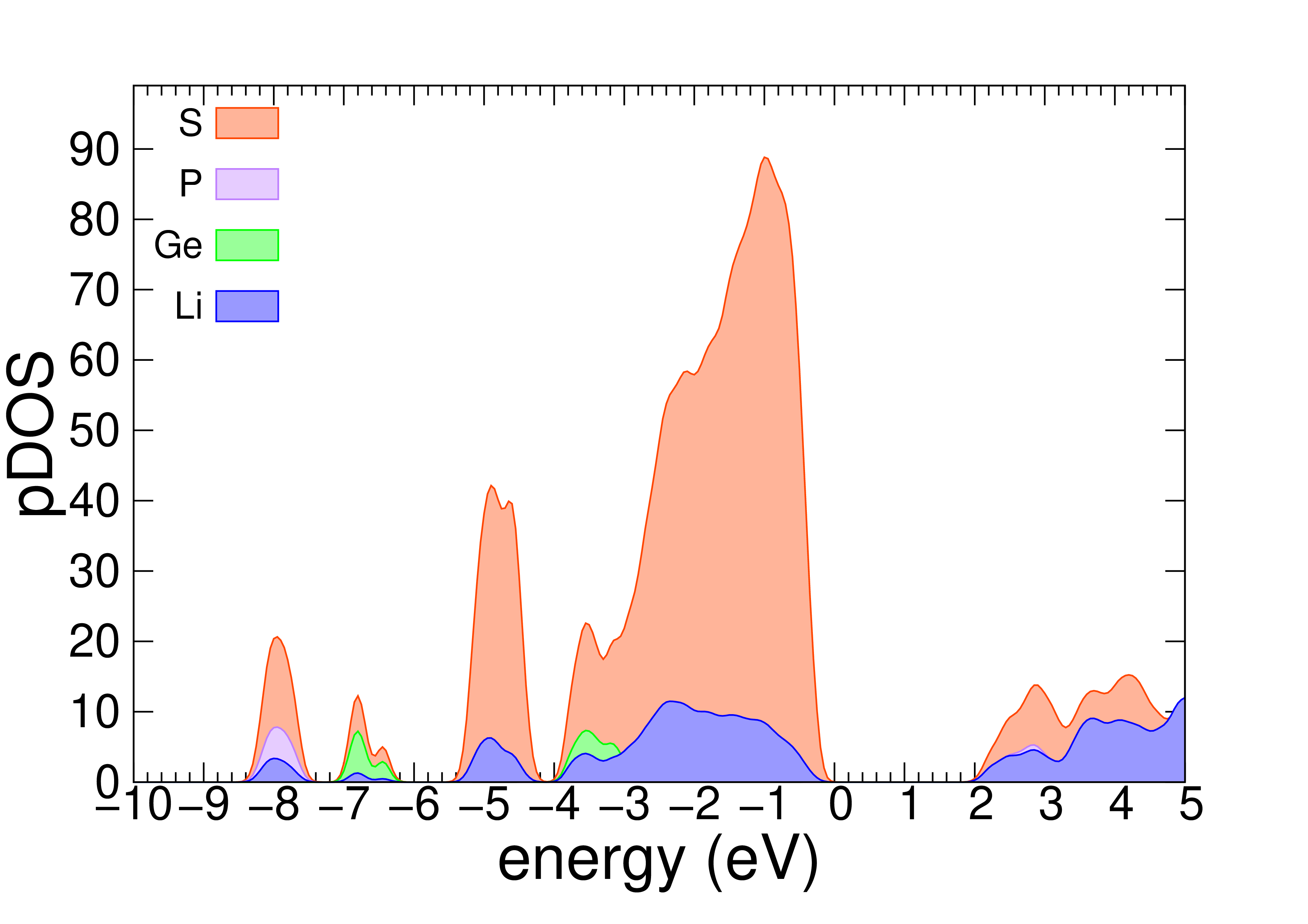}
        \caption{}
         \label{subfig:DOS_LGPS_ortho}%
\end{subfigure}
\vspace{1cm}
 \caption{Same as Fig.~\ref{DOS_t-LGPO} for a) t-LGPS and b) o-LGPS. Energies are reported with respect to the highest occupied level. A Kohn-Sham band gap of $\sim$2 eV is present.}
 \label{DOS_t-LGPS}
\end{figure}

The band gap of a superionic material is a useful property that gives an estimate of its electrochemical window of stability as a solid-state electrolyte \cite{goodenough2013li}. We calculate band structures and densities of states (DOS) at the Kohn-Sham PBE level (thus, only qualitatively accurate) and evaluate band gaps for the different structures considered in this work, with the aim of shedding light on their relative electrochemical stabilities as electrolytes in a battery.
In Figs.~\ref{DOS_t-LGPO} and \ref{DOS_t-LGPS} the projected DOS (p-DOS, i.e. the DOS projected over the different {\it{n, l, m}} components and then summed for any given atom) are displayed for LGPO and LGPS respectively, both in the tetragonal and orthorhombic phases.
The p-DOS displayed are taken from the converged SCF density at the relaxed geometry with full $\mathbf{k}$-points sampling (see Table~\ref{tab:lattice_parameters}), but
very similar results were obtained from the $\Gamma$-point relaxed geometry (band gaps converged within 0.1 eV). A number of bands $\sim 20 \%$ larger than the number of filled bands is used (302 and 151 bands for the orthorhombic and tetragonal structures, respectively).
Our calculations show the sulfides, having about half the band gap with respect to the oxides, are thus expected to have significantly smaller electrochemical windows. Our findings are supported by the aforementioned experimental results reporting a 9 V electrochemical stability window for the chlorine-substituted LISICON Li$_{10}$Si$_{1.5}$P$_{1.5}$Cl$_{0.5}$O$_{11.5}$ \cite{song2015facile} and a 5 V electrochemical stability window  for the oxysulfide Li$_{3+5x}$P$_{1-x}$S$_{4-z}$O$_z$ \cite{suzuki2016synthesis} and also by previous calculations comparing t-LGPO and t-LGPS \cite{ong2013phase}.
\\
\section{\label{sec:FC-MD}Fixed-cell NVT and NVE Molecular Dynamics}
\subsection{\label{subsec:MDcomp} Details of the simulations}
We perform Car-Parrinello molecular dynamics simulations at fixed volume, in cells defined by the experimental crystallographic data  for LGPO and LGPS in the tetragonal and orthorhombic structures \cite{rabadanov2003atomic,kamaya2011lithium}, subsequently relaxed (cfr. Sec.~\ref{sec:supercells}).
Simulations are carried out in the NVT ensemble, with a time step of 4~a.u. ($\sim$ 0.1 fs, see Sec.~\ref{subsec:method}). All atoms are free to move 
and a Nose-Hoover thermostat \cite{martyna1992nose} is attached to the system for at least the first 160~ps. 
\\
In the Supplemetal Material we discuss the thermostat parameters (frequency and atomic-type specificity) and we test the effect of the thermostat on the simulated diffusion properties for t-LGPO. We choose a species-specific thermostat working at a frequency of 17 THz and we show that the thermostat used here doesn't perturb the dynamics to a significant degree.
We thus run NVT simulations for t-LGPO at 8 temperatures between 600 K and 1200 K (600, 635, 720, 771, 830, 900, 1080, 1200 K), for o-LGPO at 6 temperatures between 600 K and 1200 K (600, 720, 900, 1000, 1100, 1200 K) and for o-LGPS at 4 temperatures between 600 K and 1200 K (600, 720, 900, 1200 K).
For t-LGPS we also compare the NVE data (after 160ps thermostatting) obtained by two of us with the same method and published in Ref. \cite{marcolongo2017ionic}, the t-LGPS simulations being done at 9 temperatures between 520 and 1200 K. 
\subsection{\label{subsec:MDdiffusion}Tracer diffusion}
\label{subsec:tracer_diffusion}
\begin{figure}[t]
\vspace{1cm}
\centering
    \begin{subfigure}[t]{0.4\textwidth}
     \includegraphics[width=1.\linewidth]{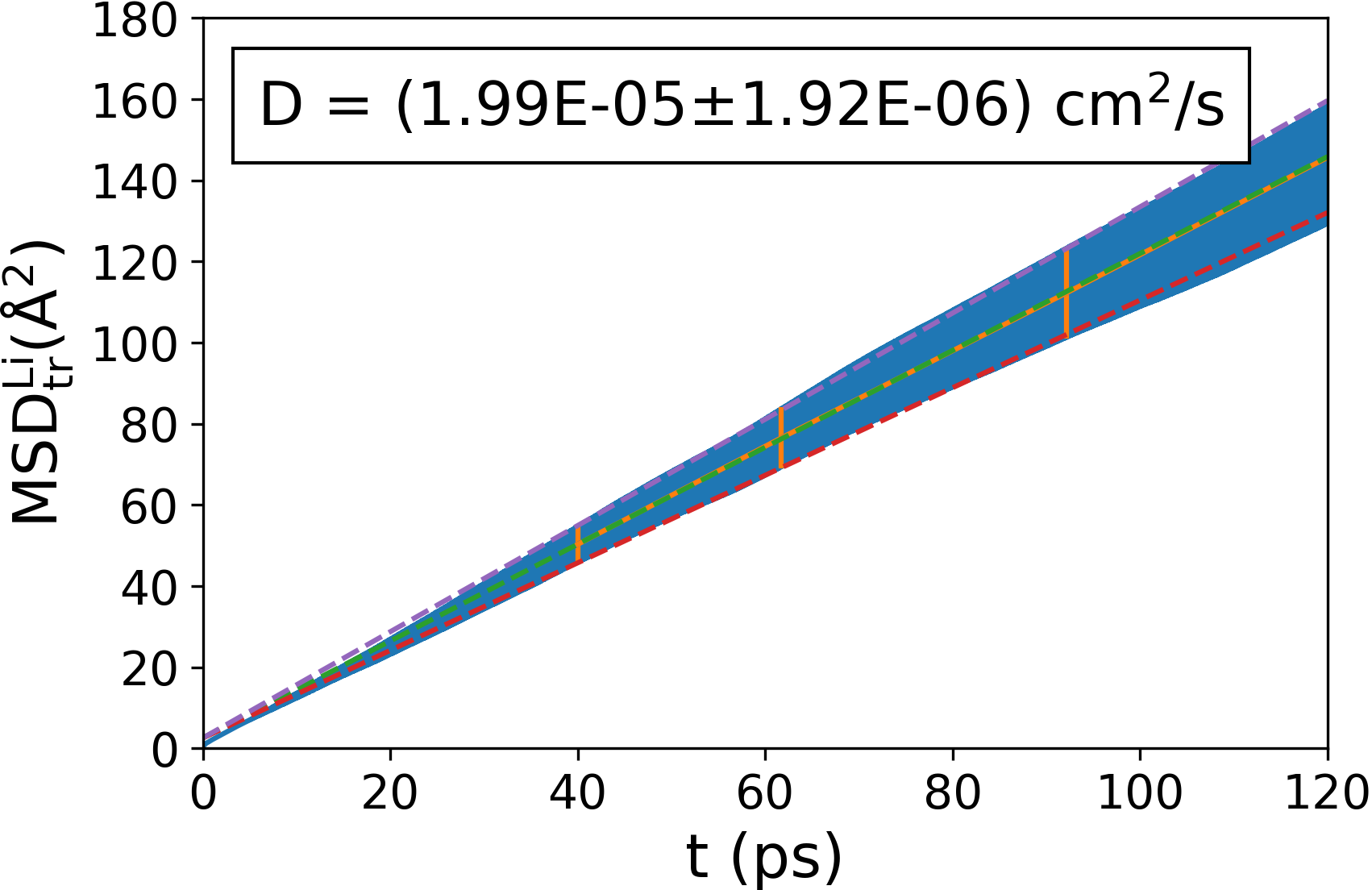}%
        \label{subfig:MSD-t-LGPS}
        \caption{t-LGPS}
      \end{subfigure}
      \hfill
          \begin{subfigure}[t]{0.4\textwidth}
          \includegraphics[width=1.\linewidth]{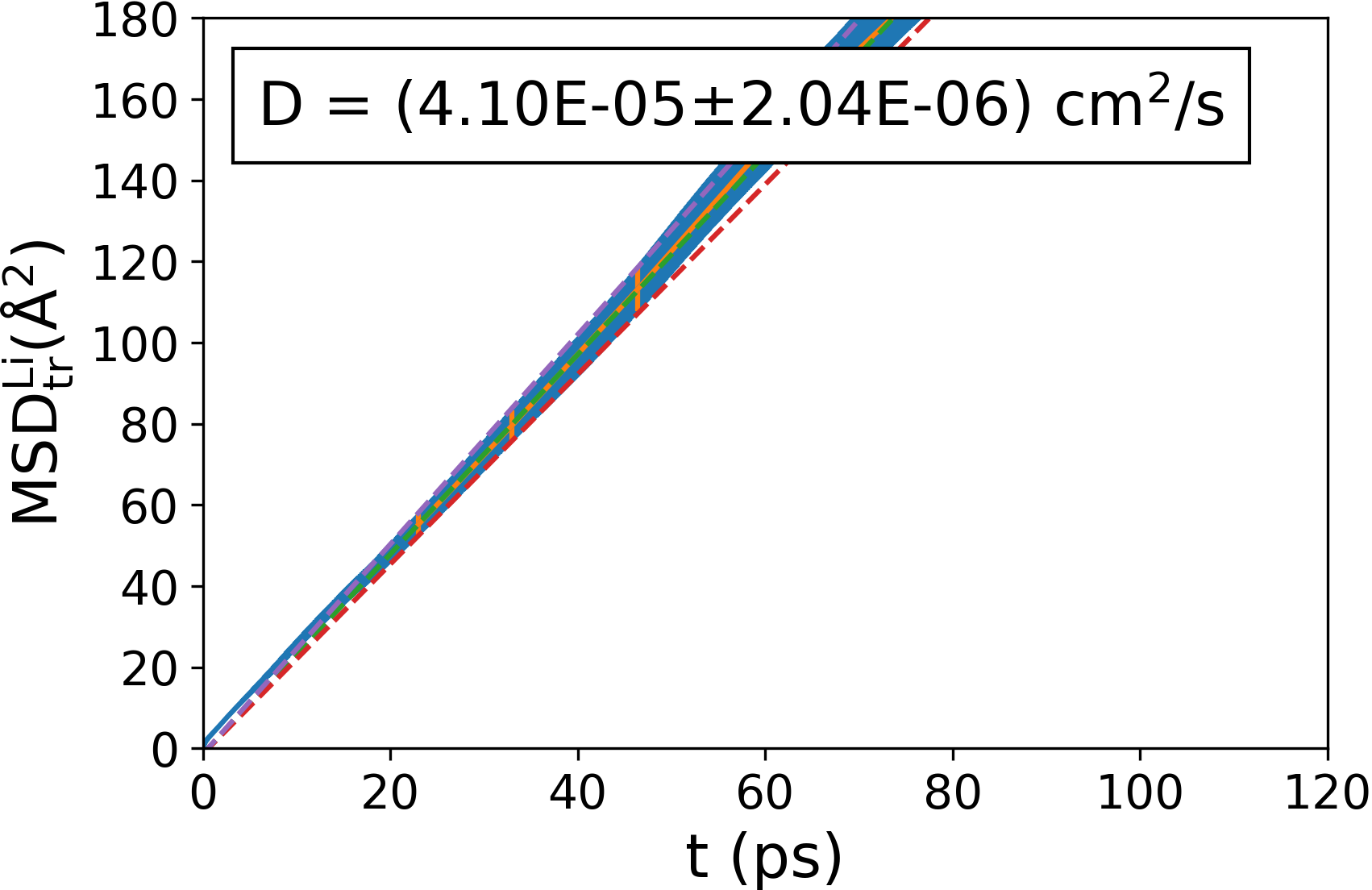}
            \label{subfig:MSD-o-LGPS}
               \vspace*{-7mm}
           \caption{o-LGPS}
    \end{subfigure}
    \vfill
        \begin{subfigure}[t]{0.4\textwidth}
            \includegraphics[width=1.\linewidth]{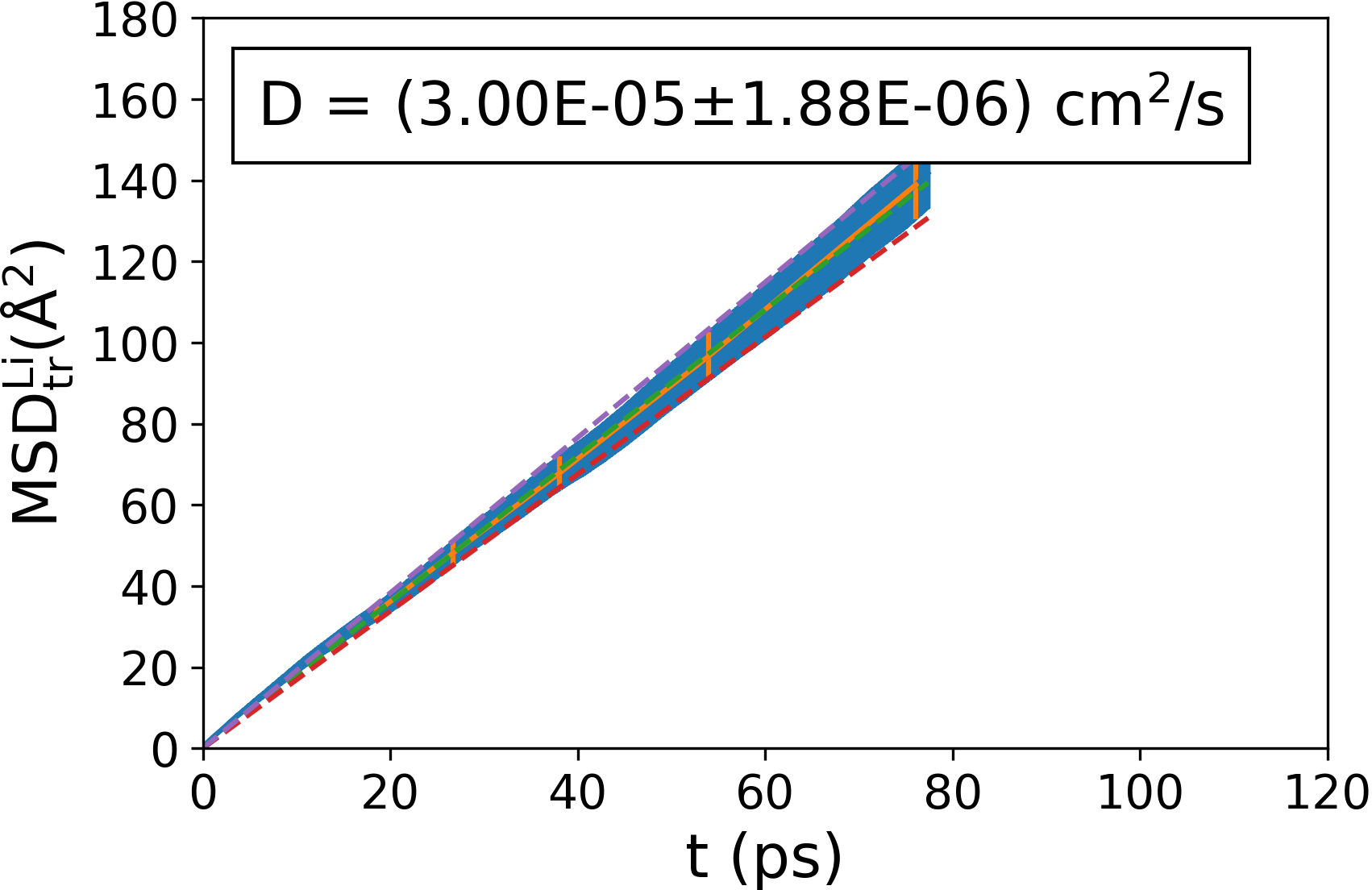}             
            \label{subfig:MSD-t-LGPO}
            \vspace*{-7mm} \caption{t-LGPO}
        \end{subfigure}
        \hfill
            \begin{subfigure}[t]{0.4\textwidth}
            \includegraphics[width=1.\linewidth]{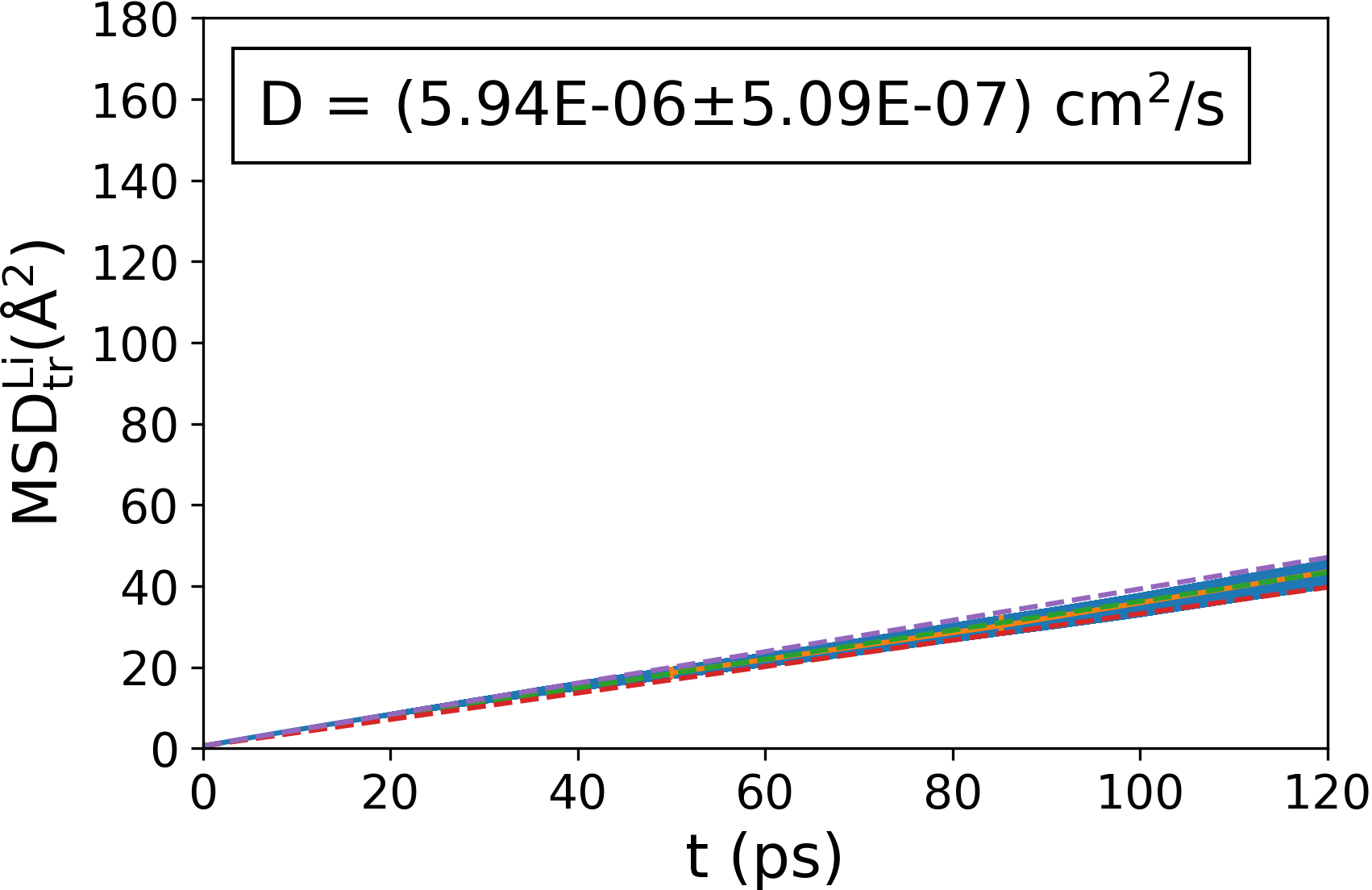}
            \caption{o-LGPO}
            \label{subfig:MSD-o-LGPO}
        \end{subfigure}
        \vspace{1cm}
           \caption{ 
Li-ion tracer mean square displacements in the four structures considered in this work, at 900K (NVE simulations for t-LGPS \cite{marcolongo2017ionic}, NVT simulations for t-LGPO, o-LGPS and o-LGPO).}
\label{fig:MSD}
\end{figure}

\begin{figure}
\vspace{1cm}
\centering
\includegraphics[width=1.\linewidth]{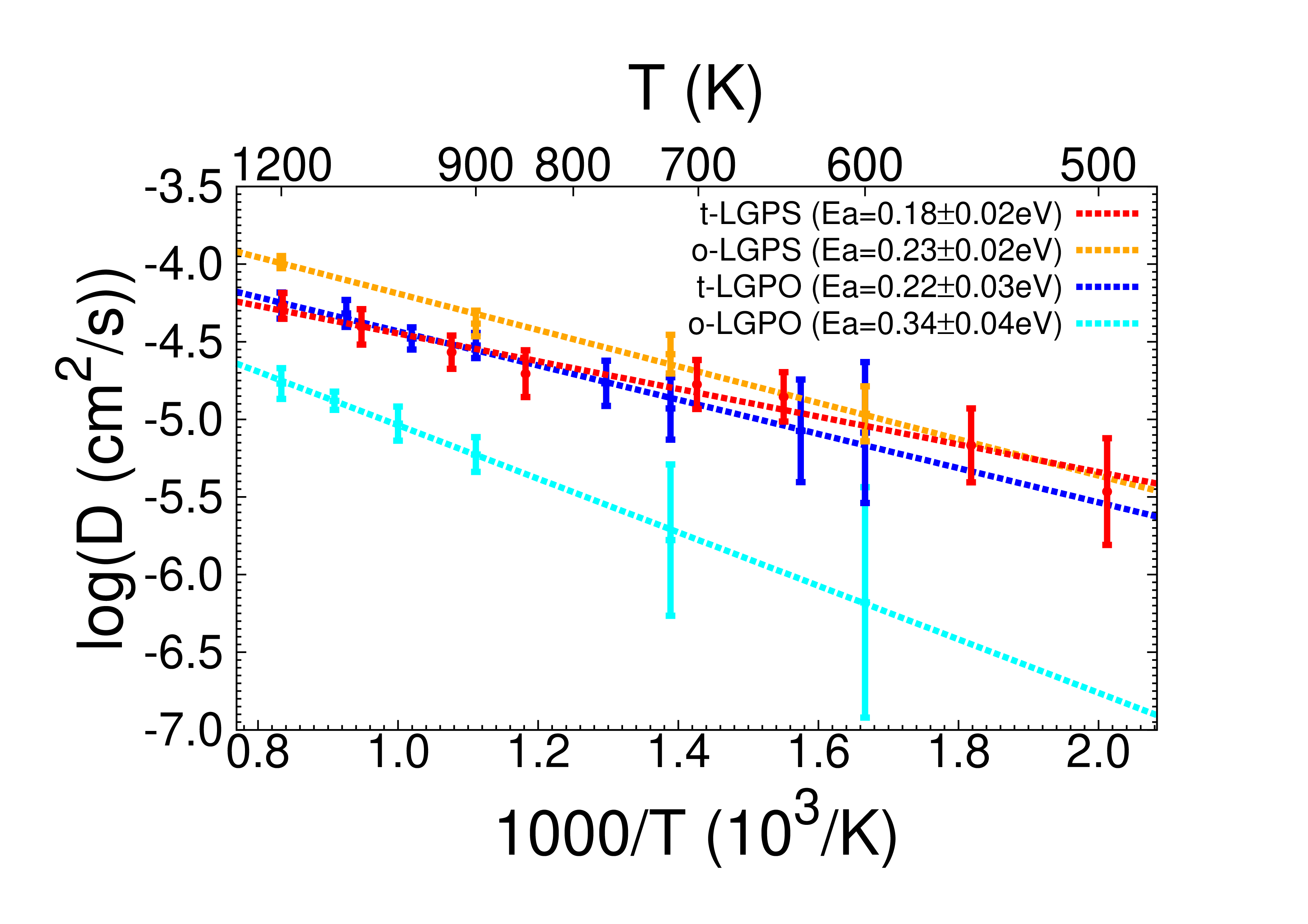}
\vspace{1cm}
    \caption{Arrhenius plots and corresponding activation energies for Li-ion tracer diffusion coefficients in the four structures considered in this work.}
    \label{fig:diff_all}
\end{figure}

The mean square displacement 
of a given ionic species is a measure of its mobility in a material and can be put in direct relationship with the self-diffusion coefficient. The latter is also known as (and will be referred to from now on) the tracer diffusion ($D_{tr}$) coefficient, as it can be compared to pulsed-field gradient 
nuclear magnetic resonance experiments \cite{price1997pulsed}. According to the Einstein relation for diffusion \cite{einstein1905molekularkinetischen,chandler1987introduction,frenkel2001understanding}, the tracer diffusion coefficient in 3 dimensions is:
\begin{equation}
D_{tr}^{Li} = \lim_{t\to\infty} \frac{1}{6} \frac{d}{dt} MSD_{tr}^{Li}(t) ,
\label{eqD}
\end{equation}
where the Li-ion tracer mean square displacement $MSD_{tr}^{Li}(t)$ in a system of $N_{Li}$ ions over a sufficiently long time $t$ (no ballistic regime) is \cite{chandler1987introduction}
\begin{equation}
MSD_{tr}^{Li}(t) = \frac{1}{N_{Li}} \displaystyle\sum_{i}^{N_{Li}} \big\langle \big|\mathbf{R}_{i}(t^\prime +  t)]-\mathbf{R}_i(t^\prime)\big|^{2}\big\rangle
\label{eqMSD}
\end{equation}
with $\mathbf{R}_{i}$ being the instantaneous position of the $i$-th Li ion
and $\big\langle...\big\rangle$ an average
over the times $t^\prime$.
\\
From the NVT simulations (Sec.~\ref{subsec:MDcomp}) 
we calculate  $MSD_{tr}^{Li}(t)$ (Eq.~(\ref{eqMSD})) and $D_{tr}^{Li}$ (Eq.~(\ref{eqD})) at all temperatures considered.  For the method employed to calculate ${MSD}_{tr}^{Li}(t)$, $D_{tr}^{Li}$ and  their statistical uncertainties \cite{he2018statistical} we refer the reader to the Supplemental Material \cite{frenkel2001understanding,allen2017computer}.
In Fig.~\ref{fig:MSD} we report  
${MSD}_{tr}^{Li}(t)$ and the corresponding $D_{tr}^{Li}$ for the four structures considered, at the representative temperature of 900 K.
At this temperature we note that diffusion is equally fast, approximately, in t-LGPO as t-LGPS and o-LGPS, while being much slower in o-LGPO.
Results for the other temperatures (not shown here) display a similar trend.
\noindent
We consider diffusion here as an activated process, that obeys an Arrhenius law \cite{atkins2013elements} 
\begin{equation}
ln D_{tr}(T) = ln A - \frac{E_{a_{D_{tr}}}}{k_B T} 
\label{eqEactivation}
\end{equation}
where the constant $A$ is related to the attempt frequency whereas the activation energy $E_{a_{D_{tr}}}$ is the energy barrier for the diffusion.
We report the Arrhenius plot for $D_{tr}^{Li}$, in the widely used form $log D_{tr}(1000/T)$, in  Fig.~\ref{fig:diff_all}.
As observed in Fig.~\ref{fig:MSD} for diffusion at one temperature, also the temperature dependence of the diffusion coefficients (Fig.~\ref{fig:diff_all}) reveals basically two different cases: highly diffusive systems (t-LGPS, o-LGPS and t-LGPO), and a less diffusive one (o-LGPO). 
This latter finding is in qualitative agreement with results from impedance spectroscopy  measurements for o-LGPO, reporting a rather high activation energy of $\sim 0.54$~eV  \cite{ivanov2003growth,gilardi2020li4}.
\\
From the tracer coefficient $D_\mathrm{tr}$ the ionic conductivity $\sigma$ (entirely ascribed to the Li ions) can be calculated according to
the Nernst-Einstein equation:
\begin{equation}
    \sigma(T) = \frac{N_{Li} Z_{Li}^2 e^2}{V k_B T} D_{tr} = \frac{N_{Li} e^2}{V k_B T} D_{tr}.
    \label{eqsigma}
    \end{equation}
In Eq.~\eqref{eqsigma} $N_{Li}/V$ is the density of the charge carriers (i.e. the Li ions), $Z_{Li} e = e$ their charge (being $e$ the elementary charge and assuming that the average Born effective charge is +1, see also Ref. \cite{grasselli2019topological}), and $D_{tr}$ is the self-diffusion coefficient from Eq.~\eqref{eqD}. 
From Eq.~\eqref{eqsigma}, assuming negligible the change of volume with temperature, the temperature dependence for $D_{tr}$ is the same as for $\sigma T$.

\subsection{\label{subsec:MDdiffusioncharge}Collective diffusion}
\begin{figure}[t]
\vspace{1cm}
\centering
\begin{subfigure}[t]{.5\textwidth}
    {%
        \includegraphics[width=1.\linewidth]{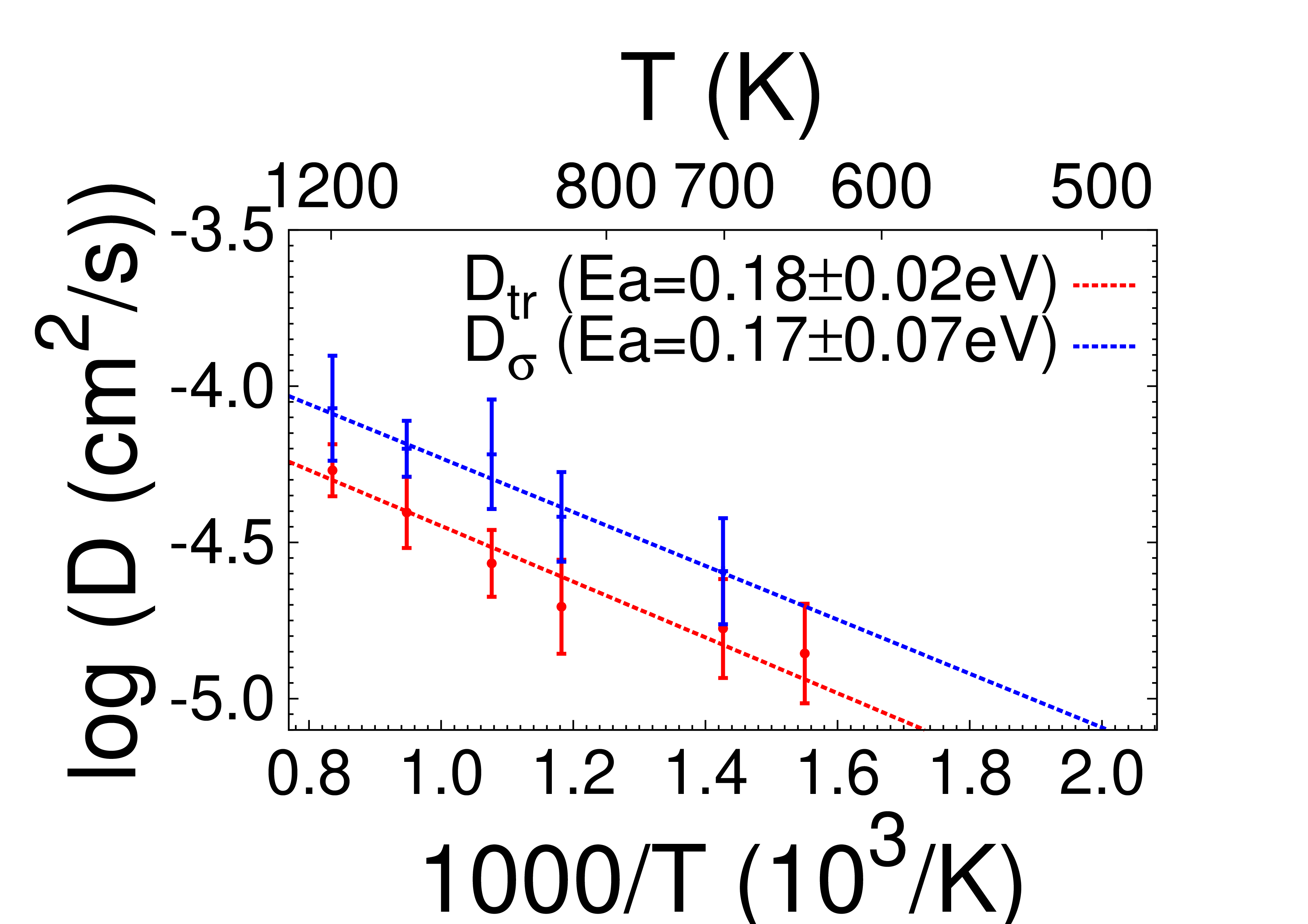}%
        \caption{t-LGPS}
        \label{subfig:HR-t-LGPS}%
        }%
\end{subfigure}\hfill
\begin{subfigure}[t]{.5\textwidth}

        \includegraphics[width=1.\linewidth]{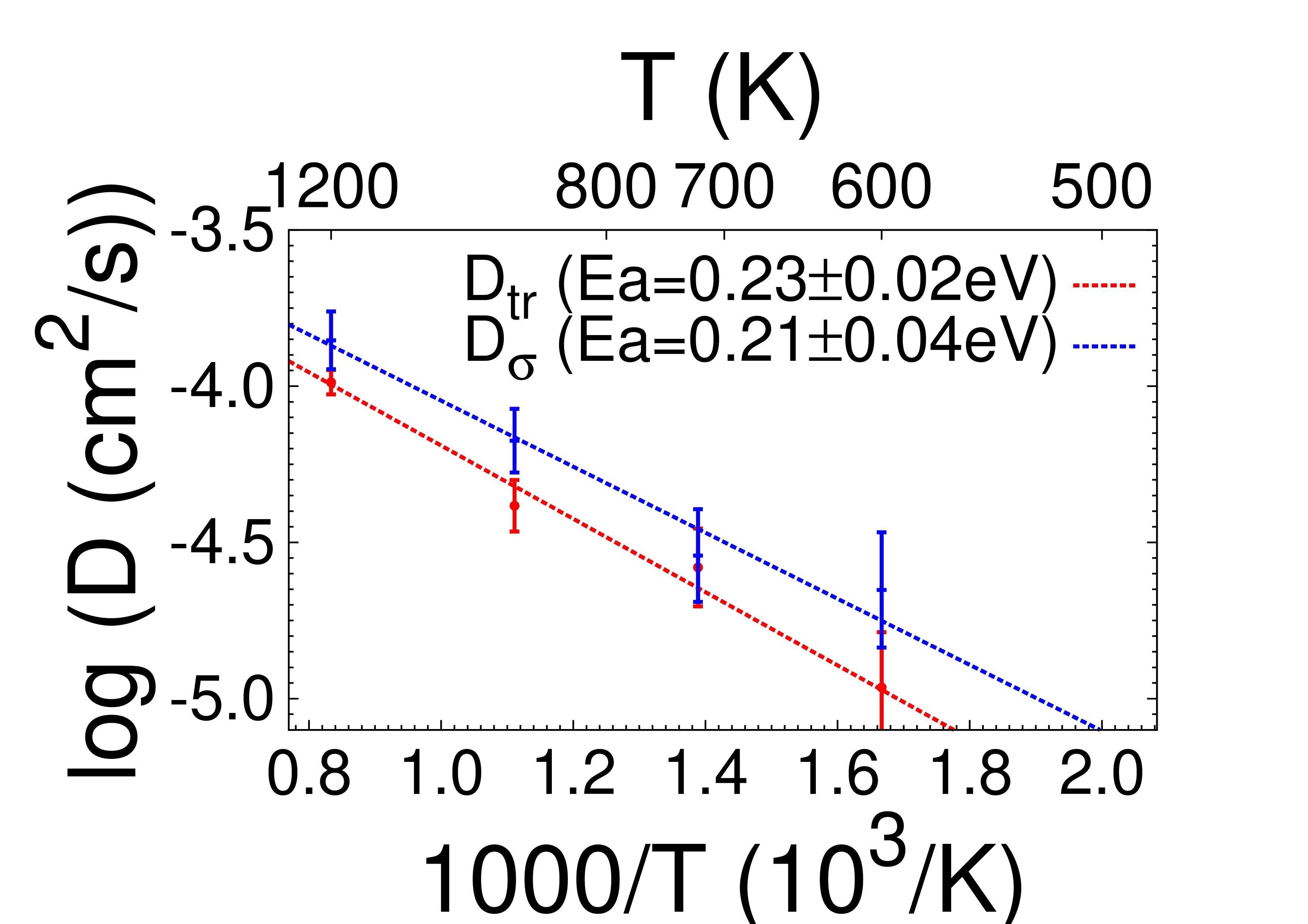}%
        \caption{o-LGPS}
        \label{subfig:HR-o-LGPS}%
\end{subfigure}
\begin{subfigure}[t]{.5\textwidth}
        \includegraphics[width=1.\linewidth]{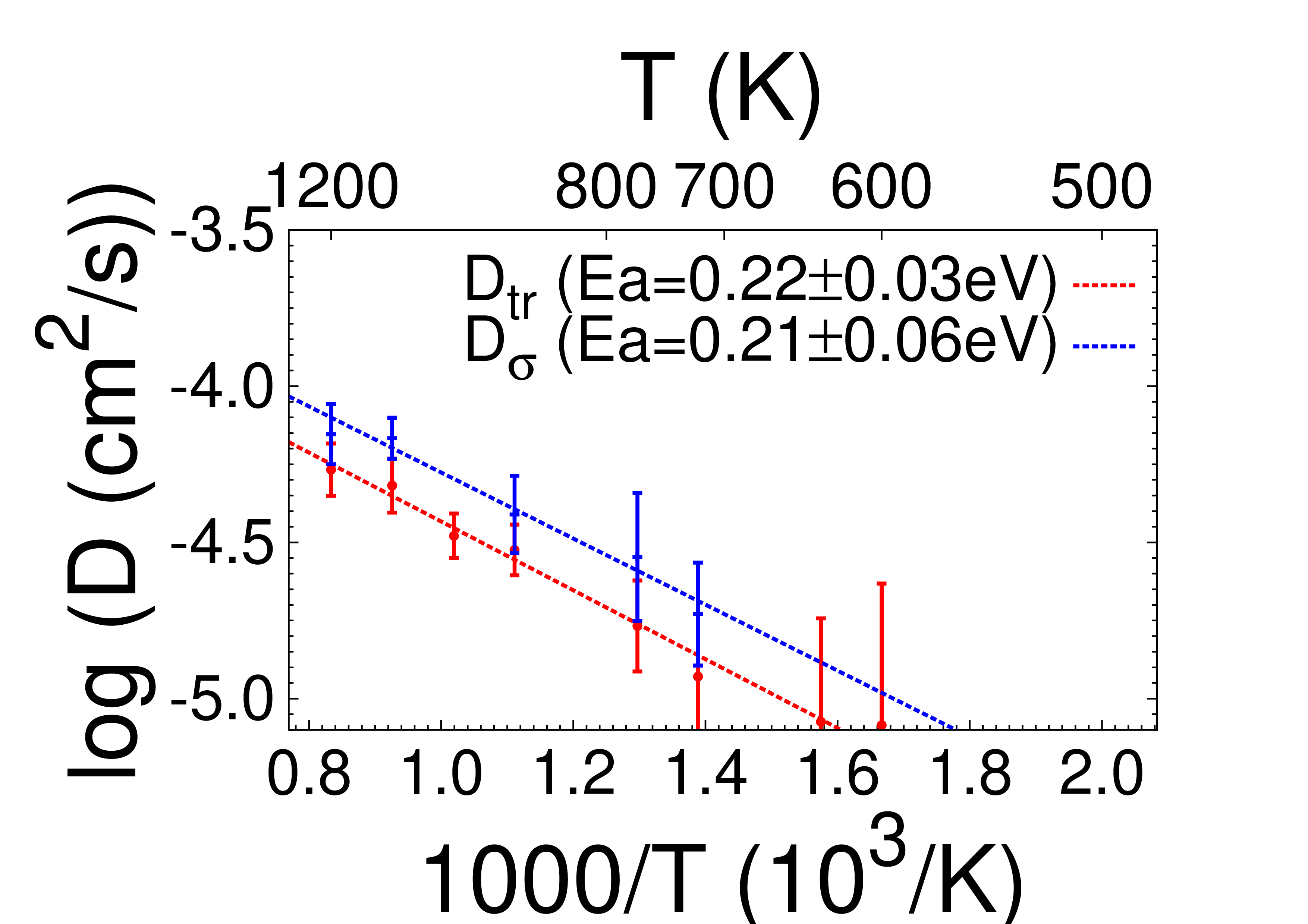}%
        \caption{t-LGPO}
        \label{subfig:HR-t-LGPO}%
\end{subfigure}\hfill
\begin{subfigure}[t]{.5\textwidth}
        \includegraphics[width=1.\linewidth]{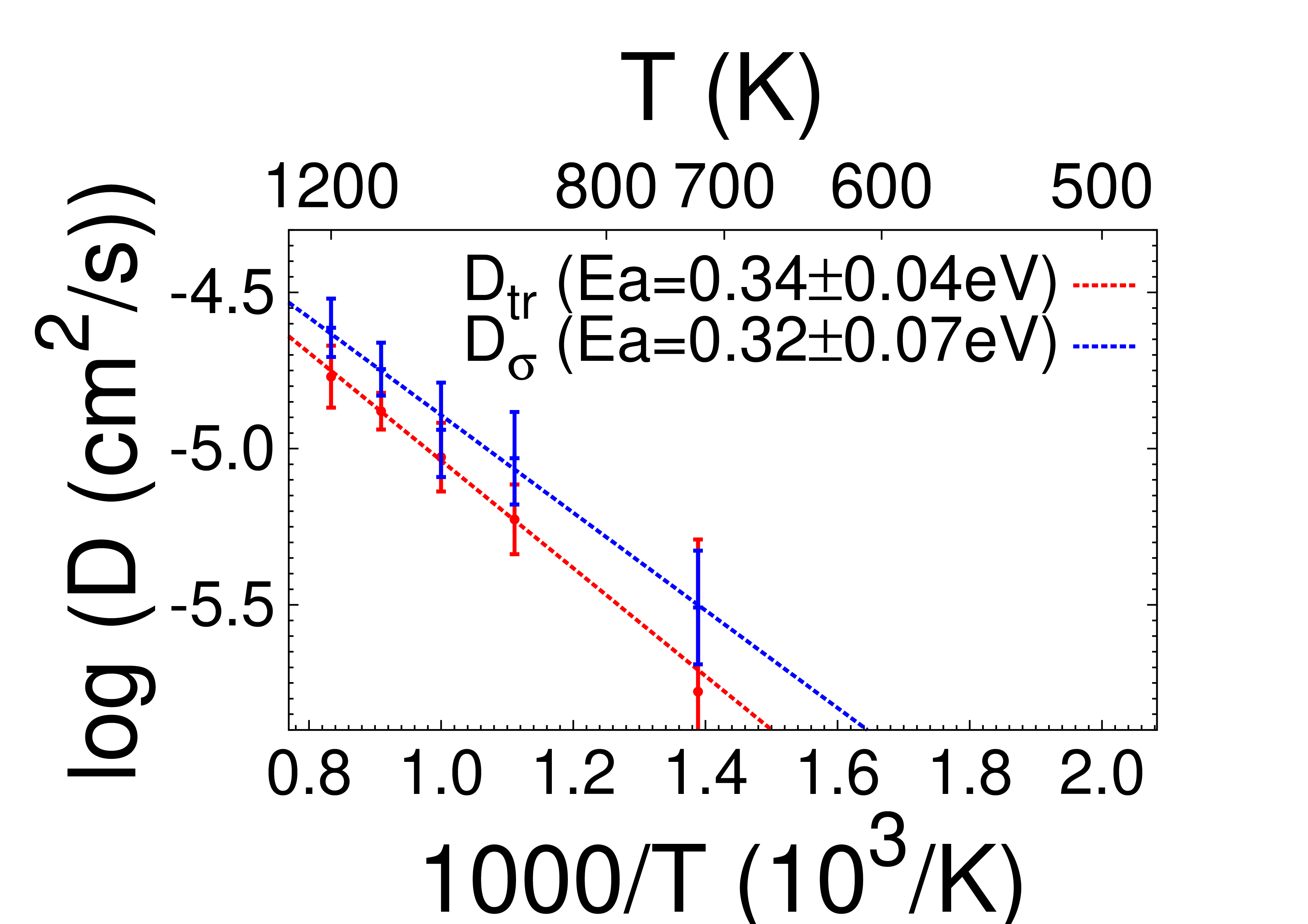}%
        \caption{o-LGPO}
        \label{subfig:HR-o-LGPO}%
\end{subfigure}
\vspace{1cm}
    \caption{Arrhenius plots for Li-ion tracer and charge ($\sigma$) diffusion in the four structures studied.} 
\label{fig:HR}
\end{figure}

As pointed out in Refs.~\cite{marcolongo2017ionic,french2011dynamical,harris2010relations}, the tracer diffusion coefficient of Eq.~(\ref{eqD}) 
assumes independent uncorrelated contributions from all the ions. The so-called charge diffusion coefficient $D_\sigma$ \cite{france2019correlations,afandak2017ion} provides a more realistic estimate, as it accounts for ion-ion correlated diffusion. For a Li ion in a system of $N_{Li}$ ions, it can be written \cite{france2019correlations,afandak2017ion}:
\begin{equation}
D_{\sigma}^{Li} = \lim_{t\to\infty} \frac{ MSD_{\sigma}^{Li}(t
)}{6t} , 
\label{eqDsigma}
\end{equation}
where
\begin{equation}
MSD_{\sigma}^{Li}(t) = 
\frac{1}{N_{Li}} \displaystyle\sum_{i,j}^{N_{Li}} \big\langle \big[\mathbf{R}_{i}(t^\prime +  t)-\mathbf{R}_i(t^\prime)\big]  \big[\mathbf{R}_{j}(t^\prime +  t)-\mathbf{R}_j(t^\prime)\big] \rangle
\label{eqMSDcharge}
\end{equation}
so that the corrected Nernst-Einstein equation \cite{marcolongo2017ionic} includes the collective $MSD_\sigma(t)$ (cfr. Eq.~\eqref{eqMSDcharge}) through $D_{\sigma}$ (cfr.Eq.~\eqref{eqDsigma}):
\begin{equation}
    \sigma(T) = \frac{N_{Li} e^2}{V k_B T} D_\mathrm{\sigma}.
    \label{eqsigmacharge}
    \end{equation}
It is worthwhile to mention that $D_{tr}$ and $D_\sigma$ have been here derived in an Einstein form (i.e. from the mean square displacement), whereas in Ref.\cite{marcolongo2017ionic} the same quantities are reported in the 
Green-Kubo formalism \cite{kubo1957statistical}.
\\
\noindent
    The Arrhenius plots for Li-ion tracer (Eq.~\ref{eqD}) and charge (Eq.~\ref{eqDsigma}) diffusion in the four systems are displayed in Fig.~\ref{fig:HR}.
    In Table \ref{tab:activation_energies} the activation energies (Eq.~(\ref{eqEactivation})) from $D_{tr}$ and $D_\sigma$ 
    are reported, together with the available experimental results for $\sigma T$ \cite{kamaya2011lithium,kanno2001lithium,ivanov2003growth} and the results from first-principles calculations in the literature \cite{ong2013phase}.
    \\
    As already noted in Ref.~\cite{marcolongo2017ionic} for t-LGPS, also for t-LGPO, o-LGPS and o-LGPO the activation energy remains approximately unchanged when including ion-ion correlations, i.e. comparing $E_{a_{D_{tr}}}$ with $E_{a_{D_\sigma}}$ 
    (Fig.~\ref{fig:HR} and Table~\ref{tab:activation_energies}). We also note that the statistical errors for $D_{\sigma}$ are systematically higher than for $D_{tr}$, due to the well known slower convergence of the former in the time averages \cite{marcolongo2017ionic}. 
   Fig.~\ref{fig:HR} and Table~\ref{tab:activation_energies} confirm that o-LGPO has moderate conducting properties, in qualitative agreement with experiments \cite{ivanov2003growth} that place this material among the candidate solid electrolytes for microbatteries \cite{gilardi2020li4}. The diffusivity  of tetragonal LGPO, however, is some orders of magnitude superior with respect to its orthorhombic allotrope \cite{rabadanov2003atomic,ivanov2003growth} at all temperatures, at variance with the results reported in Ref. \cite{ong2013phase} 
   and comparable to its more studied sulfide analogues  \cite{kamaya2011lithium,kanno2001lithium}, as partially anticipated by a previous computational study exploiting an ab-initio trained frozen-lattice potential \cite{kahle2018modeling}. The latter finding could hopefully open the way to experimental attempts aimed at synthesizing LGPO in the same phase as t-LGPS \cite{kamaya2011lithium,kuhn2013tetragonal}.

\section{Variable$-$cell NPT Molecular Dynamics}
\label{sec:VC-MD}
\subsection{\label{subsec:VCdetails}Details of the simulations}
Based on the results for diffusion, we aim here at getting more insight into the relative stability of the oxide and the sulfide systems in the orthorhombic and tetragonal structures considered in this work. For this purpose, we set up variable-cell molecular dynamics (VC-MD) simulations for each of the four systems, allowing both size and shape of the cell to change during the dynamics, within the 
Parrinello-Rahman approach, to sample the isobaric-isothermal ensemble \cite{parrinello1980crystal}.
The supercell is coupled to a barostat that keeps constant the pressure of the cell through an interaction term $-p_{ext}V$ (where $p_{ext}$ is the desired external pressure, here = 0, and $V$ is the volume of the cell), while adding to the Lagrangian a kinetic energy term that accounts for the cell motion  \cite{andersen1980molecular,parrinello1980crystal}. 
The atomic positions are expressed in scaled coordinates in the basis of the time-dependent cell vectors  \cite{parrinello1981polymorphic}. 
The mass of the barostat is chosen in such a way that the frequency of the cell volume fluctuations multiplied by $V^{1/3}$ is of the order of the free-particle sound velocity $\sqrt{(k_B T)/M}$ \cite{parrinello1981polymorphic,nose1983constant}, where $M$ is the total mass of the atoms in the cell, as suggested in the original work by Andersen  \cite{andersen1980molecular}. 
In addition, temperature is controlled during the simulations through a Nose-Hoover thermostat \cite{martyna1992nose} already discussed in Section \ref{subsec:MDcomp}; we choose the temperature of 600 K for these simulations.
 
\subsection{\label{subsec:VCenergetics}Volumes and energies}
\begin{figure}
\vspace{1cm}
\centering
\begin{subfigure}[t]{0.7\textwidth}
        \includegraphics[width=1.\linewidth]{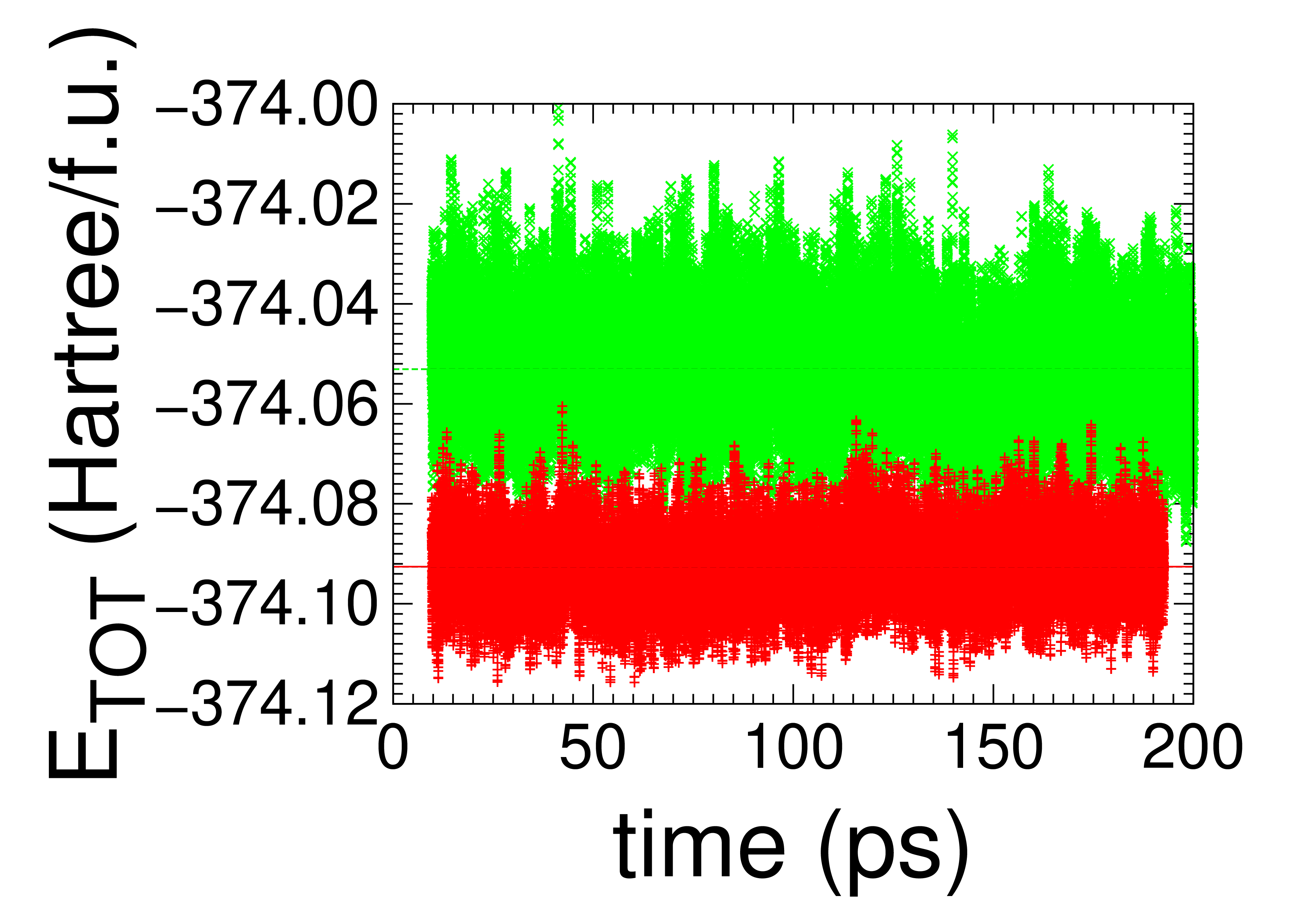}
        \caption{t- and o-LGPO}
        \label{subfig:LGPO_total_energies}
\end{subfigure}\hfill
\begin{subfigure}[t]{0.7\textwidth}
        \includegraphics[width=1.\linewidth]{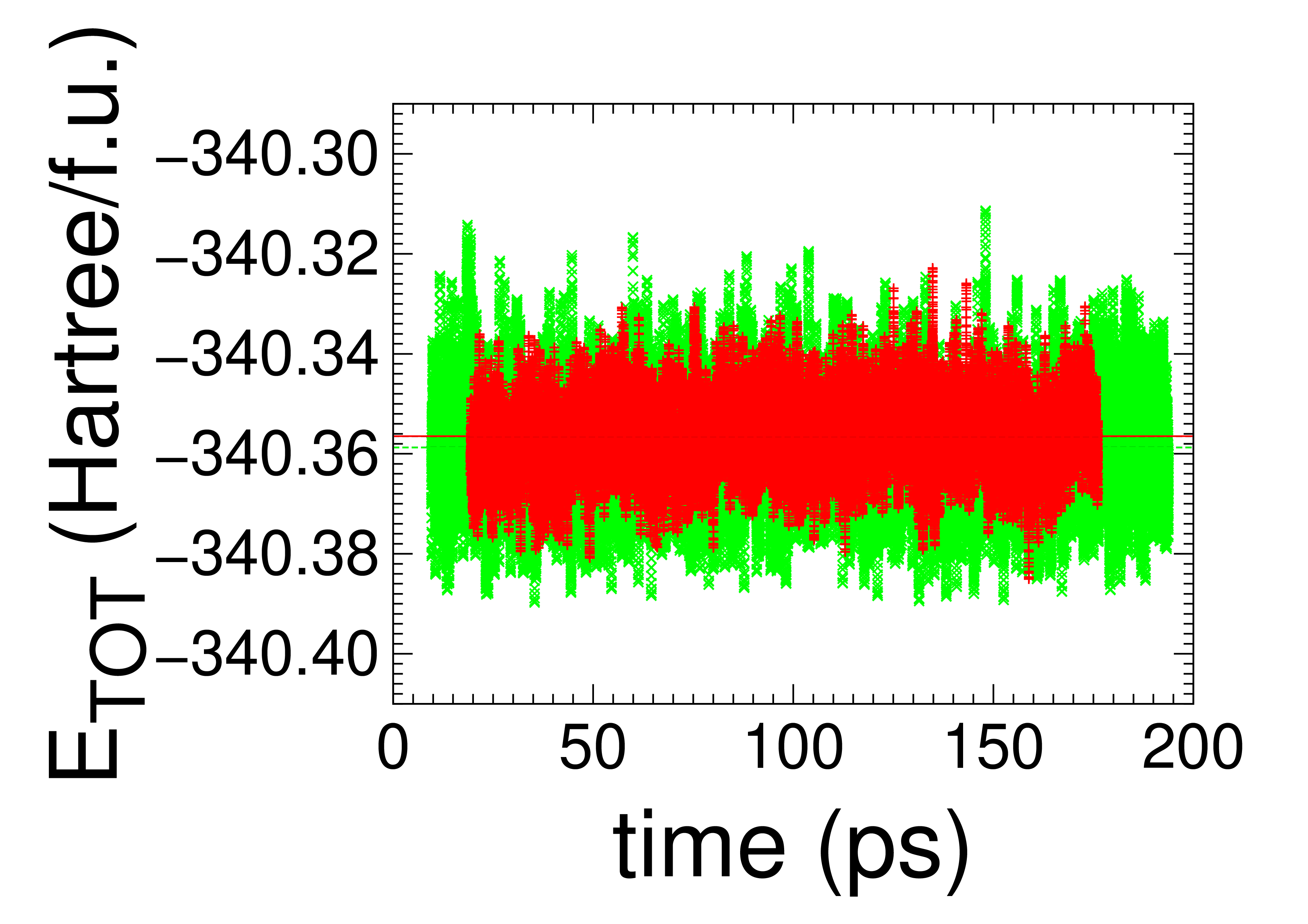}
        \caption{t- and o-LGPS}
         \label{subfig:LGPS_total_energies}
\end{subfigure}
\vspace{1cm}
        \caption{Electronic total energy per formula unit (25 atoms) from variable-cell molecular dynamics calculations (NPT at 600 K) for the orthorhombic and tetragonal phases (in red and green colour respectively) of LGPO (a) and LGPS (b). See also Table~\ref{tab:geom_vc-md}.}
 \label{fig:total_energies_vc-md}
 \end{figure}

\begin{figure}
\vspace{1cm}
\centering
        \begin{subfigure}[t]{0.7\textwidth}
            \includegraphics[width=1.\linewidth]{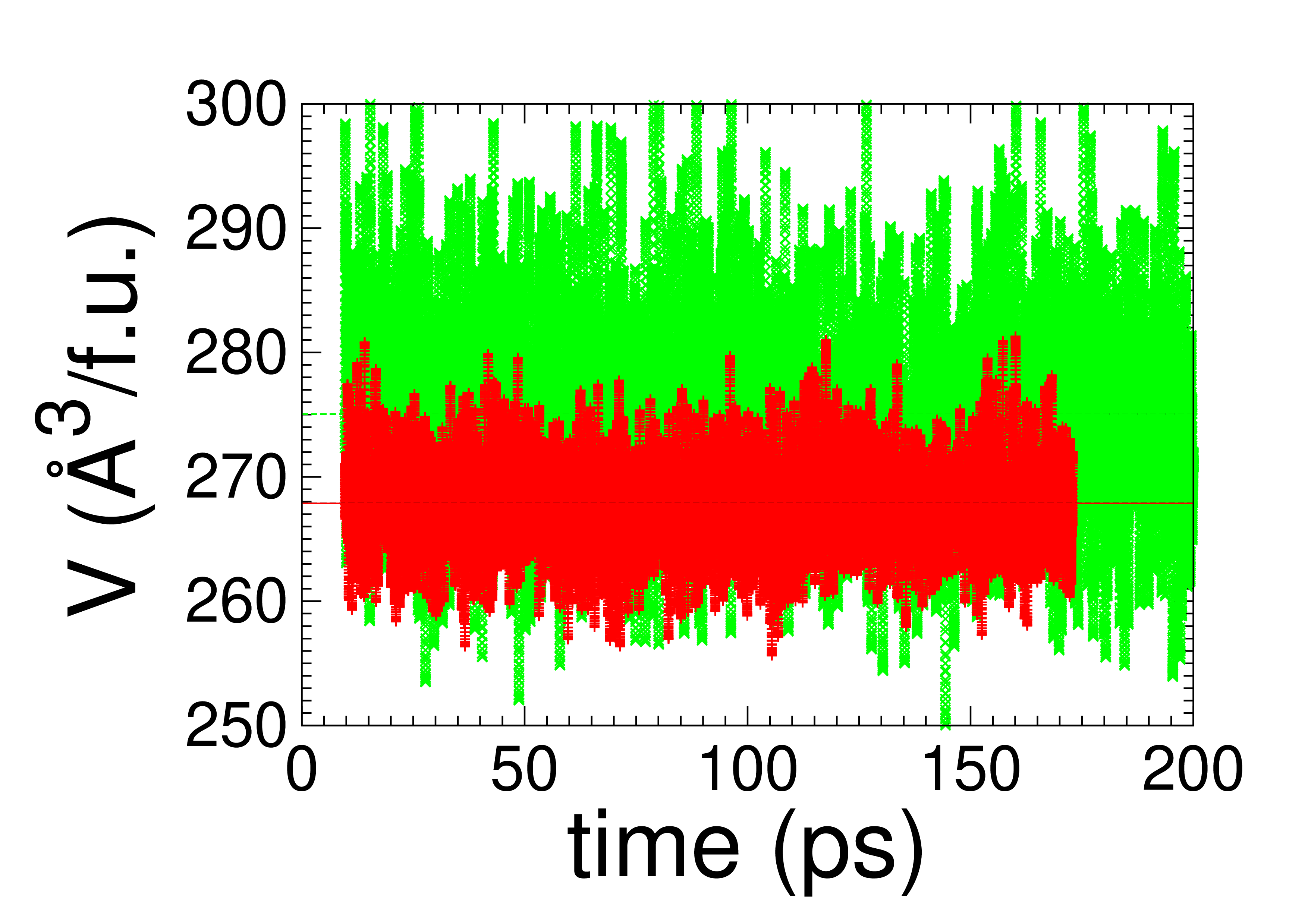}
            \caption{t- and o-LGPO}
              \label{subfig:LGPO_volume}
    \end{subfigure}\hfill
      \begin{subfigure}[t]{0.7\textwidth}
      \includegraphics[width=1.\linewidth]{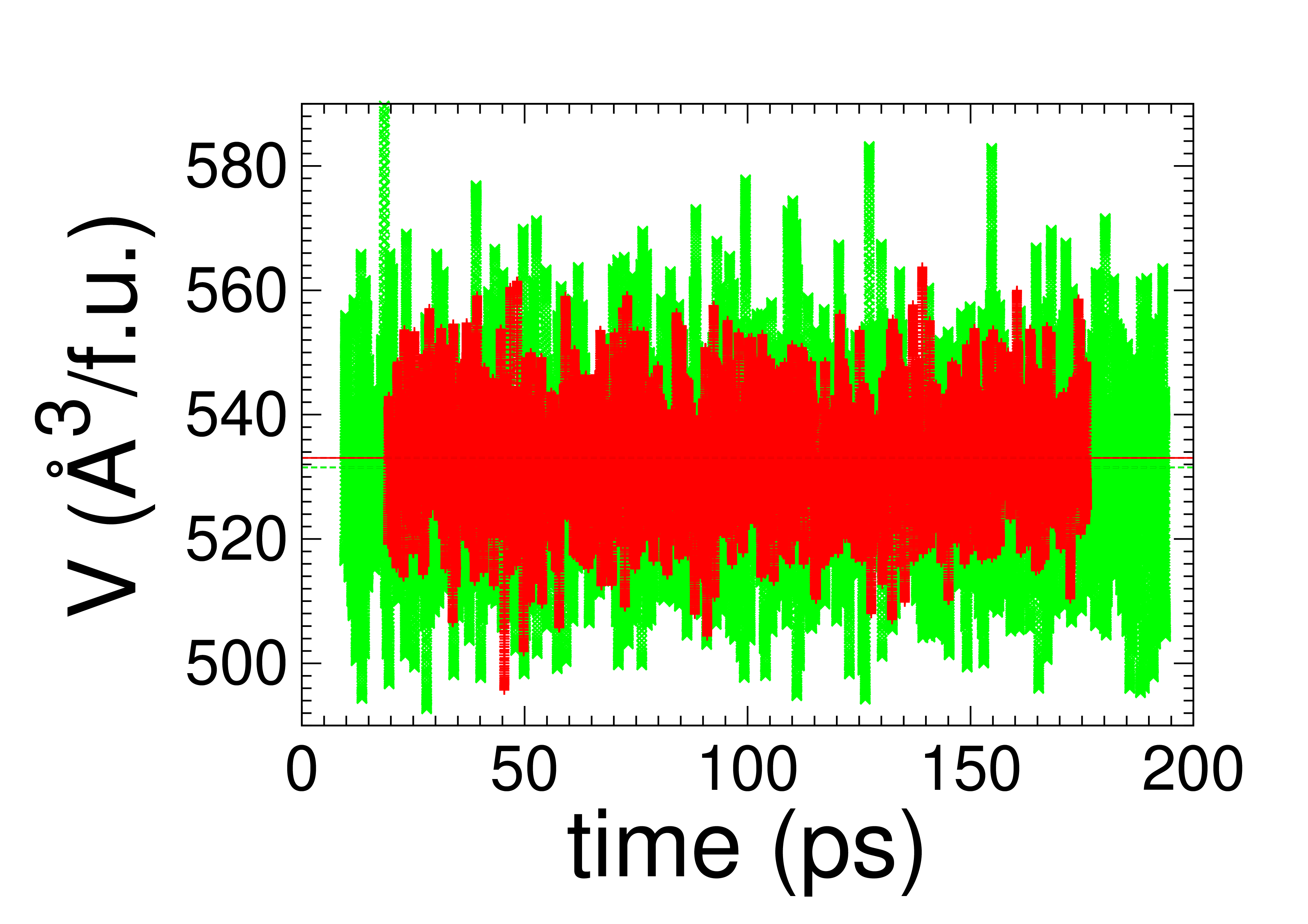}
      \caption{t- and o-LGPS}
      \label{subfig:LGPS_volume}%
      \end{subfigure}
      \vspace{1cm}
     \caption{Volume per formula unit (25 atoms) from variable-cell molecular dynamics simulations (NPT at 600 K) for the orthorhombic and tetragonal phases (in red and green colour respectively) of LGPO (a) and LGPS (b). See also Table~\ref{tab:geom_vc-md}.}
 \label{fig:volumes_vc-md}
 \end{figure}

\begin{figure}
\vspace{1cm}
\centering
     \begin{subfigure}[t]{.7\textwidth}
            \includegraphics[width=1.\linewidth]{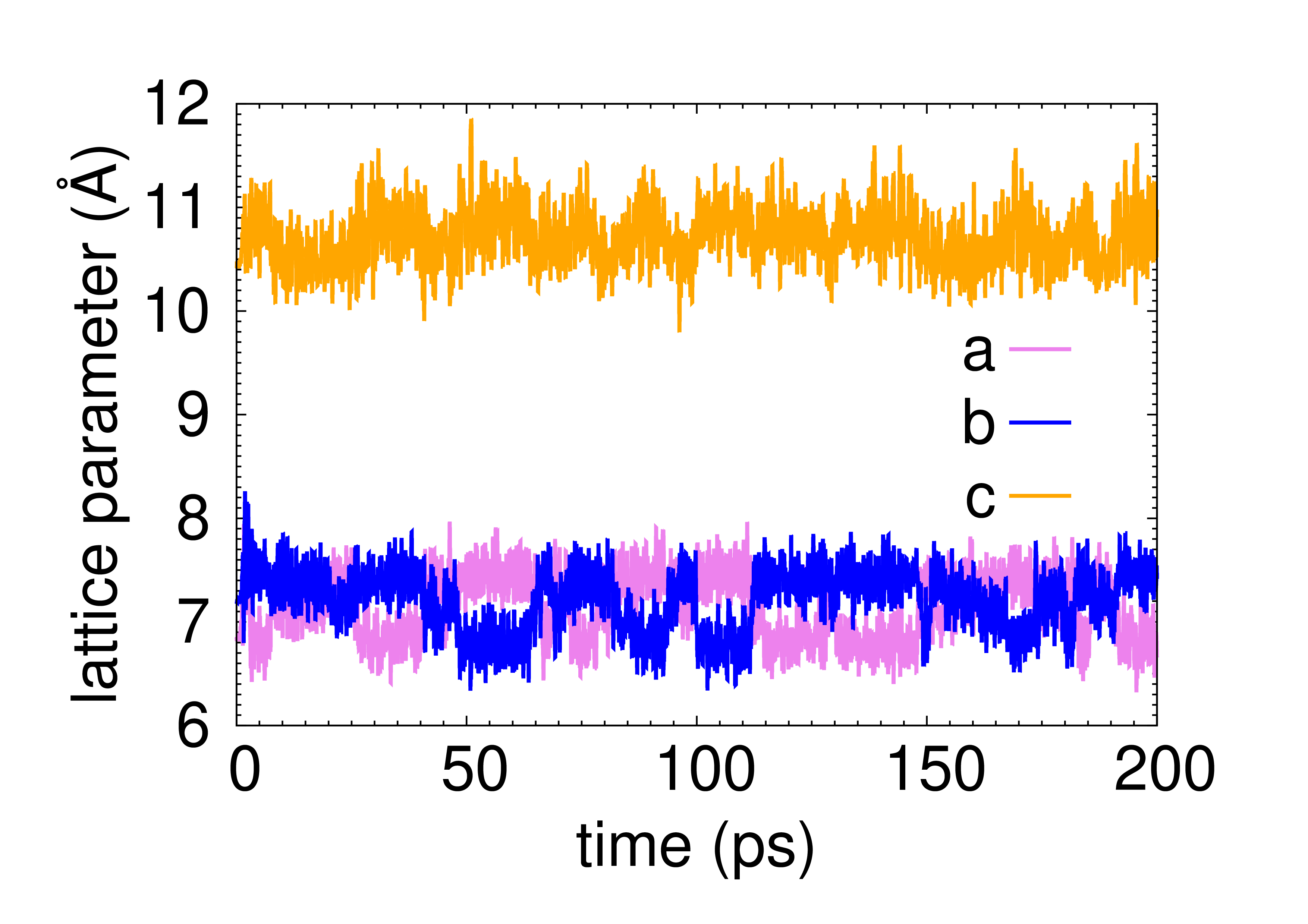}
             \caption{} \label{subfig:tLGPO_supercell_a}
    \end{subfigure}\hfill
      \begin{subfigure}[t]{.7\textwidth}
      \includegraphics[width=1.\linewidth]{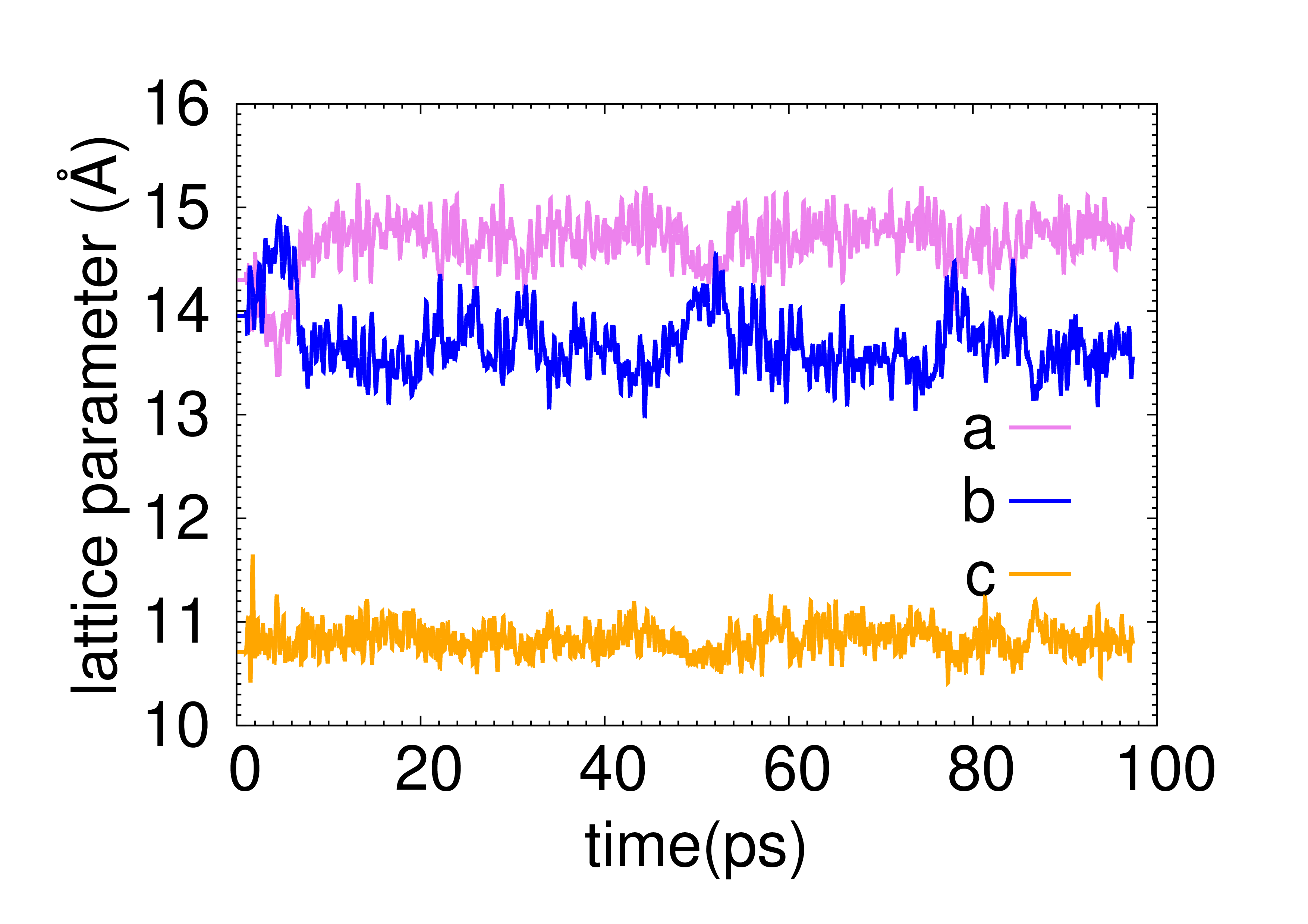}
      \caption{}
      \label{subfig:tLGPO_supercell_b}%
      \end{subfigure}
      \vspace{1cm}
     \caption{Lattice parameters for t-LGPO during the NPT simulations: a) for the 50-atom supercell (see also Table~\ref{tab:geom_vc-md}); b) for the 200-atom (2x2x1) supercell (see text).}
 \label{fig:t-LGPO_supercell}
 \end{figure}

The electronic total energy and volume fluctuations per formula unit during the NPT simulations are displayed in Fig.~\ref{fig:total_energies_vc-md} and Fig.~\ref{fig:volumes_vc-md} respectively, for the four structures considered. One formula unit here is Li$_{10}$GeP$_2$O$_{12}$ or Li$_{10}$GeP$_2$S$_{12}$, i.e.  25 atoms. In Table~\ref{tab:geom_vc-md} we report the mean values for geometries, electronic total energies and enthalpies. For all systems the average internal pressure during the dynamics 
is of the order of $10^{-2}$ GPa or less.
\\
\noindent
Each of the four structures is found to be stable during the NPT simulations,
and no hint of a phase transition is observed (Figs.~\ref{fig:total_energies_vc-md} and \ref{fig:volumes_vc-md}).
However, whereas for the sulfide system  the orthorhombic and tetragonal structures show similar energetics (Fig.~\ref{fig:total_energies_vc-md}b), for the oxide system (Fig.~\ref{fig:total_energies_vc-md}a) the orthorhombic structure is clearly lower, as also reported in Table~\ref{tab:geom_vc-md} for $\left<E_{TOT}\right>$ and $\left<H\right>$.
The trend 
$\left<E_{TOT}\right>$(o-LGPO)$<\left<E_{TOT}\right>$(t-LGPO) is observed systematically also in the NVT simulations of Sec.~\ref{sec:FC-MD}, but the NPT simulations show in addition that this trend is unchanged, at least at 600 K, even if the cell is let free to vary its volume and shape (of course, we compare here $\left< E\right>$, but the thermodynamic functional is the Gibbs free energy $ G = E - TS + PV$). 
The stability of the orthorhombic structure for the oxide 
can be seen as a further confirmation that LGPO in a tetragonal phase would be less stable and that partial sulfur substitution by oxygen was possible in tetragonal LGPS only up to $x=0.9$ in Li$_{10}$GeP${_2}$S$_{(12-x)}$O$_x$  \cite{sun2016oxygen}. 
\\
For cell parameters (Table~\ref{tab:geom_vc-md}), in the case of the oxide the volume fluctuations are larger for the tetragonal than for the orthorhombic structure, whereas they have comparable amplitude in the sulfide. Also, for the tetragonal phases one observes that the in-plane lattice parameters $a$ and $b$ display stronger oscillations 
in the oxide with respect to the sulfide, and the former shows a larger fractional difference between $\left<a\right>$ and $\left<b\right>$ than the latter (0.7\% instead of 0.09\%).
These results 
point to 
a slight deviation from tetragonality for the oxide,
with the simulations happy to switch between two slightly-non-tetragonal equivalent forms. 
In Fig.~\ref{fig:t-LGPO_supercell} we show $a$, $b$ and $c$ for the 50-atom t-LGPO supercell described above and for the 200-atom ($2\times2\times1$) t-LGPO supercell that we also simulate (NPT,  100 ps) for this purpose: while $a$ and $b$ often switch in the 50-atom supercell (Fig.~\ref{subfig:tLGPO_supercell_a}), in the 200-atom cell simulation these fluctuations are suppressed (Fig.~\ref{subfig:tLGPO_supercell_b}).
\subsection{\label{subsec:VCpressure}Stability of t-LGPO vs o-LGPO}
\begin{figure} [tb]
\vspace{1cm}
\centering
\begin{subfigure}[c]{.7\textwidth}
 \includegraphics[width=1.\linewidth]{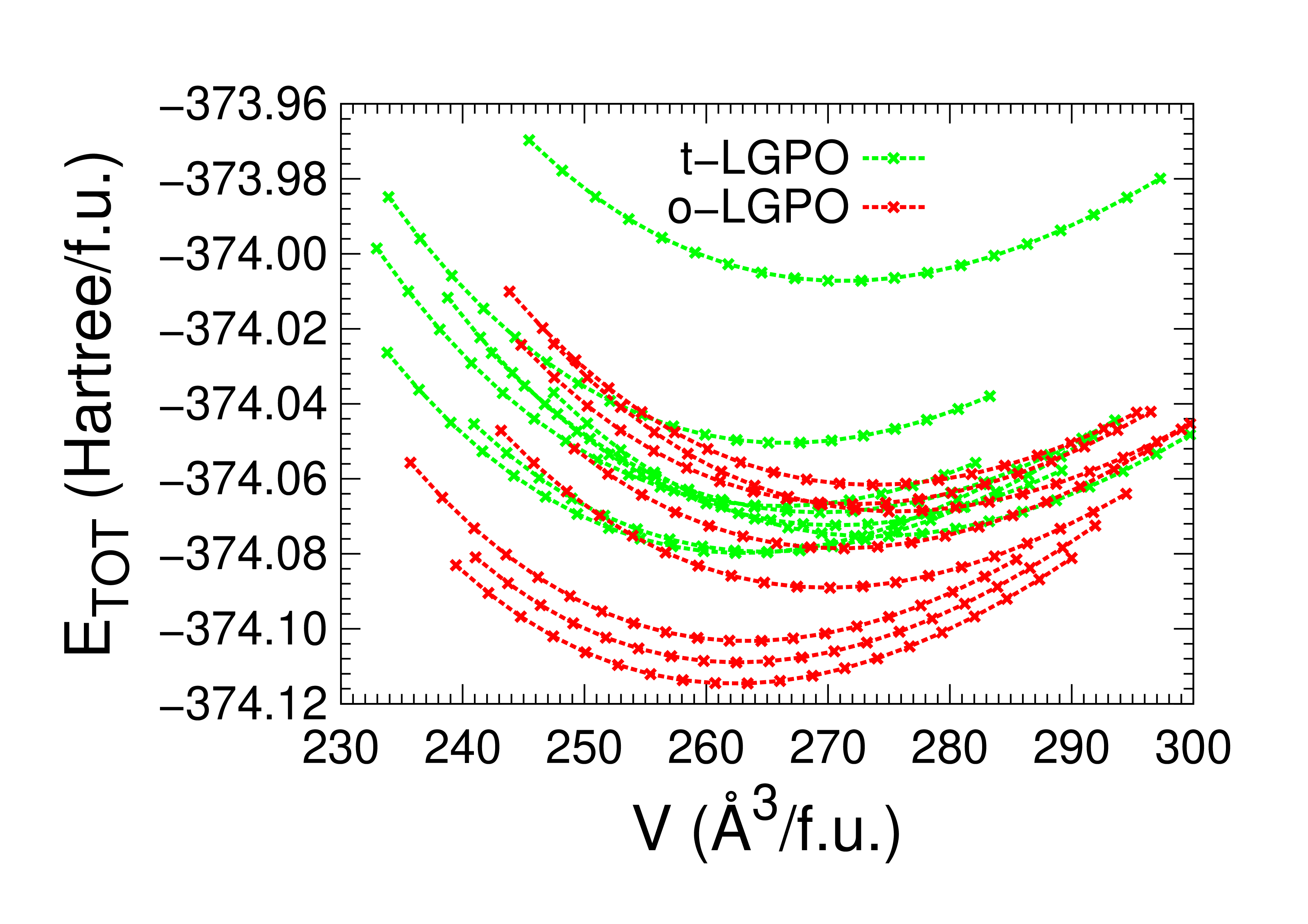}%
 \caption{}
        \label{subfig:en-vol}%
\end{subfigure}\vfill
   
    \begin{subfigure}[c]{.7\textwidth}
            \includegraphics[width=1.\linewidth]{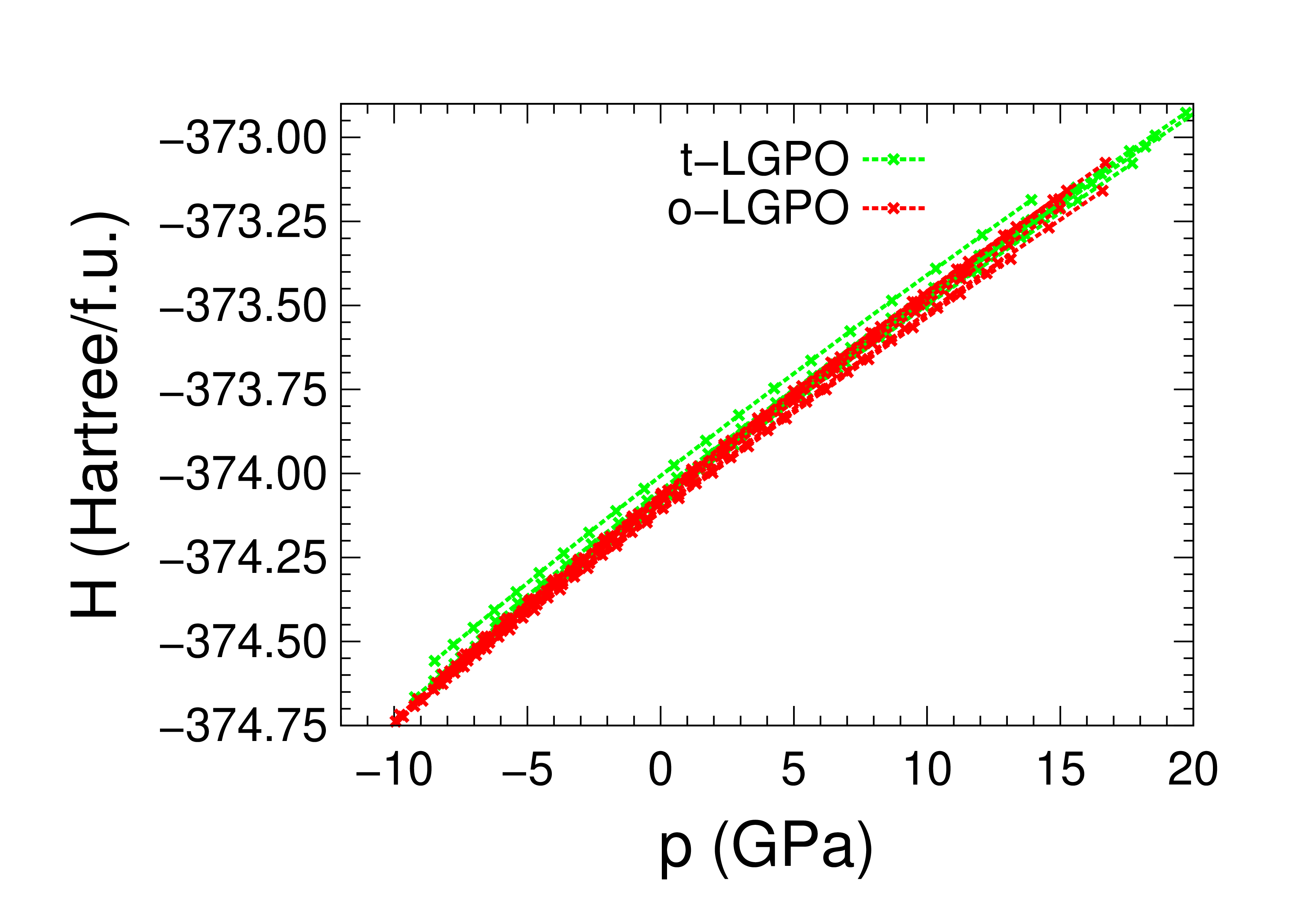}
    \caption{}
    \label{subfig:enth-press}
    \end{subfigure}
   \vspace{1cm}
    \caption{Energy-volume (a) and enthalpy-pressure (b) relations (see text) for t-LGPO and o-LGPO, reported in green and red respectively.}
\label{fig:en-vol-enth-press}
\end{figure}

\begin{figure}
\vspace{1cm}
\centering
    {%
        \includegraphics[width=1.\textwidth]{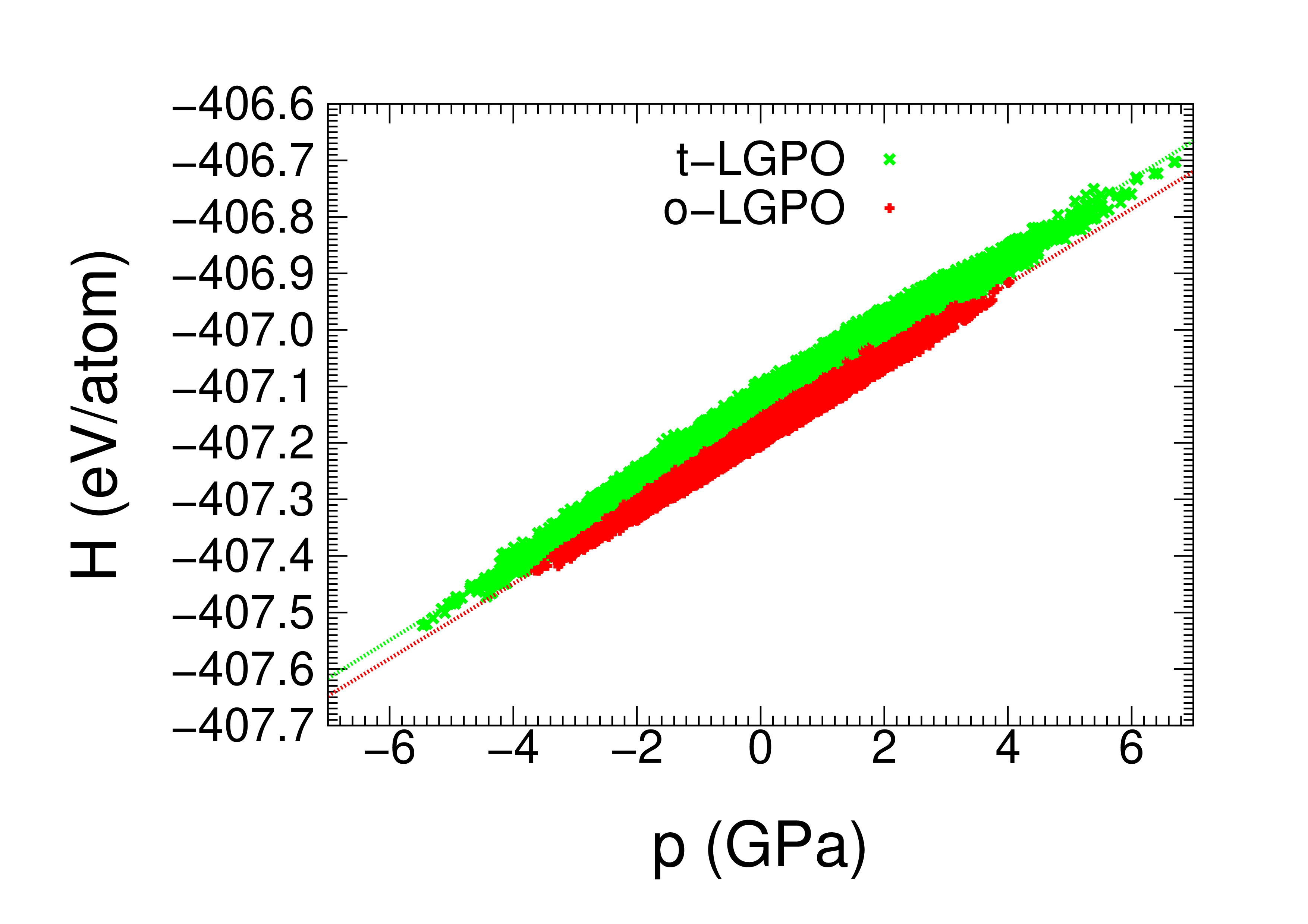}%
        \label{subfig:enthalpy}%
        }%
    \hfill%
    \vspace{1cm}
    \caption{Enthalpy-pressure relations as obtained from NPT simulations at 600 K. t-LGPO and o-LGPO results are reported in green and red, respectively.}
\label{fig:enth-press_vc-md}
\end{figure}

The NVT results for diffusion (Sec.~\ref{sec:FC-MD} and Figs.~\ref{fig:diff_all} and ~\ref{fig:HR}) show that one could in principle obtain the same conductivity in t-LGPO as in o-LGPS or t-LGPS, i.e. from the oxide analogue of t-LGPS.
However, results from the NPT simulations presented in this Section  (Fig.~\ref{fig:total_energies_vc-md} and Table~\ref{tab:geom_vc-md}) reiterate that this phase, though being dynamically stable over 200 ps, is less favoured with respect to its orthorhombic allotrope o-LGPO \cite{rabadanov2003atomic, song2018lithium}. It thus becomes appealing to investigate the possibility of a transition from the more stable and less conductive o-LGPO to the metastable and more conductive t-LGPO. 
Neglecting entropy, a first step in such investigation is to extract representative configurations from the NPT dynamics of the two LGPO structures and  construct equations of state $E_{TOT}(V)$ by isotropically compressing and expanding the cells at each of these configurations. The location of a possible crossing between a t-LGPO and an o-LGPO $E_{TOT}(V)$ would suggest whether  compression or expansion could induce a transition from  t-LGPO to o-LGPO or viceversa.
A set of 16 such equations of state (8 for t-LGPO and 8 for o-LGPO) is displayed in Fig.~\ref{subfig:en-vol}. For each phase the 8 configurations are chosen uniformly in the range of $E_{TOT}$ (Fig.~\ref{fig:total_energies_vc-md}a) and for each configuration the maximum expansion/compression of the cell volume is $\pm$10\%.
 In addition, we fit the values of $E_{TOT}(V)$ in Fig.~\ref{subfig:en-vol} to a Murnaghan equation of state \cite{tyuterev2006murnaghan,murnaghan1944compressibility,giannozzi2009quantum}, from which we calculate the pressure $p$ and the enthalpy $H$.
 The enthalpy-pressure curves are reported in Fig.~\ref{subfig:enth-press}.
 From both the energy-volume and enthalpy-pressure curves (Fig.~\ref{fig:en-vol-enth-press}) it is not straightforward to get unambiguous information about whether a cell compression or expansion could help a transition between t-LGPO and o-LGPO: in particular in Fig.~\ref{subfig:enth-press} one can see that in general the enthalpies have a similar slope, as a function of pressure, in the two phases. 
We can extract the same information directly from the VC-MD dynamics. 
In Fig.~\ref{fig:enth-press_vc-md} we report the enthalpy-pressure data sampled at every 20 time steps of the VC-MD simulations for the two phases, together with a linear fit for each of them. 
It is clear that t-LGPO has univocally a higher enthalpy than o-LGPO for an extended range of pressures. Fig.~\ref{fig:enth-press_vc-md} might show that this trend could be inverted at $p\ll 0$, so that an expansion should favour the tetragonal phase more than a compression, but in general the enthalpy cost, if ascribed entirely to the 10 Li atoms in the formula unit, is around 100 meV/atom (see Table~\ref{tab:geom_vc-md}), i.e. affordable at temperatures of $\sim$1100K.
 Eventually, considering the free energy $G=H-TS$, one should stress that the entropic contribution, stronger for the more diffusive phase, could tilt the thermodynamic balance in favour of t-LGPO, at large enough temperatures. 

\section{\label{sec:DISC} Discussion}
\begin{figure}
\vspace{1cm}
\centering
\begin{subfigure}[c]{0.65\textwidth}
 \includegraphics[width=1.\linewidth]{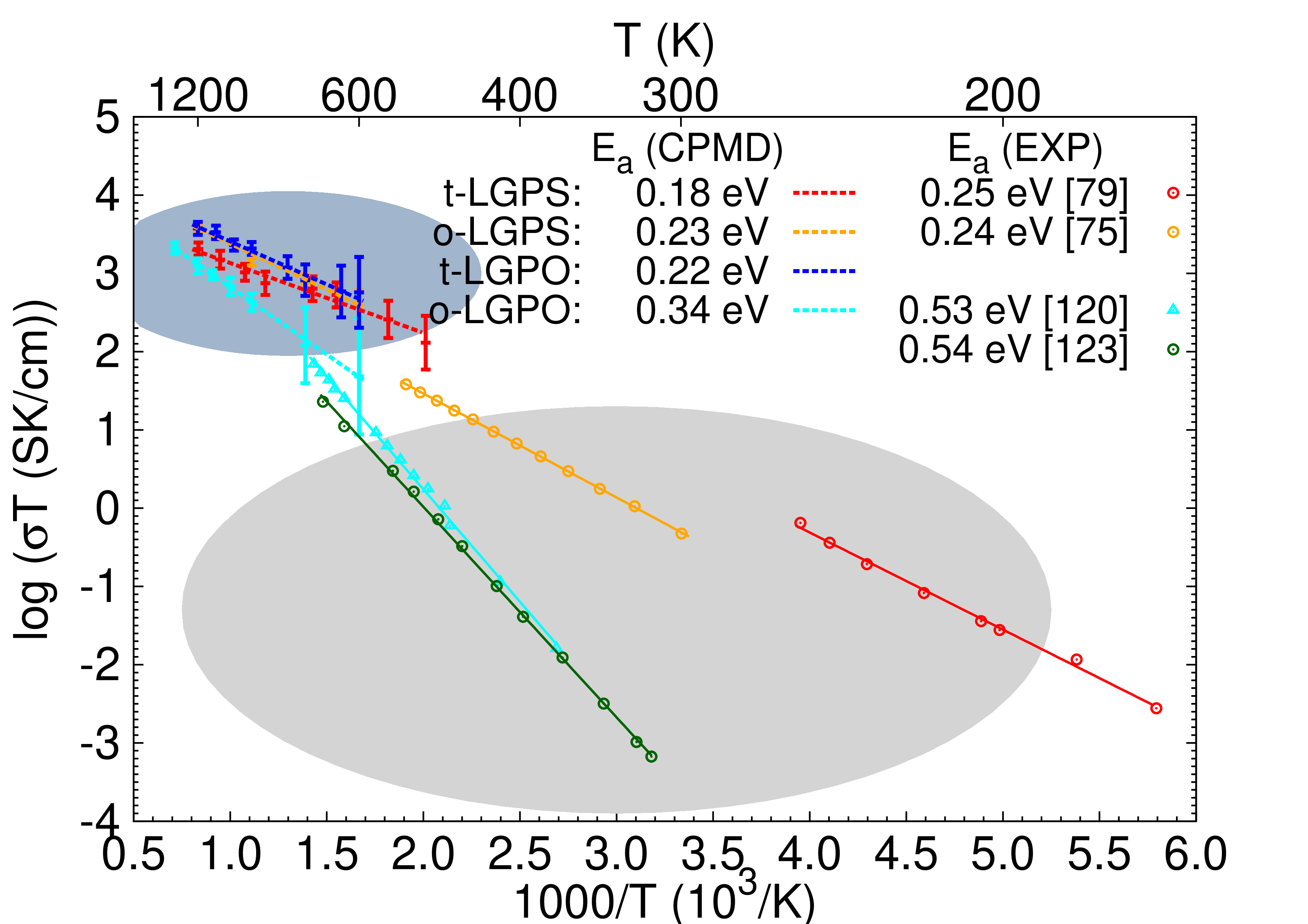}%
 \caption{}
 \label{subfig:conductivities_all}
 \end{subfigure}\vfill
 \begin{subfigure}[c]{0.65\textwidth}
 \includegraphics[width=1.\linewidth]{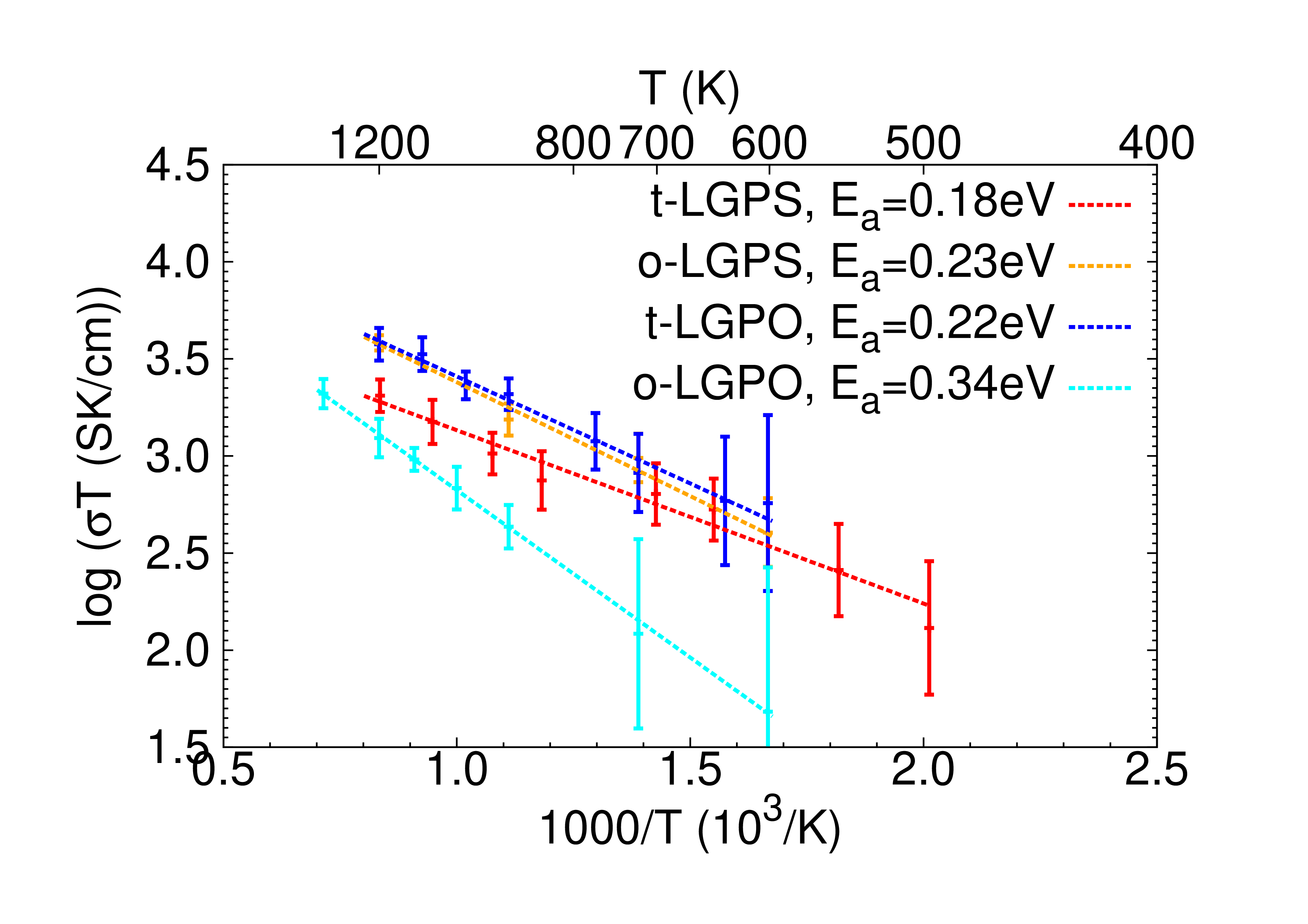}%
 \caption{}
 \label{subfig:conductivities_theory}
 \end{subfigure}\vfill
 \vspace{1cm}
 \caption{(a) Arrhenius plots for $\sigma$T (Li-ion conductivity $\sigma$ multiplied by T) from Eqs.~(\ref{eqEactivation}) and (\ref{eqsigma}) in t-LGPS, o-LGPS, t-LGPO and o-LGPO from NVT CPMD simulations, and compared with the experimental results for the sulfides and o-LGPO. CPMD simulations (here) and experiments (Refs. \cite{kuhn2013tetragonal,kanno2001lithium,ivanov2003growth,gilardi2020li4}) are highlighted in the blue-gray and light-gray ovals, respectively. (b) Enlargement of (a) for the CPMD simulations.}
 \label{fig:conductivities}
\end{figure}

Table~\ref{tab:activation_energies} 
reports the activation energies for diffusion (Eq.~(\ref{eqEactivation})) in the four LGPO and LGPS structures examined in this work, compared with the available experiments \cite{kanno2001lithium, kuhn2013tetragonal,
ivanov2003growth,gilardi2020li4}.
It should be noted first that many physical parameters can affect the bulk ionic conductivity in impedance spectroscopy measurements  \cite{uddin2018reassessing}, that the polycrystalline texture of the samples introduces many grain-boundary effects, and that on the computational side the Nernst-Einstein equation (Eq.~(\ref{eqsigma})), even including Haven ratios \cite{murch1982haven} and the effects of the ionic correlations (Eq.~(\ref{eqsigmacharge}), \cite{marcolongo2017ionic}), assumes the oxidation number of the Li atoms to be constant ($Z=+1$) during the dynamics \cite{french2011dynamical,grasselli2019topological}.
Comparison with the experimental results should thus be regarded as a guide rather than an ultimate benchmark. 
To better clarify this comparison, we report in Fig.~\ref{subfig:conductivities_all} the $\sigma T$ Arrhenius plots from our calculations (Eqs.~\eqref{eqEactivation} and \eqref{eqsigma}) 
for t-LGPS, o-LGPS and o-LGPO together with the corresponding experimental values \cite{kanno2001lithium, kamaya2011lithium, ivanov2003growth,gilardi2020li4}, and the computational one for t-LGPO.
For t-LGPS and o-LGPO we note a similar underestimation ($\sim 60-70\%$) of the  activation barrier with respect to the experiments \cite{kamaya2011lithium,ivanov2003growth,gilardi2020li4}, that might be related to the different ranges of temperatures covered - our results being at 500~K - 1200~K and the experimental results at 150~K - 700~K. 
For o-LGPS the activation barrier for conductivity is in close agreement with the experimental value, whereas the absolute values of log($\sigma T$) are systematically higher (see Fig.~\ref{subfig:conductivities_all}). 
The latter result is likely related to the differences between the cell of Ref.~\cite{kanno2001lithium} and the cell that we have used for the simulations, that is the LGPS-adapted o-LGPO cell from Refs.~\cite{rabadanov2003atomic,ivanov2003growth}, as
details of the genuine II Thio-LISICON structure \cite{kanno2001lithium} are not available (see also \cite{adams2012structural}).
We also recall that, as shown in Table~\ref{tab:lattice_parameters}, convergence of the lattice parameters is achieved with a (343) $\mathbf{k}$-points mesh, whereas, due to an already high numerical cost, we use for these simulations the $\Gamma$-sampled 100-atom supercell.
Nevertheless, extrapolating the t- and o-LGPS and the o-LGPO results at room temperature, we still reproduce the experimentally observed rank $\sigma$(o-LGPO)$\ll$ $\sigma$(o-LGPS) $<$ $\sigma$(t-LGPS) \cite{ivanov2003growth,kanno2001lithium,kamaya2011lithium,gilardi2020li4}.
    \label{fig:conductivities-t-LGPO}
We conclude that the present results, in good agreement with the available experimental literature for what concerns the absolute conductivities of each structure, reproduce well the relative conductivity trends of the experimentally known structures (Fig.~\ref{subfig:conductivities_all} and Table~\ref{tab:activation_energies}). 
\\
The agreement of theoretical and experimental conductivities for the first three cases 
can serve as the ground to discuss conductivity in the fourth case, which is, among those studied in this paper, the only one that has not been experimentally found so far, i.e. the oxide-analogue of t-LGPS (t-LGPO). 
In Fig.~\ref{subfig:conductivities_theory} an enlargement of the CPMD results displayed in Fig.~\ref{subfig:conductivities_all} is reported.
t-LGPO would be a very conductive material, with a $\sigma T$ Arrhenius behaviour very similar to the one of o-LGPS. 
This finding is at variance with the diffusion results for t-LGPO from a previous 
first-principles calculation \cite{ong2013phase}, where Li-ion diffusivity in this system is reported to be significantly lower than in t-LGPS, with a much higher activation barrier of 0.36~eV (see also Table~\ref{tab:activation_energies}). 
Interestingly enough, we find instead an activation barrier of 0.22~eV which is very close to the theoretical activation barriers of 0.18~eV and 0.23~eV for t-LGPS and o-LGPS respectively, and much smaller than the activation energy of 0.34~eV for o-LGPO (see Table~\ref{tab:activation_energies} and Fig.~\ref{subfig:conductivities_theory}). 
\\
t-LGPO, though remaining stable in a 200~ps simulation, is thermodynamically unfavoured with respect to o-LGPO (Fig.~\ref{fig:total_energies_vc-md}), and in fact only o-LGPO has been synthesized so far \cite{ivanov2003growth,rabadanov2003atomic,song2018lithium,gilardi2020li4}. 
A solid solution starting from t-LGPS (Li$_{10}$GeP${_2}$S$_{(12-x)}$O$_x$) was shown to be stable only up to $x=0.9$ \cite{sun2016oxygen}, showing that preparing t-LGPO via solid-solution synthesis doesn't seem a feasible method, due to the difference in ionic radii between oxygen and sulphur (see also Fig.~\ref{fig:volumes_vc-md}).
In the case of the solid solution Li$_{3+5x}$P$_{1-x}$S$_{4-z}$O$_z$, a highly conductive tetragonal structure has been obtained up to $z = 0.8$ thanks to rapid quenching of the molten mixture from T $>$ 700 $^\circ$C
\cite{suzuki2016synthesis}.
A totally different procedure would be to start from the experimental o-LGPO \cite{ivanov2003growth,rabadanov2003atomic,song2018lithium,gilardi2020li4} and try to induce a phase transition from orthorhombic to tetragonal by either imposing a negative hydrostatic pressure or introducing a tensile in-plane strain via epitaxial growth, as pointed out in Sec.~\ref{subsec:VCpressure} (see also \cite{donner2011epitaxial}).
Most importantly entropy, being in principle higher for the more conductive phase t-LGPO, could tilt the thermodynamic balance between the two phases. We leave the latter consideration to a further study.

\section{\label{sec:CONCL} Summary and conclusions}
Sulfide crystalline lithium ionic conductors, in particular within the thio-LISICON family \cite{kanno2001lithium,kamaya2011lithium}, have shown optimal conductive properties that are in general superior to the oxide materials originally proposed within the same family (LISICONs \cite{hong1978crystal,rodger1985li+,fujimura2013accelerated}). They also display severe practical hindrances \cite{kerman2017practical}, starting from the need of strictly controlling the atmosphere in which they are processed, in order to avoid releasing toxic H$_2$S \cite{jung2015issues} and consequent degradation in conductivity  \cite{muramatsu2011structural}. 
Recent experimental work shows that H$_2$S generation is significantly suppressed after adding Li$_2$O or P$_2$O$_5$ to the Li$_2$O-Li$_2$S-P$_2$S$_5$ glass \cite{ohtomo2013glass,minami2008electrical}.
In this respect, new investigations on the oxide-analogue crystalline materials,
potentially less conductive but displaying superior safety features, are highly desirable.
\\
The purpose of this paper is to compare ionic conductivities and relative stability of crystalline LGPS with those of its oxide analogue, LGPO. 
LGPS has been extensively studied so far, both experimentally \cite{kuhn2013tetragonal,kuhn2013single,kato2014synthesis,kwon2015synthesis,han2016electrochemical,weber2016structural,sun2017facile,wenzel2016direct,yang2015dual,zhang2017interfacial} and theoretically \cite{adams2012structural,du2014structures,hu2014insights,mo2011first,ong2013phase,wang2015design,xu2012one,zhu2016first,marcolongo2017ionic}, in its highly conductive tetragonal phase (space group $P4_2/nmc$, $\sigma=1.2\times 10^{-2}$ $\mathrm{S} \mathrm{cm}^{-1}$ at room temperature \cite{kamaya2011lithium}, here denoted t-LGPS), but it is known from experiments to exist also in a monoclinic allotrope (space group $P2_1/m$, here denoted o-LGPS), that also shows high conductivity ($\sigma \sim 10^{-3}$ $ \mathrm{S}  \mathrm{cm}^{-1}$ at room temperature) and
whose cell is related to the orthorhombic parent lattices Li$_4$GeS$_4$ and Li$_3$PS$_4$ \cite{kanno2001lithium}. 
On the other hand, LGPO is known from experiments to exist in its
moderately conducting (1.8$\times 10^{-6}$ $ \mathrm{S} \mathrm{cm}^{-1}$ at 40 $^o$C \cite{ivanov2003growth,gilardi2020li4}) orthorhombic phase (space group $Pnma$, here denoted o-LGPO) \cite{rabadanov2003atomic, song2018lithium,rodger1985li+,muy2018lithium}, that has been also studied in simulations using classical molecular dynamics \cite{muy2018lithium} and machine-learning techniques \cite{fujimura2013accelerated}, but no
experimental evidence of an analogue tetragonal phase, that we call here t-LGPO, has been reported in the literature so far. 
\\
We have calculated Li-ion diffusion and conductivity of 
t- and o-LGPS and t- and o-LGPO via first-principles molecular dynamics (Car-Parrinello, plane waves, ultrasoft pseudopotentials \cite{giannozzi2009quantum}), with $\sim$200~ps NVT simulations for the four structures at 600, 720, 900, 1000, 1100 and 1200 K.
We have also performed variable-cell NPT simulations at T = 600 K for the four cases studied, with the aim of comparing their relative stabilities. Results for the conductivity for the three experimentally studied structures agree well with the experiments. t-LGPO, i.e. the oxide-analogue of t-LGPS (that has not been synthesized so far) reveals itself to be a fast ionic conductor, with an Arrhenius behaviour for Li-ion conductivity in this system similar to the one in o-LGPS and a conductivity at room temperature comparable to the conductivities of o- and t-LGPS calculated here. In addition, t-LGPO, being an oxide, is not expected to carry the stability problems encountered in sulfide electrolytes \cite{kerman2017practical} for solid-state batteries, and could represent a relevant and important challenge for experimental synthesis and stabilization.


\section{Acknowledgments}
This research was supported by the Swiss National Science Foundation, through project 200021-159198 and the MARVEL 
NCCR. We acknowledge computational support from the Swiss National Supercomputing Centre CSCS (projects s836 and mr28). Fruitful discussions with Tommaso Chiarotti and Claire Villevieille are gratefully acknowledged.
\clearpage
\bibliography{biblio}
\clearpage
\clearpage
 \clearpage
\clearpage
\clearpage
\clearpage
\clearpage
 \clearpage
 \clearpage
 \clearpage
\clearpage
\clearpage

\clearpage
 \begin{table}[tb]
\vspace{2cm}
\centering
\centering
\begin{tabularx}{\columnwidth}{@{\extracolsep{\fill}} | C | C | C | C | C | C | C | C |}
\hline
&  & {\it a} & {\it b} & {\it c}& $\alpha$ & $\beta$ & $\gamma$ \\
\hline \hline
t-LGPS & $\Gamma$ & 8.764   & 8.781 & 12.943 & 91.17 & 91.20 & 90.42 
\\
& (444) & 8.661 & 8.819 & 13.072 & 91.21 & 90.51 & 90.33
\\
& \cite{kamaya2011lithium}& 8.718  & 8.718  & 12.634 & 90.00 & 90.00 & 90.00 
\\
\hline
t-LGPO & $\Gamma$ & 6.836 & 7.191 & 10.433 & 89.43 & 90.56 & 89.97 \\
& (444) & 6.769 & 7.214 & 10.501 & 89.57 & 90.49 & 89.95 \\
 & & & & & & & \\
\hline
o-LGPS & $\Gamma$ & 12.770 & 8.537 & 18.330 & 89.82 & 89.33 & 90.24   \\
& (343) & 13.604 &  7.888 &  18.684 & 90.27 & 90.33 & 90.17 \\
& \cite{kanno2001lithium} &13.393 & 22.947 & 18.201 & 90.00 & 92.562 & 90.00  \\
\hline
o-LGPO & $\Gamma$ & 10.604 & 6.398  & 15.052 & 89.71 & 90.50 & 90.31 \\
& (343) & 10.788 & 6.223 & 15.147 & 89.69 & 90.63 & 90.15\\
& \cite{rabadanov2003atomic} & 10.690 & 6.195 & 5.027 & 90.00 & 90.00 &  90.00 \\
\hline
\end{tabularx}
\vspace{50mm}
\caption{Cell geometries (lattice parameters and angles) for the four systems studied in this work ($\Gamma$-sampling in CP and fully converged $\mathbf{k}$-point sampling with a (444) and a (343) Monkhorst-Pack mesh for the t- and o- structures, respectively) compared with the available experimental values (see text). Lengths are in \AA\hspace{0.1cm} and angles in degrees. t-LGPS and t-LGPO cells have 50 atoms, o-LGPS and o-LGPO cells have 100 atoms. See text for Ref. \cite{kanno2001lithium}.}
\label{tab:lattice_parameters}
\end{table}
\clearpage
\begin{table}[!htb]
\vspace{2cm}
\centering
\setlength\tabcolsep{0pt} 
\centering
\renewcommand{\arraystretch}{2}
\begin{tabularx}{\columnwidth}{@{\extracolsep{\fill}}|C|C|C|C|C|}
\hline
 & $E_{a_{D_{tr}}}$ & $E_{a_{D_{tr}}}$  & $E_{a_{D_\sigma}}$  &  $E_{a_{\sigma T}}$ 
 \\
 & & (Ref.~\cite{ong2013phase}) & & (expt.)
 \vspace*{-2cm}\\
\hline \hline
t-LGPS &   0.18 $\pm$ 0.02
\hspace*{0.04cm}&  0.21 $\pm$ 0.04  
\hspace*{0.02cm}& 0.17 $\pm$ 0.07 
 & 0.26 \cite{kamaya2011lithium}
\\
\hline
o-LGPS & 0.23 $\pm$ 0.02 
\hspace*{0.04cm}& 
\hspace*{0.02cm}& 0.21 $\pm$ 0.04   &  0.24 \cite{kanno2001lithium} \\
\hline
t-LGPO & 0.21 $\pm$ 0.02
\hspace*{0.04cm}& 0.36 $\pm$ 0.05 
\hspace*{0.02cm}&  0.21 $\pm$ 0.06  &  \\
\hline
o-LGPO & 0.34 $\pm$ 0.04 
\hspace*{0.04cm}& 
\hspace*{0.02cm}&  0.32 $\pm$ 0.07 & 0.54 \cite{ivanov2003growth}\\
\hline
\end{tabularx}
\vspace{50mm}
\caption{\small Activation energies for the tracer diffusion and charge diffusion (in eV) for the four systems studied in this work (see text), compared with experiments and previous calculations.}
\label{tab:activation_energies}
\end{table}

\clearpage
\centering
\begin{sidewaystable}
\begin{tabularx}
{0.945\columnwidth}
{ | c | c | c | c | c | c | c | c | c | c | c |}
  \hline
 & $\left<a\right>$ & $\left<b\right>$ & $\left<c\right>$ & $\left<\alpha\right>$ & $\left<\beta\right>$ & $\left<\gamma\right>$ & $\left<V\right> $& $\left<E_{TOT}\right>$ & $\left<H\right>$ 
 \\
  \hline
    t-LGPS & 9.01 $\pm$ 0.18 & 9.01 $\pm$ 0.18 & 13.12 $\pm$ 0.25 & 90.0 $\pm$ 2.3 & 89.9 $\pm$ 2.3 & 90.1 $\pm$ 1.7 & 531.5 $\pm$ 13.4 & -340.359 $\pm$ 0.010 & -340.361 $\pm$ 0.072 
    \\
    \hline
     o-LGPS & 12.85 $\pm$ 0.20 & 8.86 $\pm$ 0.12     & 18.72 $\pm$ 0.27 & 90.1 $\pm$ 1.1  & 90.1 $\pm$ 1.8 & 90.1 $\pm$ 1.3 &533.0 $\pm$ 8.8 & -340.356 $\pm$ 0.007  & -340.357 $\pm$ 0.053
     \\
     \hline
     t-LGPO &  7.15 $\pm$ 0.31 & 7.20 $\pm$ 0.30   & 10.72 $\pm$ 0.27 & 90.1 $\pm$ 2.2  & 90.0 $\pm$2.1 & 90.1 $\pm$ 1.3& 275.0 $\pm$ 7.1& -374.053 $\pm$ 0.010 & -374.055 $\pm$ 0.087
     \\
     \hline
     o-LGPO & 10.69 $\pm$  0.12 & 6.50 $\pm$  0.08  &  15.42 $\pm$  0.18  & 90.0 $\pm$  0.8 & 89.9 $\pm$ 1.0 & 90.1 $\pm$ 0.9  & 267.9 $\pm$ 3.5 & -374.092 $\pm$ 0.007 & -374.093 $\pm$ 0.059 
     \\
    \hline
\end{tabularx}
\vspace{50mm}
\caption{Average lattice parameters, angles, volumes, total energies and enthalpies from NPT dynamics at T = 600 K for the four structures considered. Lengths are in \AA, angles in degrees and energies in Hartree per formula unit. Formula units contain 25 atoms: Li$_{10}$GeP$_2$S$_{12}$ for t- and o-LGPS and Li$_{10}$GeP$_2$O$_{12}$ for t- and o-LGPO (in the calculations tetragonal supercells contain 50 atoms and orthorhombic supercells contain 100 atoms). Errors are calculated as mean standard deviations in a block analysis, as done for diffusion (see Supplemental Material and Refs.~\cite{frenkel2001understanding,allen2017computer}). See also Figs.~\ref{fig:total_energies_vc-md} and ~\ref{fig:volumes_vc-md}.}
 \label{tab:geom_vc-md}
 \end{sidewaystable}


%

\end{document}


\title{Supplemental Material}
\maketitle
\section{\label{CP parameters}NVE-CPMD: tuning $\mu$ and $dt$}
\begin{figure}
    \centering
    \begin{subfigure}{0.65\textwidth}
    \includegraphics[width=1.\linewidth]
    {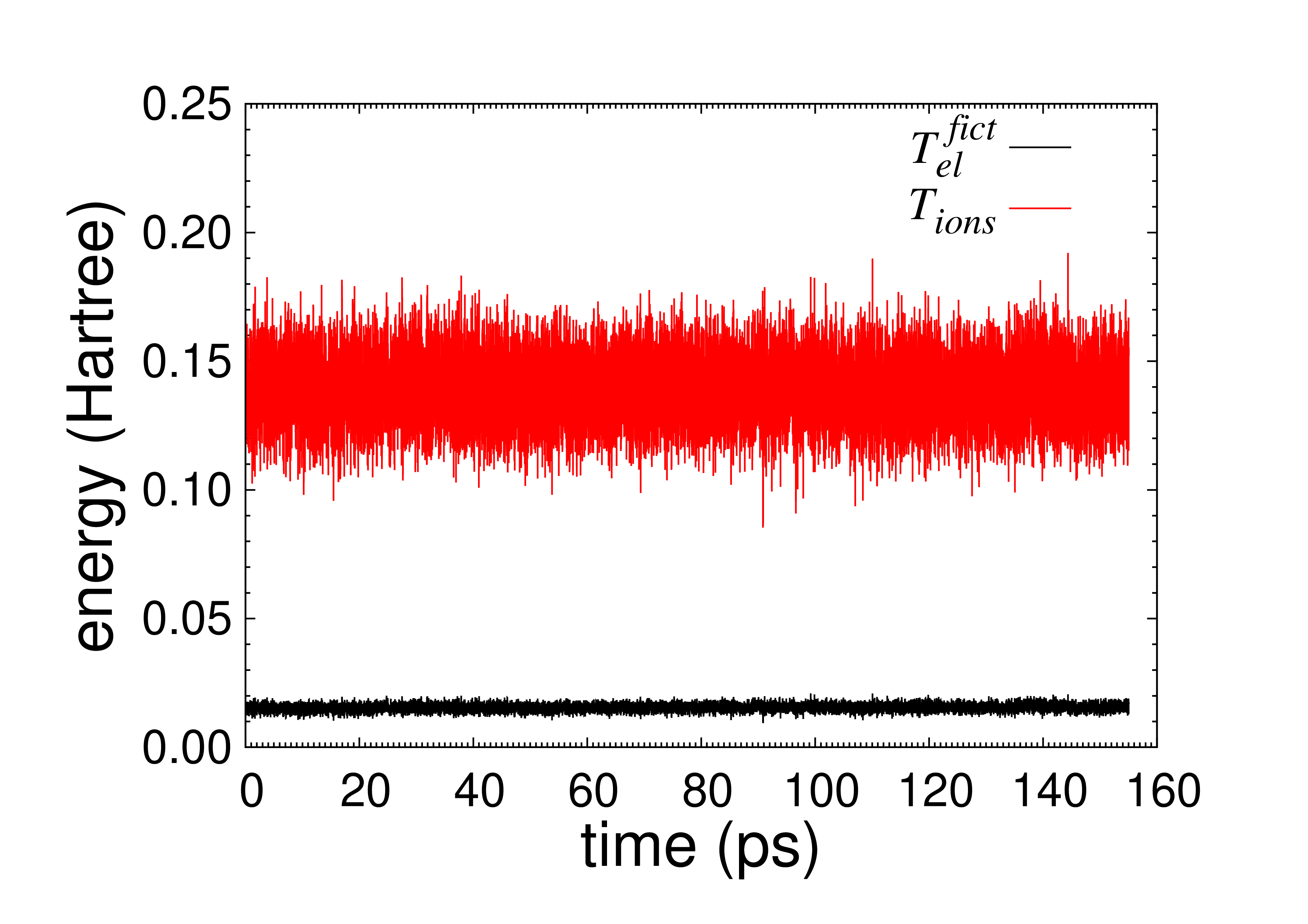}
    \caption{Comparison between the fictitious kinetic energy of the electronic wavefunctions $T^{fict}_{el}$ and the kinetic energy of the ions $T_{ions}$.}
    \label{subfig:econs-a}
    \end{subfigure}\vfill
    \begin{subfigure}{.65\textwidth}
            \includegraphics[width=1.\linewidth]
    {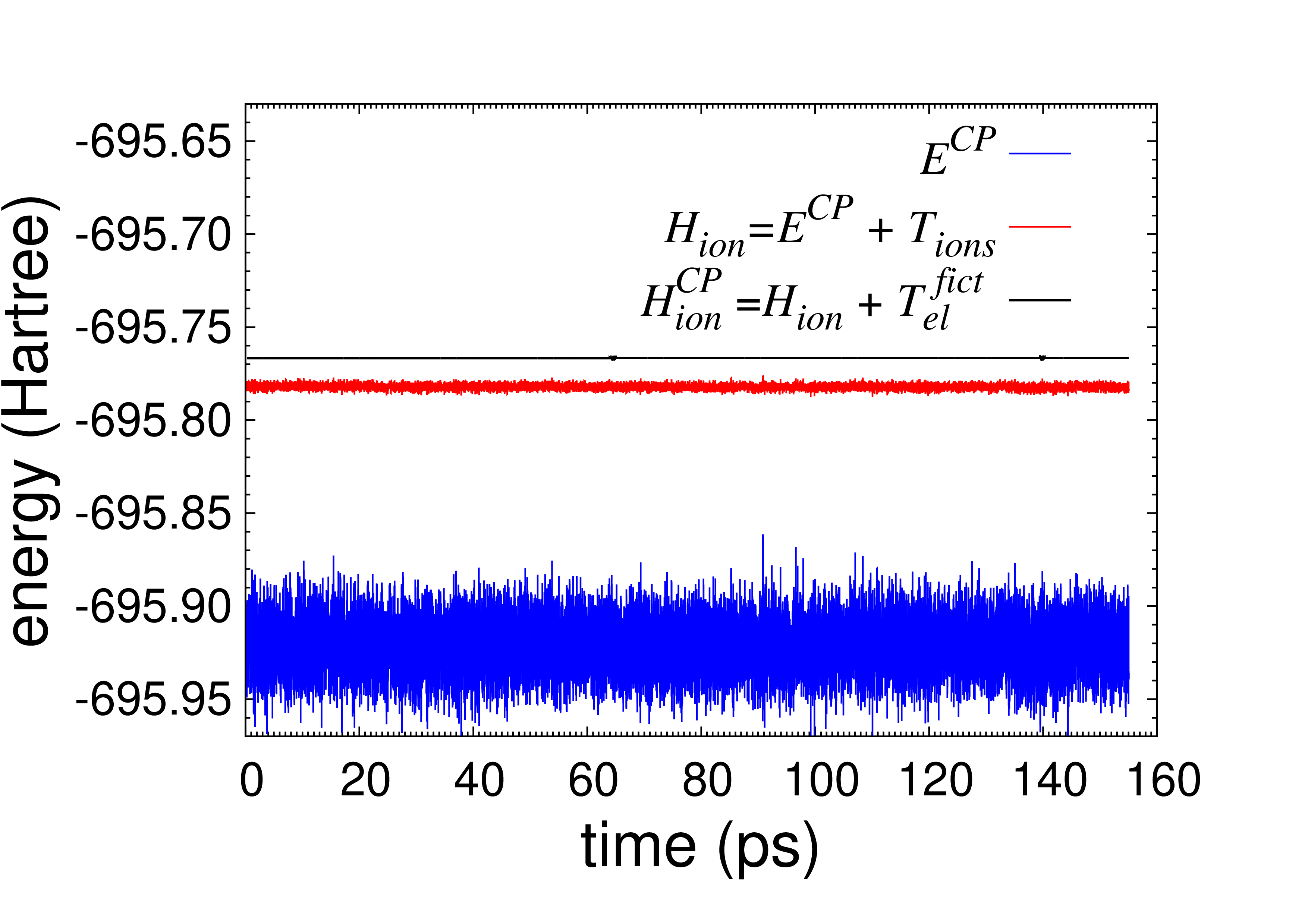}
      \caption{CP potential energy $E^{CP}$, physical constant of motion $H_{ion}$ and CP constant of motion $H^{CP}_{ion}$.}
       \label{subfig:econs-b}
    \end{subfigure}
     \vspace{1cm}
        \caption{{Details of a NVE CPMD for the benchmark system t-LGPO with $\mu= 500$ a.u. and $dt =$ 4 a.u. (for the meaning of the symbols, see text).}}
    \label{fig:econs}
\end{figure}

In the Car-Parrinello method both the mass $\mu$ associated to the electronic wavefunctions and the time step $dt$ are parameters that cannot be tuned independently  \cite{galli1993first, pastore1991theory, remler1990molecular}. $\mu$ needs to be small enough to ensure adiabaticity in the dynamics, but 
large enough to allow for sizeable $dt$ integration steps.
In Fig.~\ref{subfig:econs-a} we report the kinetic energy of the ionic system $T_{ions}$ and the fictitious kinetic energy of the electronic wavefunctions $T^{fict}_{el}$ for a 160 ps NVE simulation of t-LGPO, with $\mu=500$ a.u. and $dt = 4 $ a.u. $\simeq 0.1$ fs. It is seen that $T^{fict}_{el} \simeq \frac{1}{10}T_{ions}$ for the whole simulation, and no energy exchange occurs between the hot degrees of freedom (ions) and the cold degrees of freedom (electronic wavefunctions). Fig.~\ref{subfig:econs-b} displays the CP potential energy $E^{CP}$, the physical ionic hamiltonian $H_{ion}=T_{ions}+E^{CP}$ and the CP Hamiltonian $H_{ion}^{CP}=H_{ion}+T^{fict}_{el}$ for the same simulation (again $\mu=500$  a.u. and $dt = $ 4 a.u. $\simeq 0.1$  fs). The latter, $H_{ion}^{CP}$, should be conserved in a NVE simulation, and this is what we observe for the whole length of the dynamics. In addition, the instantaneous average value of $H_{ion}=T_{ions}+E^{CP}$ is also constant, showing that not only $T^{fict}_{el}$, but also $E^{CP}$, is not drifting during the dynamics. 
\section{\label{thermostat}NVT-CPMD: tuning the thermostat parameters}
\begin{figure}[H]
\centering
    \begin{subfigure}[t]{.7\textwidth}
            \includegraphics[width=1.\linewidth]{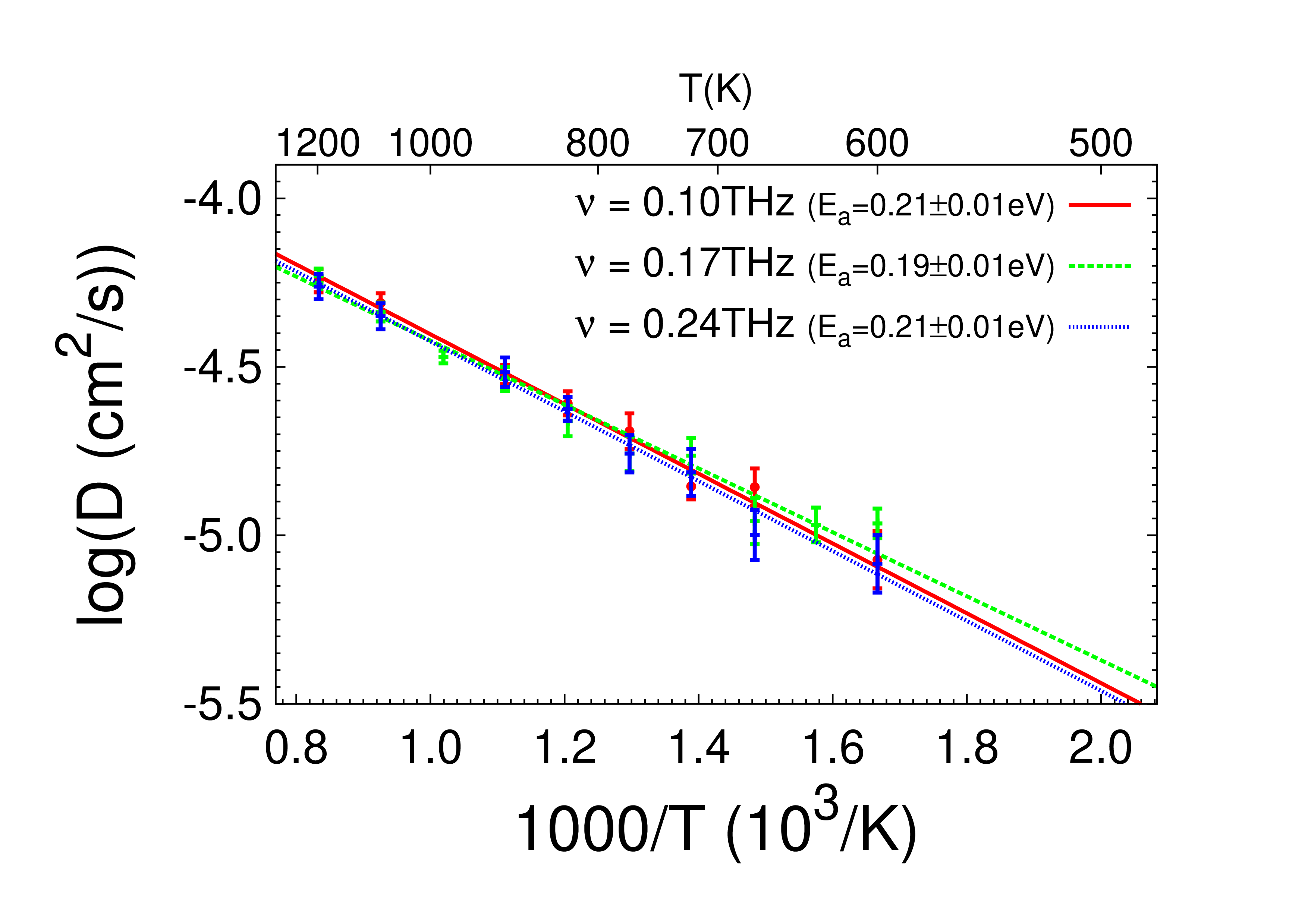}
            \caption{}
            \label{subfig:test_t-LGPO_nosep}
    \end{subfigure}\vfill
       \begin{subfigure}[t]{.7\textwidth}
            \includegraphics[width=1.\linewidth]{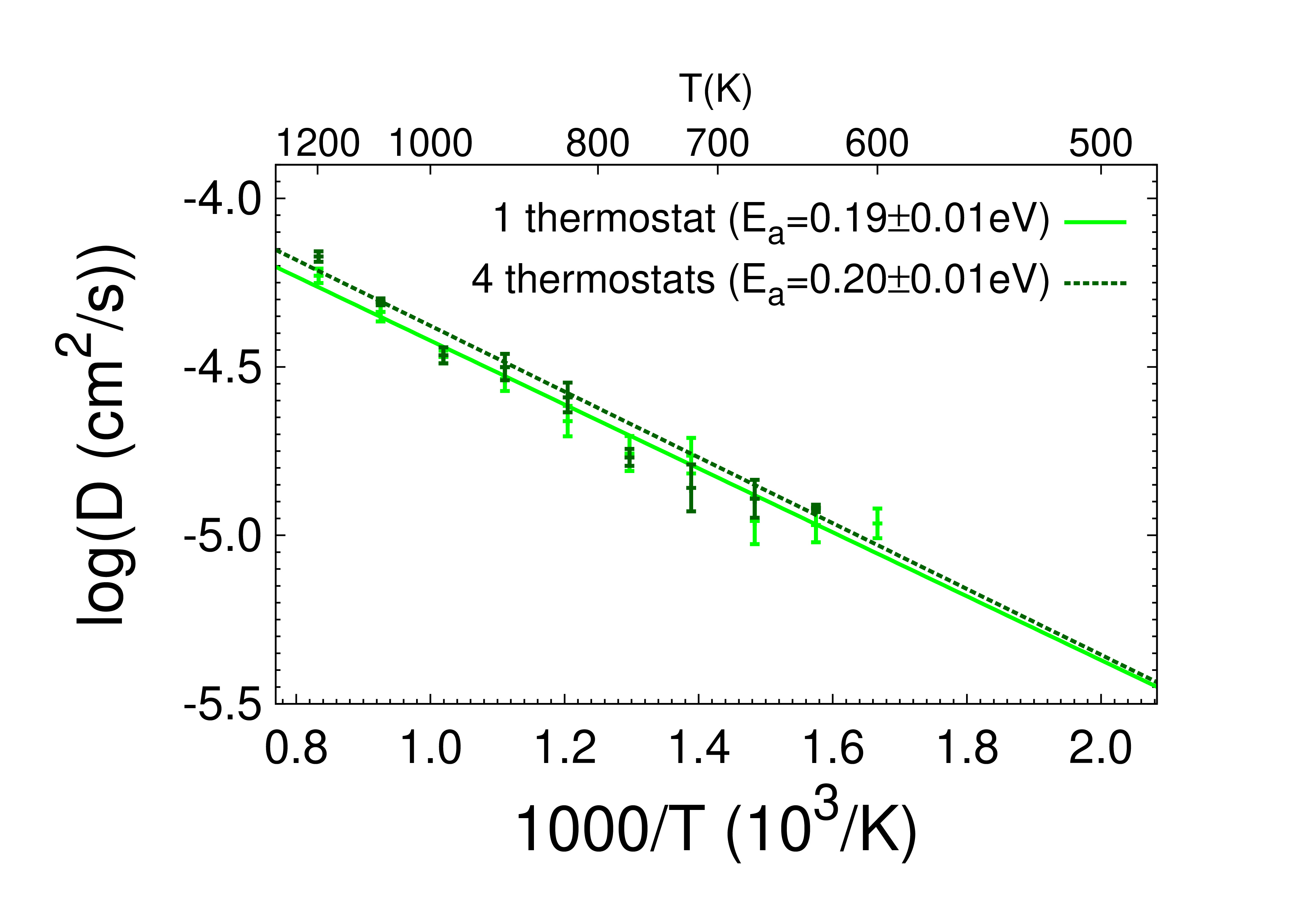}
            \caption{}
            \label{subfig:test_t-LGPO_nhptyp}
    \end{subfigure}
    \vspace{1cm}
        \caption{a) Arrhenius plots and corresponding activation energies for Li diffusion in t-LGPO from NVT simulations at different thermostat frequencies, with only one thermostat for the whole system; b) the result at $\nu = 17$  THz in (a) (light-green) is compared with the same result obtained by using one thermostat for each atomic species (dark-green).}
\label{fig:diff_NVT_frequencies}
\end{figure}

\begin{figure}[H]
\centering
    \begin{subfigure}[t]{.7\textwidth}
            \includegraphics[width=1.\linewidth]{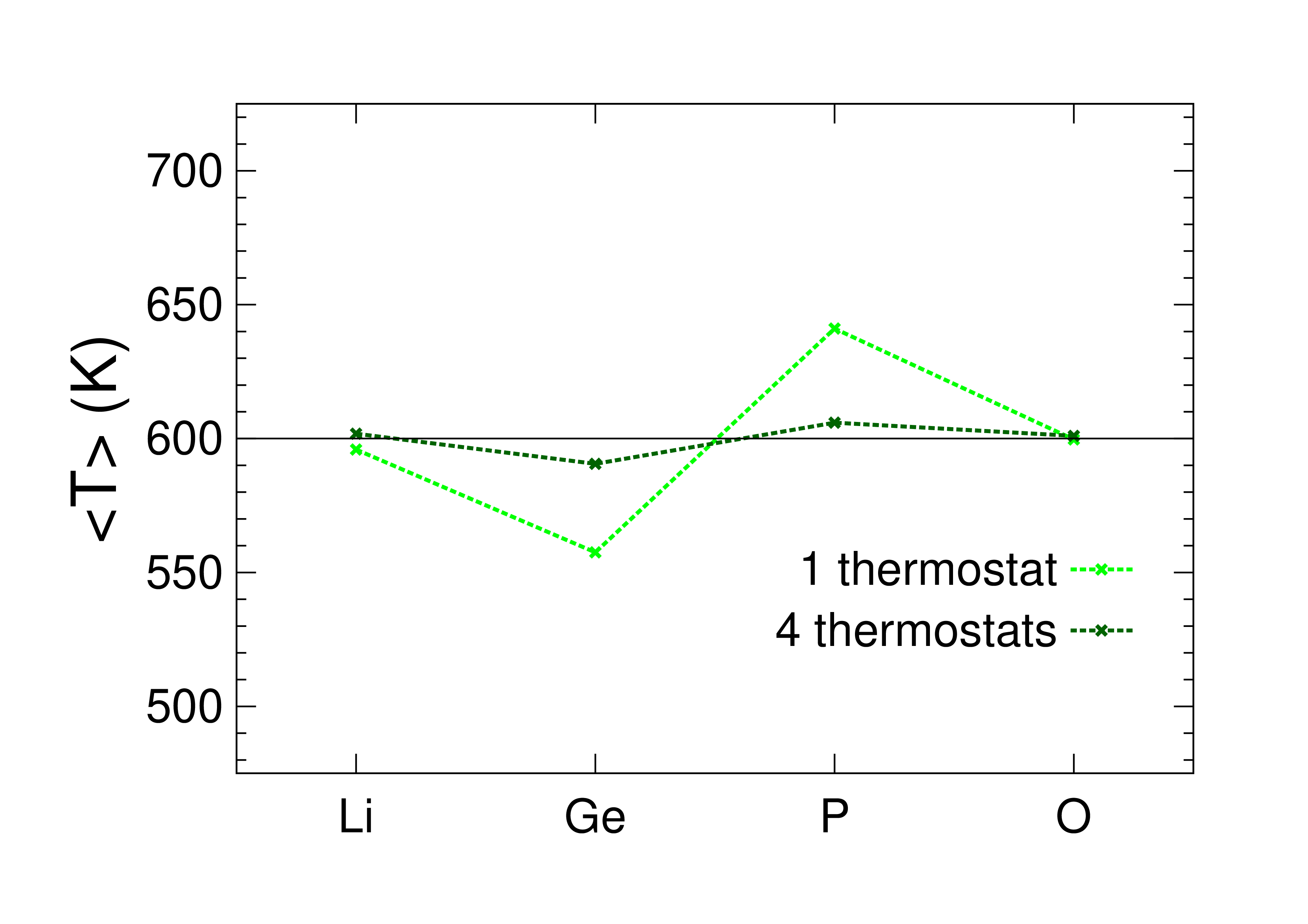}
            \caption{}
            \label{subfig:partial_temperatures_a}
    \end{subfigure}\vfill
       \begin{subfigure}[t]{.7\textwidth}
            \includegraphics[width=1.\linewidth]{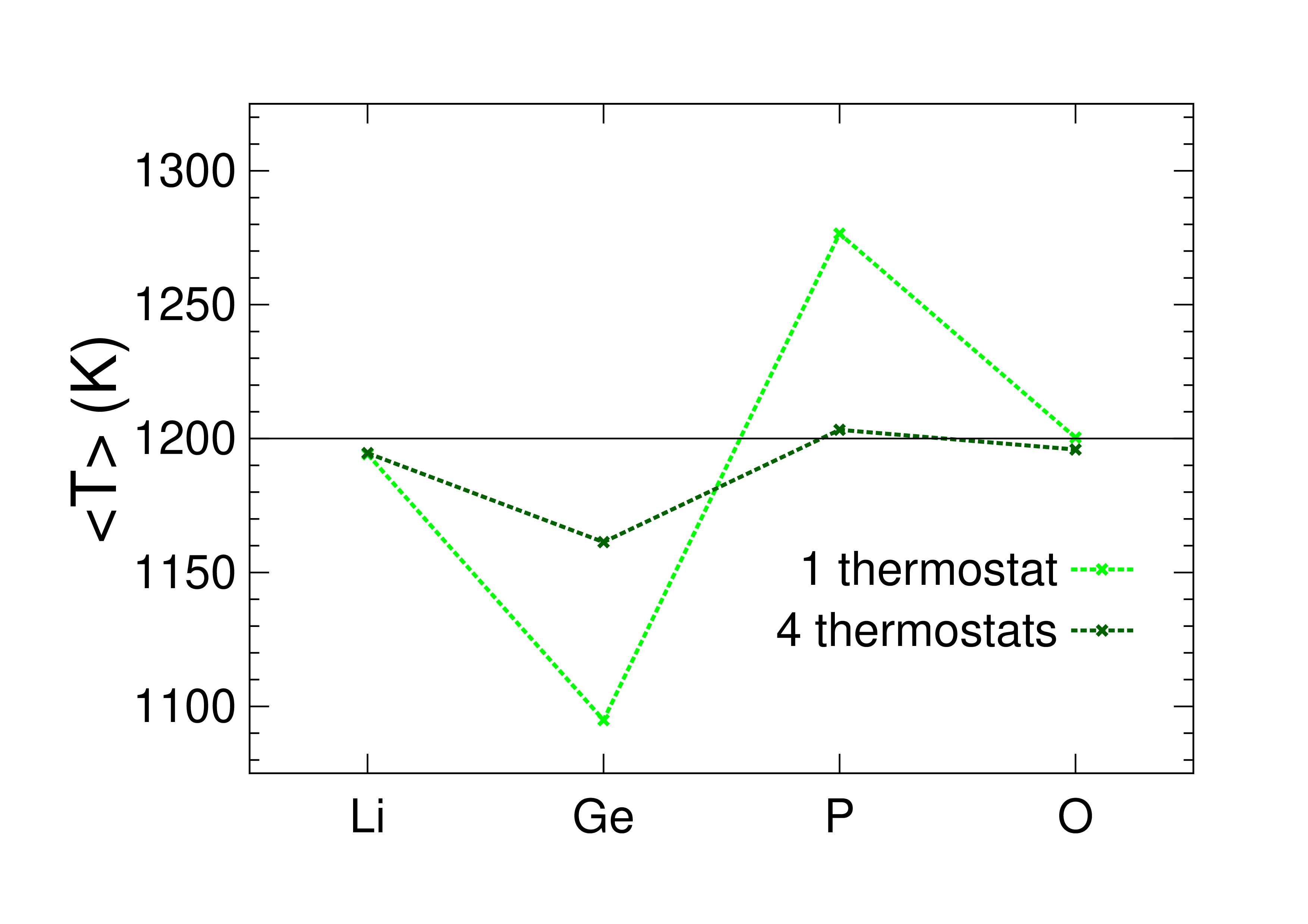}
            \caption{}
            \label{subfig:partial_temperatures_b}
    \end{subfigure}
    \vspace{1cm}
        \caption{Average temperatures of the atomic species during NVT simulations at the two ending temperatures 600 K (a) and 1200 K (b). In each case results are reported with one thermostat for the whole system (light-green) and with one thermostat for each atomic species (dark-green).}
\label{fig:partial_temperatures}
\end{figure}

\begin{figure} [H]
    \centering
    \includegraphics[width=1.1\linewidth]{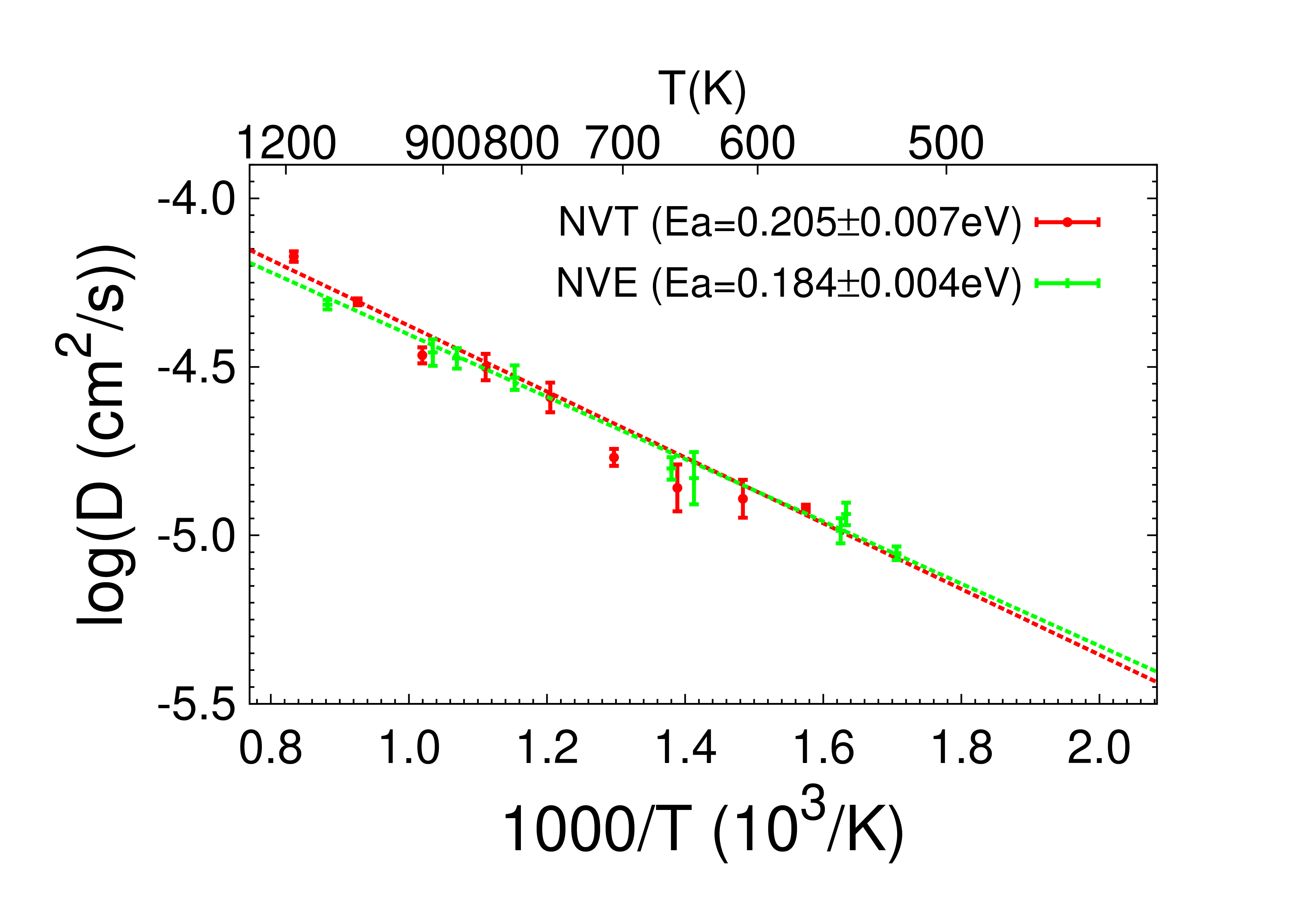}
    \caption{Arrhenius plots and corresponding activation energies for Li diffusion in t-LGPO from NVT and NVE simulations (see text).}
    \label{fig:diff_NVT_NVE}
\end{figure}

We test the reliability of the NVT simulations for the parameters chosen. The benchmark system is again t-LGPO, and we select 10 temperatures between 600 K and 1200 K (600, 635, 674, 720, 771, 830, 900, 981, 1080, 1200~K).
To test the Nose-Hoover (NH) thermostat that we use, we set up two groups of simulations, the first using one thermostat for all atoms and the second group with a different thermostat per each atomic species. 
For each of these simulations the thermostat is set to three different frequencies \cite{martyna1992nose,hunenberger2005thermostat} (10, 17 and 24~THz), ranging from maximum to zero overlap with the Li power spectra obtained from the calculations \cite{marcolongo2017ionic}.
Both thermostat's frequency and atomic-type specificity are found not to affect significantly the diffusion results, as reported in Fig.~\ref{fig:diff_NVT_frequencies}.
However, as shown in Fig.~\ref{fig:partial_temperatures}, we note that thermostatting each atomic species individually is a more efficient technique for bringing the heavier species (Ge and P) to the desired temperature. A possible explanation is that the thermostat is tuned to the typical frequencies of Li (i.e. the lightest atom in the system), thus, if used universally for the whole system, it is expected to be more effective for Li (and O) than for P and Ge thermalization (Fig.~\ref{fig:partial_temperatures}).
From the above tests we choose a species-specific thermostat, working at $\nu = 17$~THz.
Subsequently, to test the effect of the thermostat on the diffusion results for the t-LGPO system, we perform a $\sim$160 ps microcanonical evolution after each NVT run at the chosen frequency and specificity,
in a similar set up as in Ref. \cite{marcolongo2017ionic}.
 Comparison between NVT and NVE simulations for t-LGPO, reported in Fig.~\ref{fig:diff_NVT_NVE} for the diffusion coefficients, 
 shows that the thermostat doesn't influence the simulations to a significant degree. 
 \section{\label{Diffusion}Determination of diffusion coefficients and their statistical variances from simulations}
\subsection{Determination of diffusion coefficients}
The tracer diffusion coefficient $D_{tr}$ of a mobile species is the time derivative of its tracer mean square displacement $MSD_{tr}$ over a sufficiently long period of time (Einstein relation \cite{einstein1905molekularkinetischen}):
\begin{equation} 
D_{tr} = \lim_{t\to\infty} \frac{1}{6} \frac{d}{dt}{ MSD_{tr}(t)} =
\lim_{t\to\infty} \frac { MSD_{tr}(t)}{6t} .
\label{eqD}
\end{equation}
The time interval has to be long enough
for the system to be effectively out of the ballistic regime and to have entered a diffusive regime, so that we can fit $MSD_{tr}(t)$ to a line as in eq.~(\ref{eqD})
 \cite{chandler1987introduction,frenkel2001understanding,he2018statistical}.
 \\
The expression for $MSD_{tr}(t)$ is a mean square displacements  (over the atoms) that is also averaged over the possible starting times $t^\prime$ allowed by the elapsed time $t$:
\begin{equation}
MSD_{tr}(t) = \frac{1}{N} \displaystyle\sum_{i}^{N} \big\langle \big|\mathbf{R}_{i}(t^\prime +  t)]-\mathbf{R}_i(t^\prime)\big|^{2}\big\rangle .
\label{eqMSD}
\end{equation}
In eq.(\ref{eqMSD}) $N$ is the number of mobile particles,  $\mathbf{R}_{i}$ is the instantaneous position of the $i$-th particle
and $\big\langle...\big\rangle$ the average
over the times $t^\prime$.
This average over $t^\prime$ is also called the statistical average, and is related to the use of molecular dynamics simulations as phase space sampling of microscopic quantities through trajectory averages \cite{allen2017computer}. 
\\
To extract $D_{tr}$ following eqs.~(\ref{eqD}) and (\ref{eqMSD}),  we first need to assess the linearity region of $MSD_{tr}(t)$ in eq.~(\ref{eqD}), avoiding on one side (low elapsed times $t$) the ballistic regime, and on the other side (high  elapsed times $t$) the statistical inaccuracy \cite{he2018statistical}. The ballistic regime is usually over, in the systems studied in this work, at $t>2$ ps, whereas the length of the here calculated CP  trajectories allows us to consider a maximum elapsed time ranging from $~80$ ps to $~250$ ps, depending on the particular system and temperature considered.
\subsection{Determination of the statistical variance of diffusion coefficients}
\begin{figure}
\centering
\includegraphics[width=1.\linewidth]{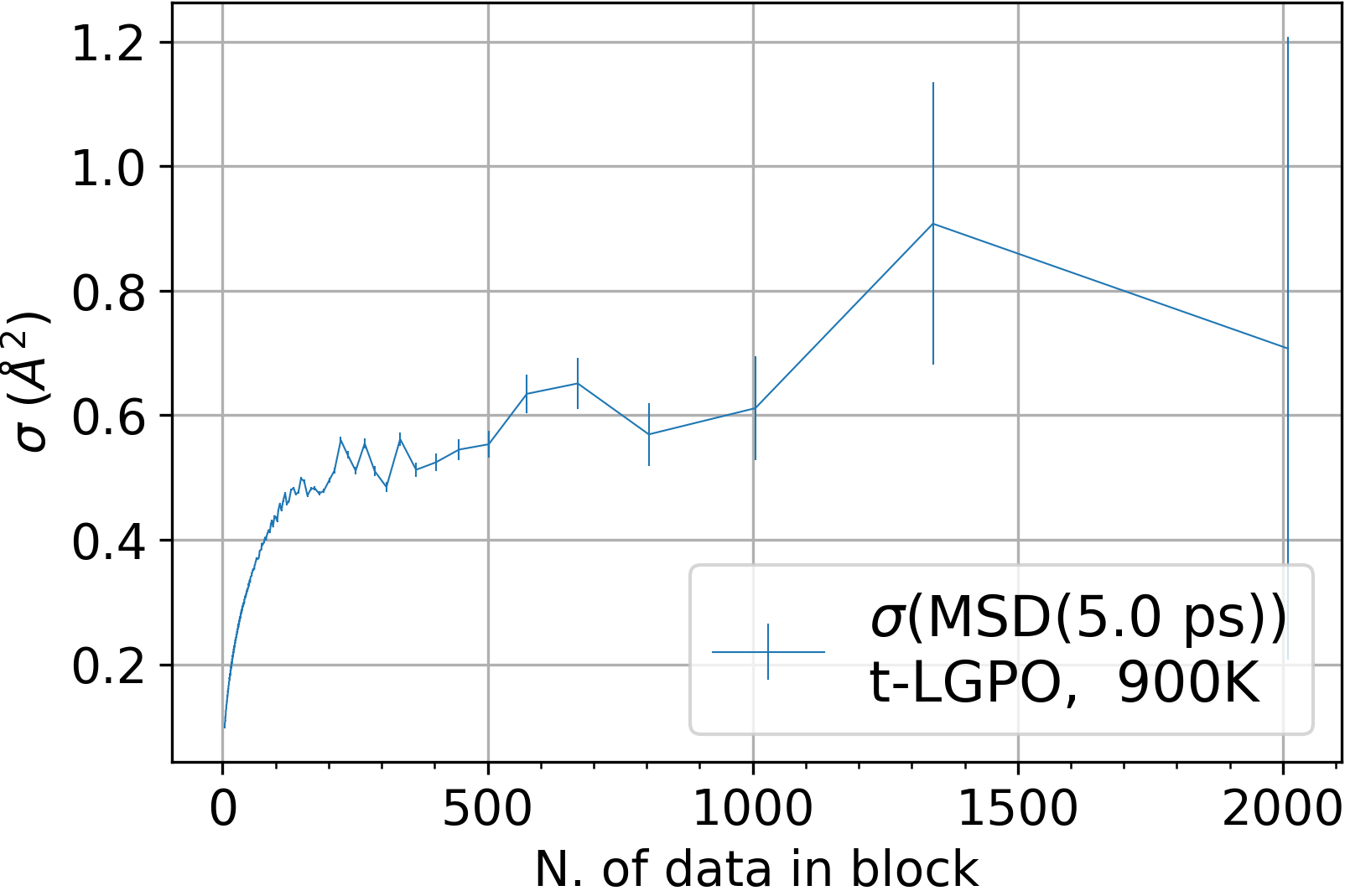}
\vspace{1cm}
    \caption{Dependence of the standard deviation $\sigma$ on the number of data in block for the Li $MSD_{tr}(t)$ at $t=$ 5 ps (eq.~(\ref{eqMSD})) of t-LGPO at 900 K.}
    \label{fig:sigma_berend}
\end{figure}

\begin{figure}
\centering
\includegraphics[width=1.\linewidth]{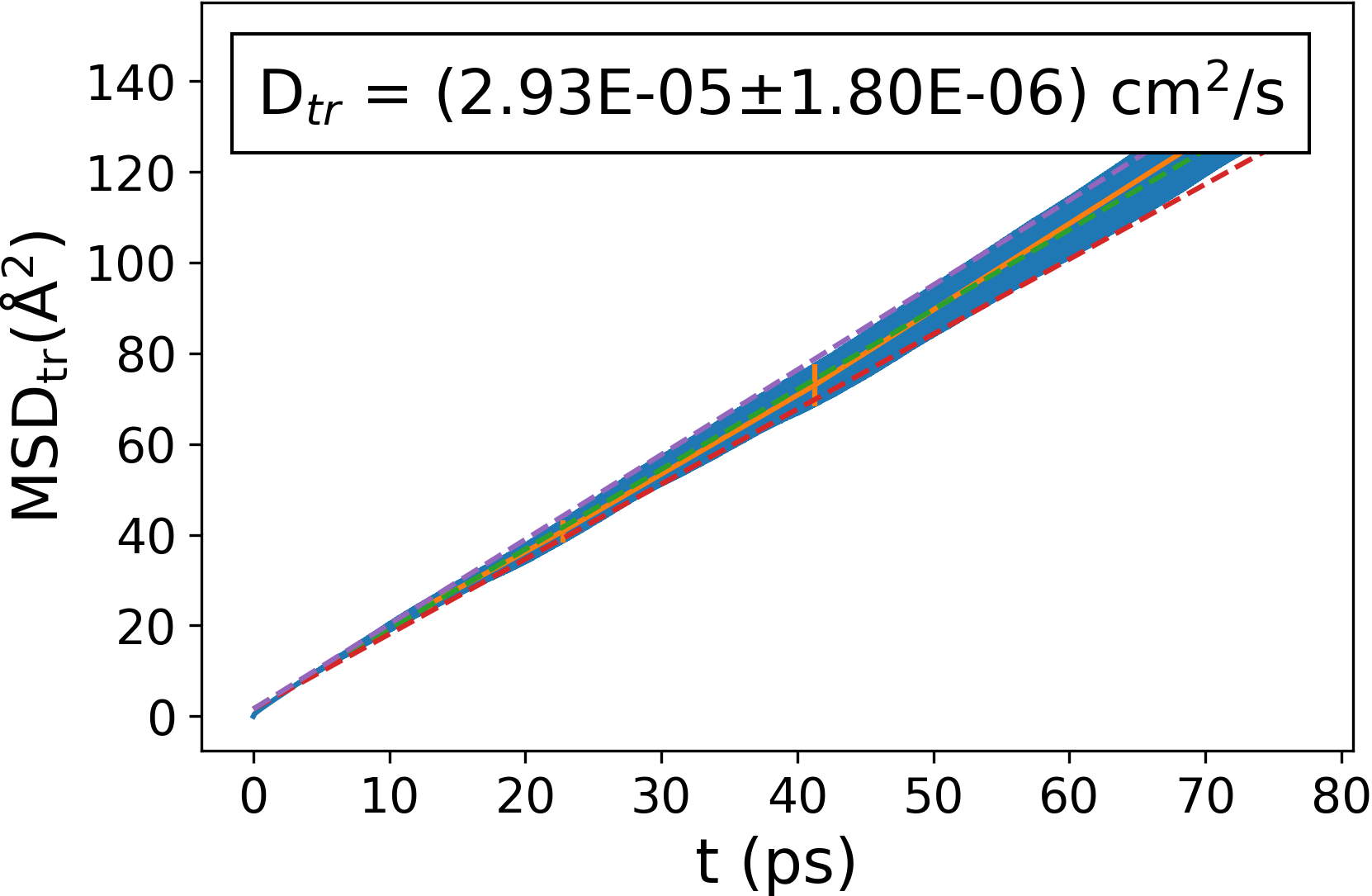}
\vspace{1cm}
    \caption{$MSD_{tr}(t)$ for t-LGPO at 900 K and extracted $D_{tr}$ from decorrelated data of this simulation.} 
    \label{fig:MSD_uncorrelated}
\end{figure}

The usual practice in estimating the statistical error on the diffusion coefficient is to 
divide the trajectory into parts, calculate $D_{tr}$ in each of them, and consider the RMS of the set $\{D_{tr,m}\}$ \cite{kahle2020high,he2018statistical}.
Here we follow a slightly different approach, considering the trajectory as a whole and estimating errors and correlation times from a block analysis.
The idea behind any block analysis is that a set of correlated data can be recast to a smaller set of uncorrelated data, each of which is the mean of a subset (block) of data of the whole set \cite{frenkel2001understanding}. 
The way to choose the number of blocks (or alternatively the number of data $N_{data}$ in each block) can be determined by the need to have a stable variance of the desired quantity over $N_{data}$  (ref.~\cite{frenkel2001understanding} pp. 98-100). If the blocks are too small the data are still too correlated (the variance goes to zero and is not trustable), whereas if the blocks are too large we waste trustable experiments, or, in other words, we overestimate the error on the variance.  
In Fig.~\ref{fig:sigma_berend} the standard deviation of $MSD_{tr}(t)$ at $t=5$ ps is reported, for t-LGPO at 900 K, as a function of the number of data in each block, where the highest number of data in block corresponds to 2 blocks and the statistical window considered is $\sim 80$ ps. As mentioned above \cite{frenkel2001understanding}, for very small blocks $\sigma(MSD_{tr})$ goes to zero, whereas for very large blocks the variance has large errors. At intermediate values ($100<N_{data}<500$)  $\sigma(MSD_{tr})$ is roughly a constant, and one should choose these block sizes in order to have a meaningful $MSD_{tr}$ variance.
Let us call $\overline{\sigma}$ this constant value and  $\overline{N}_{data}$ the  lowest  value of $N_{data}$ that corresponds to it.  
\\
Since $MSD_{tr}(t)$ depends on time, one should in principle repeat the analysis of Fig.~\ref{fig:sigma_berend} to determine the size of the blocks and the variance of $MSD_{tr}$ at each elapsed time $t$. However, one can observe that $\overline{\sigma}(MSD_{tr}(t))$  is a linear function of $t$ as is $MSD_{tr}(t)$ itself, so that one can extrapolate  $(\overline{\sigma}, \overline{N}_{data}$)
at all times $t$ by knowing it at 2 times. We can use this information as a way to build a set of uncorrelated $\{MSD_{tr, i}\}$ data, where $MSD_{tr, i}$ has variance $\overline{\sigma}_i$ and is separated from ${MSD_{tr, i+1}}$ by $\overline{N}_{data, i}$.
In Fig.~\ref{fig:MSD_uncorrelated} we report $MSD_{tr}(t)$ for t-LGPO at 900 K after such a study, where the number of uncorrelated data $\{MSD_{tr, i}\}$, taken at intervals of $\overline{N}_{data, i}$ from each other, is 4. These 4 data of $MSD_{tr}(t)\pm\sigma(t)$ were used to perform a linear regression and obtain the diffusion coefficient as in eq.~(\ref{eqD}), giving the value reported in Fig.~\ref{fig:MSD_uncorrelated}.
\\
The same procedure applies to the charge diffusion coefficients (eqs. (7) and (8) in the main text).
\bibliography{biblio}

\clearpage

\clearpage
\clearpage
\clearpage
\clearpage
\clearpage
